\documentclass{statsoc}

\usepackage{algorithm}
\usepackage{algorithmic}
\usepackage{amscd}
\usepackage{amsfonts}
\usepackage{amsmath}
\usepackage{comment}
\usepackage{amssymb}
\let\proof\relax

\usepackage{amsthm}
\usepackage{array}
\usepackage{bbm}
\usepackage{booktabs}
\usepackage{calc}
\usepackage{cite}
\usepackage{enumerate}
\usepackage{enumitem}
\usepackage{epsf}
\usepackage{etoolbox}
\usepackage[T1]{fontenc}
\usepackage[a4paper]{geometry}
\usepackage{graphicx}
\usepackage[space]{grffile}
\usepackage[colorlinks=false]{hyperref}
\usepackage{leftidx}
\usepackage{longtable}
\usepackage{multirow}
\usepackage{placeins}
\usepackage{psfrag}
\usepackage{relsize}
\usepackage[group-separator={,}]{siunitx}
\usepackage{subfigure}
\usepackage{tabulary}
\usepackage{url} 
\usepackage{xcolor}
\usepackage{xpatch}
\usepackage{xspace}

\makeatletter
\patchcmd{\@makecaption}
  {\parbox}
  {\advance\@tempdima-\fontdimen2} 
  {}{}
\makeatother 

\newtheoremstyle{seriesb}
  {}
  {2.5em}
  {\itshape}
  {}
  {\bfseries}
  {.}
  {.5em}
  {}
\theoremstyle{seriesb}

\xpretocmd{\proof}{\setlength{\parindent}{0pt}}{}{}

\newcommand{\noop}[1]{}

\newtheorem{theorem}{\bf Theorem}

\newtheorem{condition}{\bf Condition}

\newtheorem{corollary}{\bf Corollary}

\newtheorem{apx_lemma}{\bf Lemma}[section]

\newtheorem{apx_definition}{\bf Definition}[section]
\newtheorem{apx_condition}{\bf Condition}[section]
\newtheorem{apx_assumption}{\bf Assumption}[section]

\newtheorem{apx_corollary}{\bf Corollary}[section]

\newcommand{\target}[2]{f_{#1}\!\left(#2\right)} 
\makeatletter
\renewcommand{\boxed}[2][\fboxsep]{{%
  \setlength{\fboxsep}{#1}\fbox{\m@th$\displaystyle#2$}}}
\makeatother

\newcommand{\brown}[1]{\mathbb{W}_{#1}} 
\newcommand{\brownbar}{\bar{\mathbb{W}}} 
\newcommand{\fusion}{\mathbb{F}} 
\newcommand{\proposal}{\mathbb{P}} 
\newcommand{\layer}{{\cal R}} 
\newcommand{\lawK}{\mathbb{K}} 
\newcommand{\lawU}{\mathbb{U}} 

\newcommand{\cores}{^{C}_{c=1}} 
\newcommand{\partition}{{\cal P}}
\newcommand{\Crange}{c\in\{1,\dots,C\}}
\newcommand{\Trange}{t \in [0,T]}

\newcommand{\tX}{\ensuremath{\mathfrak{X}}\xspace} 
\newcommand{\x}{{\boldsymbol x}} 
\newcommand{\X}[2]{{\x}_{#1}^{\ifthenelse{\equal{#2}{}}{}{(#2)}}} 
\newcommand{\vecX}[1]{{ \vec {\x}}_{#1}} 
\newcommand{\barX}[1]{{ \bar {\x}}_{#1}} 
\newcommand{\cx}{{\boldsymbol X}} 
\newcommand{\cX}[2]{{ \cx}_{#1}^{\ifthenelse{\equal{#2}{}}{}{(#2)}}} 
\newcommand{\cvecX}[1]{{ \vec {\cx}}_{#1}} 
\newcommand{\cbarX}[1]{{ \bar {\cx}}_{#1}} 

\newcommand{\cw}{{\boldsymbol W}} 
\newcommand{\cW}[2]{{ \cw}_{#1}^{\ifthenelse{\equal{#2}{}}{}{(#2)}}} 
\newcommand{\cvecW}[1]{{ \vec {\cw}}_{#1}} 
\newcommand{\y}{{\boldsymbol y}} 
\newcommand{\bolda}{{\boldsymbol a}} 
\newcommand{\boldxi}{{\boldsymbol \xi}} 
\newcommand{\boldtheta}{{\boldsymbol \theta}} 
\newcommand{\boldeta}{{\boldsymbol \eta}} 
\newcommand{\boldz}{{\boldsymbol z}} 
\newcommand{\boldbeta}{{\boldsymbol \beta}} 
\newcommand{\boldv}{{\boldsymbol v}} 
\newcommand{\boldzero}{{\boldsymbol 0}} 
\newcommand{\boldOmega}{{\boldsymbol \Omega}} 

\newcommand{\bM}{{\boldsymbol M}} 
\newcommand{\cM}[2]{{ \bM}_{#1}^{\ifthenelse{\equal{#2}{}}{}{(#2)}}} 
\newcommand{\cvecM}[1]{{ \vec {\bM}}_{#1}} 
\newcommand{\bV}{{\boldsymbol V}} 
\newcommand{\cV}[1]{{ \bV}_{#1}} 
\newcommand{\identity}[1]{\mathbf{I}_{#1}} 
\newcommand{\onemat}[1]{\mathbf{J}_{#1}} 
\newcommand{\bSigma}{\mathbf{\Sigma}}

\newcommand{\phifn}[2]{\phi_{#1}\!\left(#2\right)} 
\newcommand{\Phifn}[1]{\Phi_{#1}} 
\newcommand{\phihatfn}[2]{\hat{\phi}_{#1}\!\left(#2\right)} 

\newcommand{\shl}{\ensuremath{\text{SH}(\lambda)}\xspace} 
\newcommand{\ssh}{\ensuremath{\text{SSH}(\gamma)}\xspace} 
\newcommand{\ssj}[1]{\ensuremath{\text{CESS}_{#1}}\xspace} 

\newcommand{\gpea}{UE-$a$}
\newcommand{\gpeb}{UE-$b$}
\newcommand{\gpeas}[1]{\ensuremath{\tilde{\rho}^{(a)}_{#1}\xspace}}
\newcommand{\gpebs}[1]{\ensuremath{\tilde{\rho}^{(b)}_{#1}\xspace}}

\newcommand{\mvbar}{\middle\vert} 
\newcommand\rnd{Radon-Nikod{\'y}m derivative\xspace} 
\newcommand{\ud}{\,\mathrm{d}} 
\newcommand{\iid}{\text{iid}}
\newcommand*\laplace{\mathop{}\!\mathcal{4}}
\newcommand{\Rone}{\ensuremath{\mathbbm{R}}} 
\newcommand{\Rd}{\ensuremath{\mathbbm{R}^d}} 
\newcommand{\bigO}{{\cal O}} 
\newcommand{\expect}{\mathbb{E}} 
\newcommand{\prob}{\mathbbm{P}} 
\newcommand{\normal}{{\cal N}}
\newcommand{\uniform}{{\cal U}}

\newcommand{\algref}[1]{\hyperref[#1]{Algorithm \ref*{#1}}}
\newcommand{\stepref}[1]{\hyperref[#1]{Step \ref*{#1}}}
\newcommand{\algstref}[2]{\hyperref[#2]{Algorithm \ref*{#1} Step \ref*{#2}}}
\newcommand{\figref}[1]{\hyperref[#1]{Figure \ref*{#1}}}
\newcommand{\tabref}[1]{\hyperref[#1]{Table \ref*{#1}}}
\newcommand{\apxref}[1]{\hyperref[#1]{Appendix \ref*{#1}}}
\newcommand{\apxrefpl}[2]{Appendices \hyperref[#1]{\ref*{#1}} and \hyperref[#2]{\ref*{#2}}}
\newcommand{\secref}[1]{\hyperref[#1]{Section \ref*{#1}}}
\newcommand{\propref}[1]{\hyperref[#1]{Proposition \ref*{#1}}}
\newcommand{\prinref}[1]{\hyperref[#1]{Principle \ref*{#1}}}
\newcommand{\conref}[1]{\hyperref[#1]{Condition \ref*{#1}}}
\newcommand{\resref}[1]{\hyperref[#1]{Result \ref*{#1}}}
\newcommand{\defnref}[1]{\hyperref[#1]{Definition \ref*{#1}}}
\newcommand{\thmref}[1]{\hyperref[#1]{Theorem \ref*{#1}}}
\newcommand{\lemref}[1]{\hyperref[#1]{Lemma \ref*{#1}}}
\newcommand{\corref}[1]{\hyperref[#1]{Corollary \ref*{#1}}}
\newcommand{\remref}[1]{\hyperref[#1]{Remark \ref*{#1}}}
\newcommand{\assref}[1]{\hyperref[#1]{Assumption \ref*{#1}}}

\title[Bayesian Fusion]{Bayesian Fusion: Scalable unification of distributed statistical analyses}
\author[Dai, Pollock and Roberts]{Hongsheng Dai\footnote{\emph{Address for correspondence:} Hongsheng Dai, Department of Mathematical Sciences, University of Essex, Wivenhoe Park, Colchester, CO4 3SQ, U.K. \emph{Email:} \texttt{hdaia@essex.ac.uk}.}}
\address{University of Essex, Colchester, U.K.}
\author[Dai, Pollock and Roberts]{Murray Pollock}
\address{Newcastle University, Newcastle-upon-Tyne, \emph{and} The Alan Turing Institute, London, U.K.}
\author[Dai, Pollock and Roberts]{Gareth O. Roberts}
\address{University of Warwick, Coventry, \emph{and} The Alan Turing Institute, London, U.K.}
\begin{document}
\maketitle


\begin{abstract}
There has recently been considerable interest in addressing the problem of unifying distributed statistical analyses into a single coherent inference. This problem naturally arises in a number of situations, including in big-data settings, when working under privacy constraints, and in Bayesian model choice. The majority of existing approaches have relied upon convenient \emph{approximations} of the distributed analyses. Although typically being computationally efficient, and readily scaling with respect to the number of analyses being unified, approximate approaches can have significant shortcomings -- the quality of the inference can degrade rapidly with the number of analyses being unified, and can be substantially biased even when unifying a small number of analyses that do not concur. In contrast, the recent \emph{Fusion} approach of \citet{jap:dpr19} is a rejection sampling scheme which is readily parallelisable and is \emph{exact} (avoiding any form of approximation other than Monte Carlo error), albeit limited in applicability to unifying a small number of low-dimensional analyses. In this paper we introduce a practical \emph{Bayesian Fusion} approach. We extend the theory underpinning the Fusion methodology and, by embedding it within a sequential Monte Carlo algorithm, we are able to recover the \emph{correct} target distribution. By means of extensive guidance on the implementation of the approach, we demonstrate theoretically and empirically that Bayesian Fusion is robust to increasing numbers of analyses, and coherently unifying analyses which do not concur. This is achieved while being computationally competitive with approximate schemes.
\end{abstract}


{\it Keywords: Bayesian inference; Distributed data; Fork-and-join; Langevin diffusion; Sequential Monte Carlo}


\section{Introduction} \label{sec:intro}

There has recently been considerable interest in developing methodology to combine \emph{distributed} statistical inferences, into a single (Bayesian) inference. This distributed scenario can arise for a number of practically compelling reasons. For instance, it can arise by construction in large data settings where, to circumvent the memory constraints on a single machine, we split the available data set across $C$ machines (which we term \emph{cores}) and conduct $C$ separate inferences \citep{ijmsem:setal16}. Other modern instances appear when working under confidentiality constraints, where pooling the underlying data would be deemed a data privacy breach (for instance, \citet{sc:ye19}), and in model selection \citep{arxiv:bar20}. More classical instances of this common scenario appear in Bayesian meta-analysis (see for example \citet{smmr:f93,sim:sst95}), and in constructing priors from multiple expert elicitation \citep{bk:b80,ss:gz86}.

In this article we present general statistical methodology to address this \emph{fusion problem}. We term each of the $C$ inferences across $C$ cores that we wish to unify a \emph{sub-posterior}, denoted by $f_c(\x)$ for $\Crange$. The natural manner to unify the sub-posteriors is by considering the product pooled posterior density (which we term the \emph{fusion density}), 
\begin{align}
\target{}{\x} \propto \target{1}{\x} \dots \target{C}{\x}. \label{eq:prod}
\end{align}

Our goal is to produce a Monte Carlo sample from \eqref{eq:prod}. For convenience, and common to many existing approaches \citep{ijmsem:setal16,arxiv:n13,sc:xl19,arxiv:wd13}, we will assume in this article that independent samples from each sub-posterior are readily available and it is possible to evaluate each sub-posterior point-wise. As discussed later, neither of these are limiting factors for our methodology. 

Specific applications, such as those we used to introduce the fusion problem, have a number of specific constraints and considerations unique to them. For instance, in the large data setting particular consideration may be given to latency and computer architectures \citep{ijmsem:setal16}, whereas in the confidentiality setting of \citet{sc:ye19} one may be constrained in the number and type of mathematical operations conducted. Indeed, the majority of the current literature addressing the fusion problem has been developed to address specific applications. Our focus in this paper will not concern any particular application, but rather on methodology for the general fusion problem, which in principle could be applied and adapted to to the statistical contexts we describe. Some general discussion on particular applications is given in \secref{sec:practical}, following the introduction of our methodology. 

The methodologies proposed in the literature to address the fusion problem are mostly approximate, often supported by underpinning theory which ensures their limiting unbiasedness in an appropriate asymptotic limit. While these methods are often computationally efficient and generally effective, it is generally difficult to assess the extent of the biases introduced by these method, and equally difficult to correct for these biases. One of the earliest, and most widely used method for dealing with the fusion problem is the Consensus Monte Carlo (CMC) method \citep{ijmsem:setal16,bjps:s17}. This method weights samples from individual sub-posteriors in a way which would be completely unbiased if each sub-posterior was indeed Gaussian. This is attractive in the large data context which motivated their work. On the other hand, outside the Gaussian context CMC can be very biased \citep{neurips:wghd15,aistats:sctd16}. An alternative method involving aggregation techniques based on Weierstrass transforms to each sub-posterior was proposed in \citet{arxiv:wd13}. In comparison to CMC, the Weierstrass sampler is computationally more expensive, although it tends to produce less biased results in the context of non-Gaussian sub-posteriors. We shall use these two methods as benchmarks to compare the methodology we propose here.

Much of the existing approximate literature has been focused on distributed large data settings, and as a consequence there has been particular attention on developing \emph{embarrassingly parallel} procedures, where communication between cores is limited to a single unification step. Often termed as \emph{divide-and-conquer} approaches (although strictly speaking \emph{fork-join} approaches), recent contributions include \citet{arxiv:n13} who constructs a kernel density estimate for each sub-posterior to reconstruct the posterior density. Other approaches which construct approximations directly from sub-posterior draws include \citet{icml:ssld14,aistats:sctd16,neurips:wghd15,ieee:sa13,neurips:ad11,arxiv:n13,sc:xl19} and \citet{arxiv:wd13}. Alternative non-embarrassingly parallel approaches are discussed extensively in  \citet{jasa:jly18} and \citet{neurips:xltzz14}. Within a hierarchical framework \citet{jcgs:rjlw20} (and subsequently \citet{ieee:vdc19}) introduce a methodology in which a smoothed approximation to \eqref{eq:prod} can be obtained if increased communication between the cores is permitted.

In contrast to approximate methods, the \emph{Monte Carlo Fusion} approach recently introduced by \citet{jap:dpr19} provides a theoretical framework to sample independent draws from \eqref{eq:prod} \emph{exactly} (without any form of approximation). Monte Carlo Fusion is based upon constructing a rejection sampler on an auxiliary space which admits \eqref{eq:prod} as a marginal. However, unlike approximate approaches there are considerable computational challenges with Monte Carlo Fusion. In particular, the scalability of the methodology in terms of the number of sub-posteriors to be unified, increasing dis-similarity in the sub-posteriors, and the dimensionality of the underlying fusion target density, all inhibit the practical adoption of the methodology. The challenge that we address successfully in this paper is to devise a methodology which shares the consistency properties of Monte Carlo Fusion while sharing the scalability behaviour of of the approximate alternatives. 

In this paper we substantially reformulate the theoretical underpinnings of the auxiliary construction used in \citet{jap:dpr19} to support the use of scalable Monte Carlo methodology. There are a number of substantial and novel contributions which we list here for clarity.
\begin{itemize}[leftmargin=*]
\item We show that it is possible to sample from \eqref{eq:prod} by means of simulating for the probability measure of a forward stochastic differential equation (SDE).
\item Based upon the SDE formulation we further develop a Sequential Monte Carlo (SMC) algorithm to sample consistently from \eqref{eq:prod}, in a methodology which we term \emph{Bayesian Fusion}.
\item We develop theory to show that Bayesian Fusion is robust in the large $C$ and increasingly discrepant sub-posteriors scenarios, and as a consequence considerably more efficient when used in practical Bayesian settings.
\item For practitioners we provide practical guidance for setting algorithm hyperparameters, which will (approximately) optimise the efficiency of our approach.
\item Finally, we provide extensive pedagogical examples and real-data applications to contrast our methodology with existing approximate and exact approaches, and to study empirically the scaling properties of our approach and verify it attains that given by our theoretical guidance. 
\end{itemize}

In the next section we present the theory that underpins Bayesian Fusion, together with methodology and pseudo-code for its implementation in \secref{sec:methodology}. We provide guidance on implementing Bayesian Fusion in \secref{sec:guidance}, which includes selection of user-specified parameters in Sections \ref{sec:Tguide} and \ref{sec:nPguide}, studies of the robustness of the algorithm with respect to how similar the sub-posteriors are in Sections \ref{sec:shl_case} and \ref{sec:ssh_case}, and extensive discussion of practical considerations in Sections \ref{sec:intervals} and \ref{sec:practical}. \secref{sec:comparisons} studies the performance of our methodology in comparison to competing methodologies for idealised models and a synthetic data set, and in \secref{sec:examples} its performance in a number of real data set applications. We conclude in \secref{sec:conclusions} with discussion and future directions. We suppress all proofs from the main text, which are instead collated in the appendices. The appendices also include some discussion of the underlying diffusion theory and assumptions (\apxref{apx:bfproof}), theory to support implementations for distributed environments in \apxref{apx:parallel}, and discussion on the application of the methodology to large data settings in \apxref{apx:scale}, and are referenced as appropriate in the main text. 


\section{Bayesian Fusion} \label{sec:theory}

Consider the $d$-dimensional posterior density $\target{}{\x}$ described in \eqref{eq:prod}. As motivated in the introduction, we want to sample from $\target{}{\x}$ by means of sampling and evaluating functionals of the available \emph{sub-posterior} densities $\target{c}{\x}$ ($\Crange$). $\target{}{\x}$ can be obtained as a marginal of a suitably chosen extended target \emph{fusion measure} on an extended state space, which we present in \thmref{thm:key}.

\begin{figure}[ht]
\begin{center}
\includegraphics[width=0.55\textwidth]{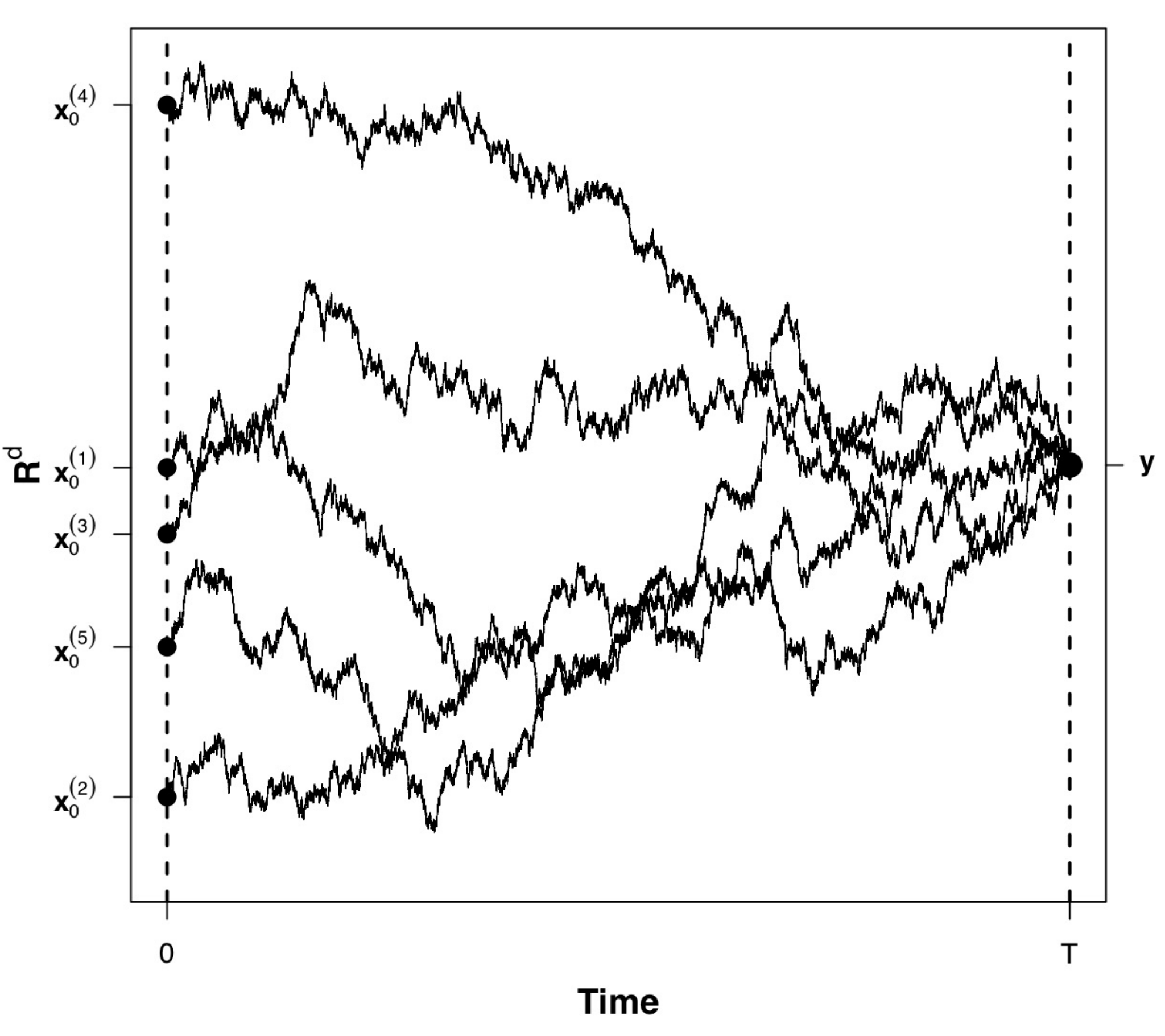}
\caption{A typical realisation of $\tX$ ($C$ interacting Markov processes)} \label{fig:original}
\end{center}
\end{figure}

To introduce the \emph{fusion measure}, we first present some notation and terminology. We term the \emph{proposal measure}, $\proposal$, to be the probability law induced by $C$ interacting $d$-dimensional parallel continuous-time Markov processes in $[0,T]$, where each process $\Crange$ is described by the following $d$-dimensional SDE,
\begin{align}
\ud \cX{t}{c} = \frac{\cbarX{t} - \cX{t}{c}}{T-t} \ud t + \ud \cW{t}{c}, \qquad \cX{0}{c} := \X{0}{c} \sim f_c, \quad \Trange, \label{eq:SDE}
\end{align}
where $\{\cW{t}{c}\}\cores$ are independent Brownian motions, and $\cbarX{t} := C^{-1}\sum\cores \cX{t}{c}$. Typical realisations of the proposal measure are denoted as $\tX := \{\vecX{t},\Trange\}$, where $\vecX{t}: = \X{t}{1:C}$ is the $dC$-dimensional vector of all processes at time $t$, with one such realisation illustrated in \figref{fig:original}. 

Interaction of the $C$ processes in a realisation of $\tX$ occurs through their average at a given time marginal ($\cbarX{t}$), and note that we have \emph{coalescence} at time $T$ ($\X{T}{1} = \dots = \X{T}{C} =: \y$). We describe in detail in \secref{sec:methodology} how to simulate from $\proposal$, but note that (critically) initialisation of the proposal measure at $t=0$ only requires independent draws from the $C$ available sub-posteriors.

Now we define the {\em fusion measure}, $\fusion$, to be the probability measure induced by the following \rnd,
\begin{align}
 \frac{\ud \fusion}{\ud\proposal}\left(\tX\right)
 & \propto \rho_0(\vecX{0}) \cdot  \prod\cores \left[\exp \left\{ - \int_0^T \phifn{c}{\X{t}{c}} \ud t\right\} \right], \label{eq:fusion_reg}
\end{align}
where $\{\X{t}{c}, \Trange\}$ is a Brownian bridge from  $\X{{0}}{c}$ to $ \X{T}{c}$, $\phifn{c}{\x}:= \laplace \target{c}{\x} / 2\target{c}{\x}$ (where $\laplace$ is the Laplacian operator), and
\begin{align}
\rho_0 := \rho_0(\vecX{0}) = \exp\left\{-\sum\cores \frac{\|\X{0}{c}-\barX{0}\| ^2}{2T}\right\}\in(0,1], \quad \text{where } \barX{0} = C^{-1}\sum\cores\X{0}{c}. \label{eq:rho0}
\end{align} 

We now establish that we can access the fusion density $f$, by means of the temporal marginal of $\fusion$ given by common value of the $C$ trajectories at time $T$.
\begin{theorem} \label{thm:key} 
Under Assumptions \ref{ass:cont_diff} and \ref{ass:lower_bound} given in \apxref{apx:bfproof}, then with probability $1$ we have that under the fusion measure, $\fusion$, $\y:= \X{T}{1} = \dots = \X{T}{C}$ and this common value has density $f$.
\proof See \apxref{apx:bfproof}. \qed
\end{theorem}


\subsection{Simulation of \texorpdfstring{$f$}{f} by means of simulating from the fusion measure \texorpdfstring{$\fusion$}{F}} \label{sec:methodology}

As suggested by \thmref{thm:key} we could simulate from the desired $f$ in \eqref{eq:prod} by simulating  $\tX\sim\fusion$ and simply retaining its time $T$ marginal, $\y$. However, direct simulation of $\fusion$ will typically not be possible, and so we now outline general methodology to simulate $\fusion$ indirectly (and so by extension $f$). In particular, we show that we can simulate from $\fusion$ by means of a rejection sampler with proposals $\tX\sim\proposal$ which are accepted with probability proportional to the \rnd given in \eqref{eq:fusion_reg}. 

For the purposes of the efficiency of the methodology we will subsequently develop, we will consider the simulation of $\proposal$ and $\fusion$ at discrete time points given by the following auxiliary temporal partition,
\begin{align}
\partition &=\{t_0, t_1, \dots, t_n: 0=: t_0 <t_1<\dots<t_n : =T\}, \label{eq:partition}
\end{align}
noting that ultimately we only require the time $T$ marginal corresponding to the $n$th temporal partition. For simplicity we will suppress subscripts when considering the Markov processes at times coinciding with the partition, denoting $\X{t_j}{c}$ as $\X{j}{c}$, and $\vecX{t_j}$ as $\vecX{j}$. We further denote $\Delta_j:=t_j-t_{j-1}$. 

We begin by considering simulating exactly $\tX\sim\proposal$ at the points given by the temporal partition, $\partition$. To do so we simply note that the SDE given in \eqref{eq:SDE} is linear and therefore describes a Gaussian process, and its finite-dimensional distributions are explicitly available. 

\begin{theorem} \label{thm:bf} 
If $\tX$ satisfies \eqref{eq:SDE} then under the proposal measure, $\proposal$, we have:
\begin{enumerate}
\item For $s<t$
\begin{align}
\left.\cvecX{t} \mvbar \left(\cvecX{s} = \vecX{s}\right) \right.
& \sim \normal\left(\cvecM{s,t}, \cV{s,t} \right),
\end{align}
where $\normal$ is a multivariate Gaussian density, $\cvecM{s,t}= (\cM{s,t}{1}, \dots \cM{s,t}{C})$ with 
\begin{align}
\label{eq:bfM}
    \cM{s,t}{c} & ={\frac{T-t}{T-s}\X{s}{c} +  \frac{t-s}{T-s}\barX{s} },
\end{align}
and where $\cV{s,t} = \bSigma \otimes \identity{d\times d}$ 
with $\bSigma = (\Sigma_{ij})$ being a $C\times C$ matrix given by
\begin{align}
\Sigma_{ii}
= \frac{(t-s)\cdot (T-t)}{T-s} + \frac{(t-s)^2}{C(T-s)}, \qquad 
\Sigma_{ij} = \frac{(t-s)^2}{C(T-s)}. \label{eq:bfS2}
\end{align}
\item
For every $\Crange$, the distribution of $ \{\cX{u}{c}, s\leq u\leq t\}$ given endpoints $\cX{s}{c}=\X{s}{c}$ and $\cX{t}{c}=\X{t}{c}$ is a Brownian bridge, so that
\begin{align}
\left. \cX{u}{c} \mvbar \left(\X{s}{c}, \X{t}{c}\right) \right.
& \sim \normal\left(\frac{(t-u) \X{s}{c} + (u-s) \X{t}{c}}{t-s}, \frac{(u-s)(t-u)}{t-s}\identity{d\times d}\right).
\end{align}
\end{enumerate}
\proof See \apxref{apx:bfproof}. \qed
\end{theorem}

\begin{figure}[ht]
\begin{center}
\includegraphics[width=0.55\textwidth]{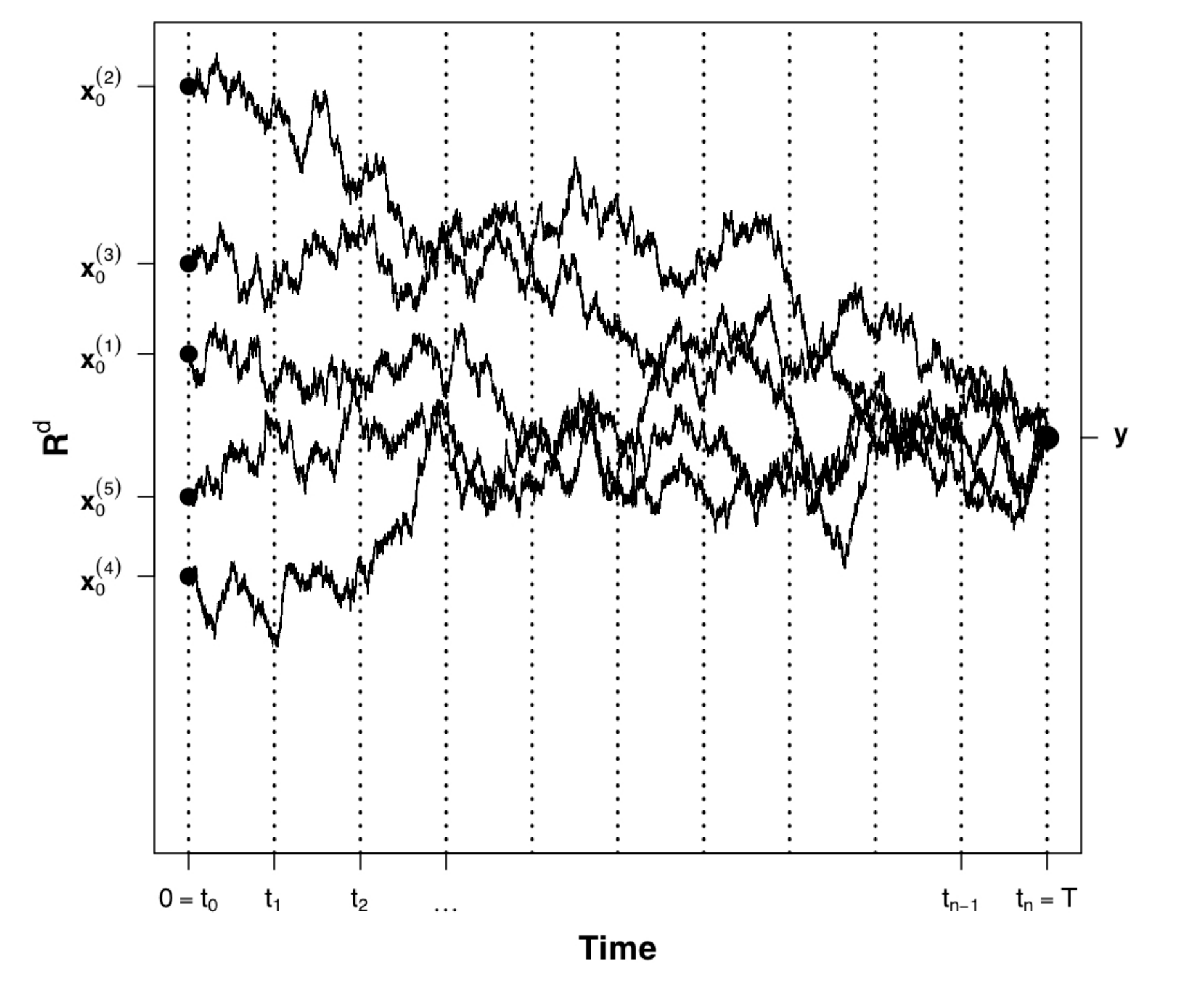}
\caption{Illustration of the $d(nC+1)$-dimensional density corresponding to the time marginals of a typical realisation of $\tX$ given by the temporal partition $\partition$.} \label{fig:partition_object}
\end{center}
\end{figure}

To simplify the presentation of the methodology, we now restrict our attention to the $d(nC+1)$-dimensional density of the $C$ $d$-dimensional Markov processes at the $(n+1)$ time marginals given by the temporal partition under $\proposal$. An illustration of this is given in \figref{fig:partition_object}. As a consequence of \thmref{thm:bf} we have,
\begin{align}
h(\vecX{0}, \dots, \vecX{n-1}, \y) & \propto 
 \prod\cores \left[f_c\big(\X{0}{c}\big) \right] \cdot \prod_{j=1}^n \normal\left( \vecX{j}; \cvecM{j}, \cV{j} \right), \label{eq:prop_bf}
\end{align}
where to simplify notation we have $\cvecM{j}:=\cvecM{t_{j-1},t_j}$ and $\cV{j}:=\cV{t_{j-1},t_j}$.

By factorising \eqref{eq:fusion_reg} according to the temporal partition $\partition$, the equivalent $d(nC+1)$-dimensional density under $\fusion$ is simply, 
\begin{align}
g(\vecX{0}, \dots, \vecX{n-1}, \y) & \propto 
 h(\vecX{0}, \dots, \vecX{n-1}, \y) \cdot \prod_{j=0}^n \rho_j, \label{eq:bf}
\end{align}
where $\rho_0$ is as given in \eqref{eq:rho0}, for $j\in\{1,\dots,n\}$,
\begin{align*}
\rho_{j} := \rho_{j} \left(\vecX{j-1}, \vecX{j} \right)  & = \prod\cores{\expect}_{\brown{j,c}} \left[ \exp \left\{ -   \int_{t_{j-1}}^{t_j} \left( \phifn{c}{\X{t}{c}}  - \Phifn{c} \right) \ud t \right\} \right] \in(0,1], 
\end{align*}
and where $\brown{j,c}$ is the law of a Brownian bridge $\{\X{t}{c}, t \in (t_{j-1}, t_j)\}$ from  $\X{{j-1}}{c}$ to $ \X{{j}}{c}$, and $\Phifn{c}$ is a constant such that $\inf_{\x} \phifn{c}{\x}\geq \Phifn{c} > -\infty$. Discussion on $\Phifn{c}$ can be found in \apxref{apx:bfproof}. 

As we are interested in sampling from the \emph{fusion density} $f$ (corresponding to the time $T$ marginal of the $d(nC+1)$-dimensional density $g$), it is sufficient to simulate $g$ rather than the more complicated object $\tX\sim \fusion$. As suggested by \eqref{eq:bf}, this can be achieved by rejection sampling by first simulating a proposal from the density $h$, and accepting this proposal with probability equal to $\prod^n_{j=0}\rho_j$ .

Simulation of a proposal from $h$ is straightforward following \thmref{thm:bf} and \eqref{eq:prop_bf}. In particular, we can do so by first simulating a single draw from each sub-posterior and composing them to obtain a proposal at the time $0$ marginal of the temporal partition $\partition$ (in particular $\vecX{0} := \X{0}{1:C}$, where for $\Crange$, $\X{0}{c} \sim f_c$). This initial draw can then be iteratively propagated $n$-times using Gaussian transitions (as given in \eqref{eq:prop_bf} to compose the entire draw from $h$.

Now, considering the computation of the acceptance probability of the proposal, note that although $\rho_0$ is computable the direct computation of $\rho_1,\dots,\rho_n$ is not possible as it would require the evaluation of path integrals of functionals of Brownian motion. However, it is possible to construct unbiased estimators of these intractable quantities, and then simulate them using variations of established techniques. We denote the estimators we use by $\hat{\rho}_1,\dots,\hat{\rho}_n$, and are given by
\begin{align}
\hat{\rho}_j := \prod\cores \frac{\Delta_j^{\kappa_c}\cdot e^{- (U_{j}^{(c)}-\Phifn{c})\Delta_j}}{\kappa_c! \cdot p(\kappa_c|R_c)} \prod^{\kappa_c}_{k_c=1}\left(U_{j}^{(c)}- \phifn{c}{\X{\chi_{c,k}}{c}}\right)\ , \label{eq:rhohat}
\end{align}
where $R_{c}$ is a function of the Brownian bridge sample path $\X{}{c}\sim\brown{j,c}$ which determines the compact subset of $\Rd$ in which it is constrained. $U^{(c)}_j$ is a constant such that $\phifn{c}{\X{t}{c}} \leq U^{(c)}_j$ for all $\X{t}{c} \sim \brown{j,c}|R_{c}$, $\kappa_c$ is a discrete random variable with conditional probabilities $\prob[\kappa_c=k_c|R_{c}] := p(\kappa_c|R_c)$, and $\{\chi_1,\dots,\chi_{\kappa_c}\} \overset{\iid}{\sim}\uniform[t_{j-1},t_j]$. The validity of \eqref{eq:rhohat} is established the follows:
\begin{theorem} \label{thm:pe}
For every $1\leq j \leq n$, $\hat{\rho}_j$ is an unbiased estimator of $\rho _j$.
\proof See \apxref{apx:ue}. \qed
\end{theorem}

The construction and of estimators of the type in \eqref{eq:rhohat}, details on their specification (including the functional $R_c$), and a full proof of \thmref{thm:pe} are deferred to \apxref{apx:ue}. 

As it is possible to construct $[0,1]$ unbiased estimators of $\rho_1,\dots,\rho_n$, we now have an implementable rejection sampler: upon simulating the proposal from $h$ we can simply simulate $\prod_{j=0}^n \hat{\rho}_j \in(0,1]$ and accept with with this probability. The validity of this can be established by simply noting that as $\hat{\rho}_1,\dots,\hat{\rho}_n\in[0,1]$, then the rejection based algorithms resulting from their use are algorithmically equivalent to the original constructions of the algorithm had the intractable quantities been available. Furthermore, as a consequence of this there are no detrimental effect from the use of the estimators (such as decreased acceptance probabilities, or inflated variance).

Rejection sampling based algorithms suffer from a number of inefficiencies in this setting. For instance, it is clear that from \eqref{eq:bf} the acceptance probability of rejection sampling will likely decay geometrically with increasing $C$, and exponentially with increasing $T$. \citet{jap:dpr19} introduced a variant of this rejection sampling approach based upon methodology developed from a substantial simplification of \thmref{thm:bf} without the auxiliary temporal partition, $\partition$. Although theoretical sound, and being the first \emph{exact fusion} approach, the \emph{Monte Carlo Fusion} approach introduced in \citet{jap:dpr19} is impractical in many settings due to this lack of robustness with increasing numbers (and heterogeneity or lack of similarity) of sub-posteriors. Some further discussion of this approach and these shortcomings are given in \secref{sec:comparisons}. 

An immediate extension of the rejection sampling approach of \citet{jap:dpr19} would be an importance sampling approach, in which importance weights are assigned to each of the proposals from $h$ corresponding to the acceptance probability. This would however ultimately suffer from similar inefficiencies to the rejection sampling approach manifested by variance in the importance weights. A drawback of both rejection and importance sampling approaches, are the computational complications from the simulation of diffusion bridges (required in \eqref{eq:rhohat}) which have computational cost which does not scale linearly in $T$ -- this is one of the motivations for introducing the temporal partition, $\partition$. 

The key novelty of \thmref{thm:bf} is that the auxiliary temporal partition $\partition$ which has been introduced allows $g$ to be simulated using a sequential Monte Carlo (SMC) approach. This mitigates the robustness drawbacks of the \emph{Monte Carlo Fusion} approach of \citet{jap:dpr19}, and allows us to leverage the results and approaches available within the SMC literature. In particular, and as suggested by \eqref{eq:bf}, one could initialise an algorithm by simulating $N$ particles from the time $0$ marginal of $h$ in \eqref{eq:bf}, $\vecX{0,1},\dots,\vecX{0,N}$ (recalling that $\vecX{0} := \X{t}{1:C}$, where for $\Crange$ $\X{t}{c} \sim f_c$), and assigning each an un-normalised importance weight $w'_{0,\cdot}:=\rho_0(\vecX{0,\cdot})$. This initial particle set (which constitutes an approximation of the time $0$ marginal of $g$ in \eqref{eq:bf}), can then be iteratively propagated $n$ times by interlacing Gaussian transitions of the particle set over the $j$th partition of $\partition$ (with mean vector $ \cvecM{j}$ and covariance matrix $ \cV{j}$ as given in \eqref{eq:bf}), and updating the particle set weightings by a factor of $\hat{\rho}_j(\vecX{j-1,\cdot},\vecX{j,\cdot})$. The weighted particle set obtained after the final ($n$th iteration of the algorithm (which is an approximation of the time $T$ marginal of $g$), can then be used as a proxy for the desired $f$ (as supported by \thmref{thm:bf}). 

We term the SMC approach outlined above \emph{Bayesian Fusion}, and present pseudo-code for it in \algref{alg:bf}. Note that in this setting (unlike the rejection sampling setting) we need to further consider the construction of the unbiased estimator for $\rho_j$ and its variance, which is fully considered in \apxref{apx:ue}. 

\algref{alg:bf} outputs a weighted particle set at the end of each iteration which are \emph{re-normalised}. As standard within the SMC literature, we monitor for weight degeneracy by monitoring the importance sampling weights, and if appropriate \emph{resampling}. In particular, we compute the \emph{effective sample size} (ESS) \citep{jasa:klw94} of the particle set, and if the ESS falls below a lower user-specified threshold then the next iteration of the algorithm is instead initialised by (re-)sampling $N$ times from the empirical distribution defined by the current set of weighted particles (for simplicity we use multinomial resampling). In our particular case re-normalisation removes all contributory components of $\Phifn{1},\dots,\Phifn{C}$ from $\hat{\rho}_j$. This conveniently allows us to avoid the computation of the constants $\Phifn{1},\dots,\Phifn{C}$ which would seem to be required by \thmref{thm:bf} and \eqref{eq:rhohat}. As such in our presentation of \algref{alg:bf} we have simply replaced $\hat{\rho}_j$ by $\tilde{\rho}_j$ to exploit this, where
\begin{align}
\tilde{\rho}_j(\vecX{j-1,\cdot},\vecX{j,\cdot}) := \tilde{\rho}_j := e^{-\sum\cores\Phifn{c}\Delta_j}\hat{\rho}_j. \label{eq:rhotilde}
\end{align}

As suggested by \algref{alg:bf}, the output can be used directly as an approximation for the fusion density, $f$. Clearly the efficiency of the Bayesian Fusion approach outlined in \algref{alg:bf} will depend critically on the user-specified time horizon $T$, and the resolution of $\partition$ (and hence the number of iterations required in the algorithm). In the following section we provide guidance on selecting these tuning parameters, together with additional practical guidance on implementation.

\begin{algorithm}[ht]
	\caption{Bayesian Fusion Algorithm.} \label{alg:bf} 
    \begin{enumerate}
	\item \textbf{Initialisation Step ($j=0$)} \label{st:initial}
	\begin{enumerate}
	\item \textbf{Input:} Sub-posteriors, $f_1,\dots,f_C$, number of particles, $N$, time horizon, $T$, and temporal partition $\partition: 0=t_0<t_1<\dots<t_n =T$.
	\item For $i$ in $1$ to $N$, \label{st:partinit}
	\begin{enumerate}
	    \item \textbf{$\vecX{0,i}$:} For $c$ in $1$ to $C$, simulate $\X{0, i}{c} \sim f_c$. Set $\vecX{0,i}: = \X{0,i}{1:C}$. \label{st:init:draw}
	    \item \textbf{${w'}_{\!\!0,i}$:} Compute un-normalised weight ${w'}_{\!\!0,i}=\rho_0(\vecX{0,i})$, as per \eqref{eq:rho0}. \label{st:init:reweight}
	\end{enumerate}
	\item \textbf{$w_{0,\cdot}$:} For $i$ in $1$ to $N$ compute normalised weight $w_{0,i} = {w'}_{\!\!0,i}/\sum^N_{k=1} {w'}_{\!\!0,k}$.
	\item \textbf{$g^N_{0}$:} Set $g^N_0(\!\ud \vecX{0}) := \sum^N_{i=1} w_{0,i}\cdot\delta_{\vecX{0,i}}(\!\ud \vecX{0})$.
	\end{enumerate}
	\item \textbf{Iterative Update Steps ($j=j+1$ while $j\leq n$)} \label{st:iterate}
	\begin{enumerate}
	\item \textbf{Resample:} If the $\text{ESS}:= (\sum^{N}_{i=1} w^2_{j-1,i})^{-1}$ breaches the lower user-specified threshold, then for $i$ in $1$ to $N$ resample $\vecX{j-1,i} \sim g^N_{j-1}$, and set $w_{j-1,i} = 1/N$. \label{st:resample}
	\item For $i$ in $1$ to $N$, 
	\begin{enumerate}
	\item \textbf{$\vecX{j,i}$:} Simulate $\vecX{j,i} \sim \normal\left(\vecX{j-1,i}; \cvecM{j,i}, \cV{j} \right)$, where $\cvecM{j,i}$ and $\cV{j}$  are computed using \thmref{thm:bf}.
	\label{st:covar}
	\item  \textbf{${w'}_{\!\!j,i}$:} Compute un-normalised weight, ${w'}_{\!\!j,i} = w_{j-1,i}\cdot \tilde{\rho}_j(\vecX{j-1,i},\vecX{j,i})$ as per \algref{alg:ue} of \apxref{apx:ue}. \label{st:ue}
	\end{enumerate}
	\item \textbf{$w_{j,\cdot}$:} For $i$ in $1$ to $N$ compute normalised weight $w_{j,i} = {w'}_{\!\!j,i}/\sum^N_{k=1} {w'}_{\!\!j,k}$. \label{st:normalise}
	\item \textbf{$g^N_{j}$:} Set $g^N_j(\!\ud \vecX{j}) := \sum^N_{i=1} w_{j,i}\cdot\delta_{\vecX{j,i}}(\!\ud \vecX{j})$.
	\end{enumerate}	
	\item \textbf{Output:} $\hat{f}(\!\ud \y) := g^N_n(\!\ud \y) \approx f(\!\ud \y)$.
	\end{enumerate}
\end{algorithm}


\section{Theoretical underpinning and implementational guidance} \label{sec:guidance}

In this section we provide guidance on implementing the Bayesian Fusion algorithm. In particular, how to select the user-specified time horizon ($T$), and an appropriate resolution of the auxiliary temporal partition ($n$ and $\partition$), This is considered in Sections \ref{sec:Tguide} and \ref{sec:nPguide} respectively. The robustness of this guidance is considered by means of two extreme possible scenarios in Sections \ref{sec:shl_case} and \ref{sec:ssh_case}. We conclude in Sections \ref{sec:intervals} and \ref{sec:practical} by presenting other practical considerations for efficiently implementing \algref{alg:bf}.

We begin in developing guidance for $T$, $n$ and $\partition$, by noting that \algref{alg:bf} is an SMC algorithm for simulating the extended target density $g$ in \eqref{eq:bf}, which is achieved by approximating successive temporal marginals of $g$ (in particular, $g^{N}_j$) by means of propagating and re-weighting the previous temporal marginal ($g^{N}_{j-1}$). As such, it is natural to choose $T$, $n$ and $\partition$ to ensure the discrepancy between the sequence of proposal and target distributions is not degenerate, and so \emph{effective sample size} (ESS) is an appropriate quantity to analyse (see \citet{jasa:klw94}). However, the implementation we present in \algref{alg:bf} makes use of both weight normalisation and resampling in order to combat weight degradation. As such it is more natural in this setting to study a variant of ESS which is instead based upon the un-normalised incremental weight change within \algref{alg:bf} (recalling that we denote by $\hat{\rho}_j(\vecX{j-1,i},\vecX{j,i})=:\hat{\rho}_{j,i}$ the incremental weight change of the $i$th particle in the $j$th iteration), which we term the \emph{conditional effective sample size} (\ssj{}) (following for instance \citet{jcgs:zja16}). In particular we denote
\begin{align*}
\ssj{j} := \frac{\big(\sum_{i=1}^N \hat{\rho}_{j,i}\big)^2}{ \sum_{i=1}^N \hat{\rho}_{j,i}^2}.
\end{align*}
To develop concrete implementational guidance we consider and analyse the idealised setting of posterior distributions of large sample size $m$. In particular, we assume that the target density $f$ is multivariate Gaussian with mean vector $\bolda$ and covariance matrix $m^{-1}b\identity{}$ (for some $b>0$), and each of the sub-posterior densities $\target{c}{\x}$ ($\Crange$) are also multivariate Gaussian but with mean vector $\bolda_c$ and covariance matrix $m^{-1}C b \identity{}$ respectively. Note that we have $\bolda = C^{-1}\sum\cores \bolda_c$, and we will further reasonably assume $m>C>1$. To study the robustness of \algref{alg:bf} we further consider the quantity $\sigma_{\bolda}^2 := C^{-1}\sum\cores\|\bolda_c -\bolda\|^2 $ which gives a measure of what we term the \emph{sub-posterior heterogeneity} (the degree to which the individual sub-posteriors agree or disagree with one another).


\subsection{Guidance on selecting \texorpdfstring{$T$}{T}} \label{sec:Tguide}
Considering the selection of $T$ note from \algref{alg:bf} that its influence appears solely in the initial weighting given to each of the $N$ particles in \eqref{eq:rho0} through $\rho_0$. As such, we study the \emph{initial} conditional effective sample size.  

\begin{theorem} \label{thm:cess0}
Considering the initial conditional effective sample size (\ssj{0}), we have that as $N\rightarrow \infty$,
\begin{align*}
N^{-1} \ssj{0} 
& \rightarrow \exp\left\{ - \frac{ \frac{\sigma_{\bolda}^2 b}{m }}{\left(\frac{T}{C} + \frac{b}{m}\right)
\cdot\left(\frac{T}{C} + \frac{2b}{m}\right)} \right\} \cdot \left [1 + \frac{\left( \frac{Cb}{T m}\right)^2 }{ 1+\frac{2Cb}{T m}  } \right]^{-\frac{(C-1)d}{2}}.
\end{align*}
\proof See \apxref{apx:guidanceproofs}. \qed
\end{theorem}
\thmref{thm:cess0} shows explicitly how \ssj{0} degrades as the level of sub-posterior heterogeneity ($\sigma_{\bolda}^2$) increases. To explore this dependency we introduce the following conditions which will allow us to clearly identify regimes where \ssj{0} is well-behaved.
\begin{condition}[\shl] \label{cond:sh1}
The sub-posteriors obey the \shl condition (for some constant $\lambda >0$) if,
\begin{align*}
\sigma_{\bolda}^2 
&= \frac{b(C-1)\lambda}{m}.
\end{align*}
\end{condition}
\begin{condition}[\ssh] \label{cond:ssh}
The sub-posteriors obey the \emph{super sub-posterior heterogeneity} \ssh condition (for some constant $\gamma>0$) if,
\begin{align*}
\sigma_{\bolda}^2 
&= b\gamma.
\end{align*}
\end{condition}
Note that \conref{cond:sh1} is a very natural condition which would arise in many settings (for instance, if $(m/C)$th of the data was randomly allocated to each sub-posterior then $ \sigma_{\bolda}^2 \sim b/m \times \chi^2_{C-1}$ and thereby have mean $b(C-1)/m$). For $m/C$ large we would expect that for $\lambda >1$ the sub-posteriors would obey the \shl condition with high probability. Whereas at the other end of the spectrum, the \ssh condition of \conref{cond:ssh} captures the case where sub-posterior heterogeneity does not decay with $m$.

Considering the initial conditional effective sample size under Conditions \ref{cond:sh1} and \ref{cond:ssh} we establish the following corollary. 

\begin{corollary} \label{cor:cess0}
If for some constant $k_1>0$, $T$ is chosen such that 
\begin{align}
T 
& \ge \frac{bC^{3/2} k_1}{m}, \label{eq:T}
\end{align}
then the following lower bounds on \ssj{0} hold:
\begin{enumerate}
\item
If \shl holds for some $\lambda >0$, then
\begin{align}
\lim_{N\to \infty} N^{-1} \ssj{0} \ge \exp \left\{-\lambda k_1^{-2} -dk_1^{-2}/2 \right\}. \label{eq:shl_guide}
\end{align}
    \item 
    If \ssh holds for some $\gamma >0$, and $ T\ge k_2 C^{-3/2}$ (for some constant $k_2>0$), then
\begin{align}
\lim_{N\to \infty} N^{-1} \ssj{0} \ge \exp \left\{ -\gamma b k_1^{-1} k_2^{-1} -dk_1^{-2}/2 \right\}. \label{eq:ssh_guide}
\end{align}
\end{enumerate}
\proof
See \apxref{apx:guidanceproofs}. \qed
\end{corollary}

\corref{cor:cess0} gives explicit guidance on minimal values of $T$ which should be selected to robustly initialise \algref{alg:bf}, as measured by initial \ssj{}. In principle one could choose $T$ in excess of this minimal guidance. This however comes at either the cost of increasing the number of iterations of the algorithm required (which we will discuss in the following section), or increasing the increment size in the auxiliary temporal partition (which will lead to increased computational cost in simulating $\hat{\rho}_{\cdot}$ from \thmref{thm:pe}), or some combination of both. 


\subsection{Guidance on selecting \texorpdfstring{$n$}{n} and \texorpdfstring{$\partition$}{P}} \label{sec:nPguide}
Having selected an appropriate $T$ (using the guidance of \secref{sec:Tguide} and \corref{cor:cess0}), we are left with choosing the remaining user-specified parameters $n$ and $\partition$ (the resolution and spacing of the auxiliary temporal partition), as 
required in \algref{alg:bf}. We address this implicitly by considering how to choose the $j$th interval size (i.e. the interval $(t_{j-1}, t_j]$) of the auxiliary temporal partition, which we do so by again considering the conditional effective sample size in \thmref{thm:cessj}. 

\begin{theorem} \label{thm:cessj}
Considering the conditional effective sample size for the $j$th iteration of \algref{alg:bf} (\ssj{j}), and letting $k_3, k_4$ be positive constants, we have
\begin{align*}
\liminf \lim_{N \to \infty }N^{-1}\ssj{j} 
& \ge e^{-k_3-dk_4},
\end{align*}
where the outer $\liminf$ is taken over sequences of $t_j - t_{j-1}\to 0$ with
\begin{align}
t_j - t_{j-1} 
& \leq \min \left\{ \left( \frac{b^4k_3 C^3}{2 m^4  \sigma ^2_{t_j}}
\right)^{1/3}, \left(\frac{2k_4b^4C^3}{m^4}\right)^{1/4} \right\}, \label{eq:cessj}
\end{align}
and $\sigma_{t_j}^2 = C^{-1}\sum\cores\| \expect \left(\X{j}{c} |{\boldxi}_j\right) - \bolda_c \|^2$ (where ${\boldxi}_j$ denotes a sequence of standard Gaussian vectors as defined in \corref{cor:parallel1}).
\proof See \apxref{apx:guidanceproofs}. \qed 
\end{theorem}

We can use \thmref{thm:cessj} to develop guidance for choosing the $j$th interval size of the auxiliary temporal partition, by considering the effect of $\sigma_{t_j}^2$ in \eqref{eq:cessj}. In essence $\sigma_{t_j}^2$ describes the average variation of the $C$ trajectories of the distribution of their proposed update locations with respect to their individual sub-posterior mean (i.e. how different $\expect (\X{j}{c} |{\boldxi}_j)$ is from $ \bolda_c$). Recalling that \algref{alg:bf} is coalescing $C$ trajectories initialised independently from their respective sub-posteriors to a common end point, then $\sigma_{t_j}^2$ will largely be determined by a combination of how close the interval is to the end point $T$, how large the interval $(t_{j-1},t_j]$ we are simulating over is, and critically the degree of \emph{sub-posterior heterogeneity} as determined by variation in their mean. Intuitively one may wish to also choose the regularity of the mesh itself dependant on sub-posterior heterogeneity (in particular, one would anticipate decreasing the interval size in the partition approaching $T$ to counteract the increasing disagreement of the coalescing trajectories with their own respective means), but for algorithmic simplicity in the following we impose a \emph{regular mesh} ($\Delta_j = t_j-t_{j-1} = T/n =: \Delta$). Consequently, and as in \secref{sec:Tguide}, we develop guidance for $n$ and $\partition$ by considering sub-posterior heterogeneity and its impact on \eqref{eq:cessj} (noting that for a regular mesh we can simply set $n = \bigO\big(T \Delta_j^{-1}\big)$). We then return in \secref{sec:intervals} to consider the implication of imposing a regular mesh over an irregular mesh.

We begin by noting (see \eqref{eq:trans_new} in \corref{cor:parallel1}) that
\begin{align*}
\sigma_{t_j}^2 &= C^{-1}\sum\cores\left\|  ({\bolda} - \bolda_c) \frac{t_j}{T} + \frac{t_j}{\sqrt{CT}}\cdot \boldxi_j \right\|^2 \\
\expect\sigma_{t_j}^2 
& = C^{-1}\sum\cores\left\|  {\bolda} - \bolda_c\right \|^2 \frac{t_j^2}{T^2} + \frac{d \cdot t_j^2}{CT}
\end{align*}

\begin{enumerate}[leftmargin=*]
\item If \shl holds we have, by choosing $T =\bigO (bC^{3/2} k_1 m^{-1})$ following \corref{cor:cess0},
\begin{align}
\expect \sigma_{t_j}^2 
 =  \frac{b(C-1)\lambda }{m} \frac{t_j^2}{T^2} + \frac{d\cdot t_j^2}{CT}  \leq  \bigO\left(\frac{bC\lambda}{m} + \frac{dbC^{3/2}k_1}{Cm} \right) \label{eq:sigmatj_sh}
\end{align}
which implies that $\expect \sigma_{t_j}^2$ is bounded above by $\bigO\left(\frac{C}{m}\right)$, then from \thmref{thm:cessj} we have that \ssj{j} will be well-behaved provided we choose
\begin{align}
t_j - t_{j-1} 
&= \bigO\left(\frac{C^{2/3}}{m}\right). \label{eq:shl_delta}
\end{align} 
    \item If \ssh holds we have
\begin{align}
\expect\sigma_{t_j}^2 = \bigO \left(\frac{t_j^2}{T^2} +  \frac{t_j^2}{CT}\right). \label{eq:sigmatj_ssh}
\end{align}
Following \corref{cor:cess0} if we choose $T = \bigO \big( \max\{bC^{3/2} k_1 m^{-1},$ $k_2 C^{-3/2}\}\big)$, and recalling we have $m>C>1$, then $\expect \sigma_{t_j}^2 =\bigO\big( \frac{t_j^2}{T^2} \big( 1 + \frac{T}{C} \big) \big) = \bigO\big( \frac{t_j^2}{T^2}\big)$ which is bounded above by $\bigO(1)$. As such, from \thmref{thm:cessj} we have that \ssj{j} will be well-behaved provided we choose
\begin{align}
t_j - t_{j-1} 
&= \bigO\left(\frac{C}{m^{4/3}} \right). \label{eq:ssh_delta}
\end{align} 
\end{enumerate}

In keeping with the intuition we developed earlier, note from \eqref{eq:sigmatj_sh} and \eqref{eq:sigmatj_ssh} that in both the \shl and \ssh settings $\sigma_{t_j}^2$ will increase as $t_j \uparrow T$, and so from \eqref{eq:cessj} we would anticipate some marginal gain using an irregular mesh and decreasing the interval size as we approach $T$. However, choosing $\Delta_j = t_j - t_{j-1}$ as per \eqref{eq:shl_delta} and \eqref{eq:ssh_delta} in conjunction with the guidance for choosing $T$ in \secref{sec:Tguide}, this results in only a marginal effect and so choosing a regular mesh is advantageous from the perspective of algorithmic simplicity (we verify this statement empirically in \secref{sec:intervals}). 

As a consequence of choosing $T$ as per \secref{sec:Tguide} and imposing a regular mesh, determining appropriate choices for $n$ and $\partition$ (the remaining user-specified parameters of \algref{alg:bf}) is direct. 

Of course, choosing interval sizes ($\Delta_j$) smaller than this minimal guidance is possible (and may help computationally in the simulation of $\hat{\rho}_{\cdot}$ as per \lemref{alg:ue}) but leads to an increased number of iterations in \algref{alg:bf}. As the reader will surmise, choosing $T$ and $n$ beyond the minimal guidance given by Sections \ref{sec:Tguide} and \ref{sec:nPguide} is more of a practical computational consideration, but as shown the algorithm should be well-behaved.

Having established guidance for choosing $T$, $n$, and $\partition$ for Bayesian Fusion, we now verify that these selections lead to Bayesian Fusion being robust to increasing data size (as measured by \ssj{}). We do so by studying the guidance in idealised settings for the posterior distribution under the \shl and \ssh conditions, which we do in Sections \ref{sec:shl_case} and \ref{sec:ssh_case} respectively. Note that we consider more substantial examples and comparisons with competing methodologies in Sections \ref{sec:comparisons} and \ref{sec:examples}. 


\subsection{Sub-posteriors with similar mean} \label{sec:shl_case}

We begin by examining the guidance for $T$ and $n$ in Bayesian Fusion under the \shl setting of \conref{cond:sh1}. Recall this would be the most common setting of relatively homogeneous sub-posteriors (as characterised by variation in the sub-posterior mean), which would occur if for instance we were able to randomly allocate approximately a $C$th of the available data to each sub-posterior. To do so we consider the idealised scenario in which we wish to recover a target distribution $f$, which is Gaussian with mean $\mu=0$ and variance $\sigma^2 = m^{-1}$, by applying \algref{alg:bf} to unify $C$ sub-posteriors ($f_c$, $\Crange$), which are Gaussian with mean $\mu_c=0$ and variance $\sigma^2_c= C\sigma^2$. In this example we consider a range of data sizes from $m=\num{1000}$ to $m=\num{50000}$, with a fixed number of sub-posteriors ($C=10$), and using a particle set of size $N=\num{10000}$. In implementing \algref{alg:bf} we use \gpeb{} (\conref{con:rho_isb}) of \apxref{apx:ue} for simulating the unbiased estimator in \stepref{st:ue}.

\begin{figure}[ht]
\begin{center}
\subfigure[Fixed user-specified tuning parameters $T$ and $n$. 
\label{subfig:shl:fixed}]{\includegraphics[width=0.31\textwidth]{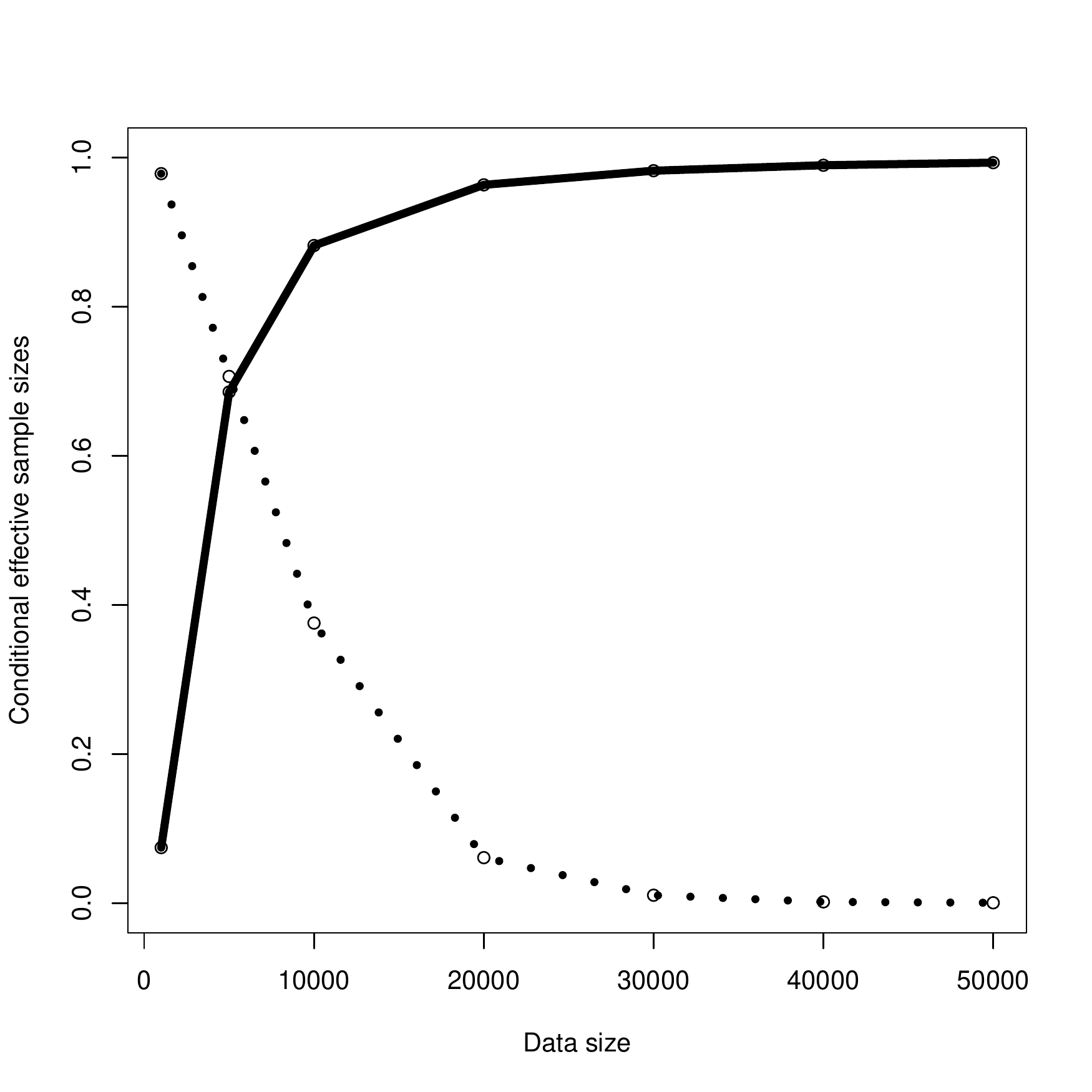}}
\hfill
\subfigure[Recommended scaling of $T$, fixed $n$. \label{subfig:shl:Tscaled}]{\includegraphics[width=0.31\textwidth]{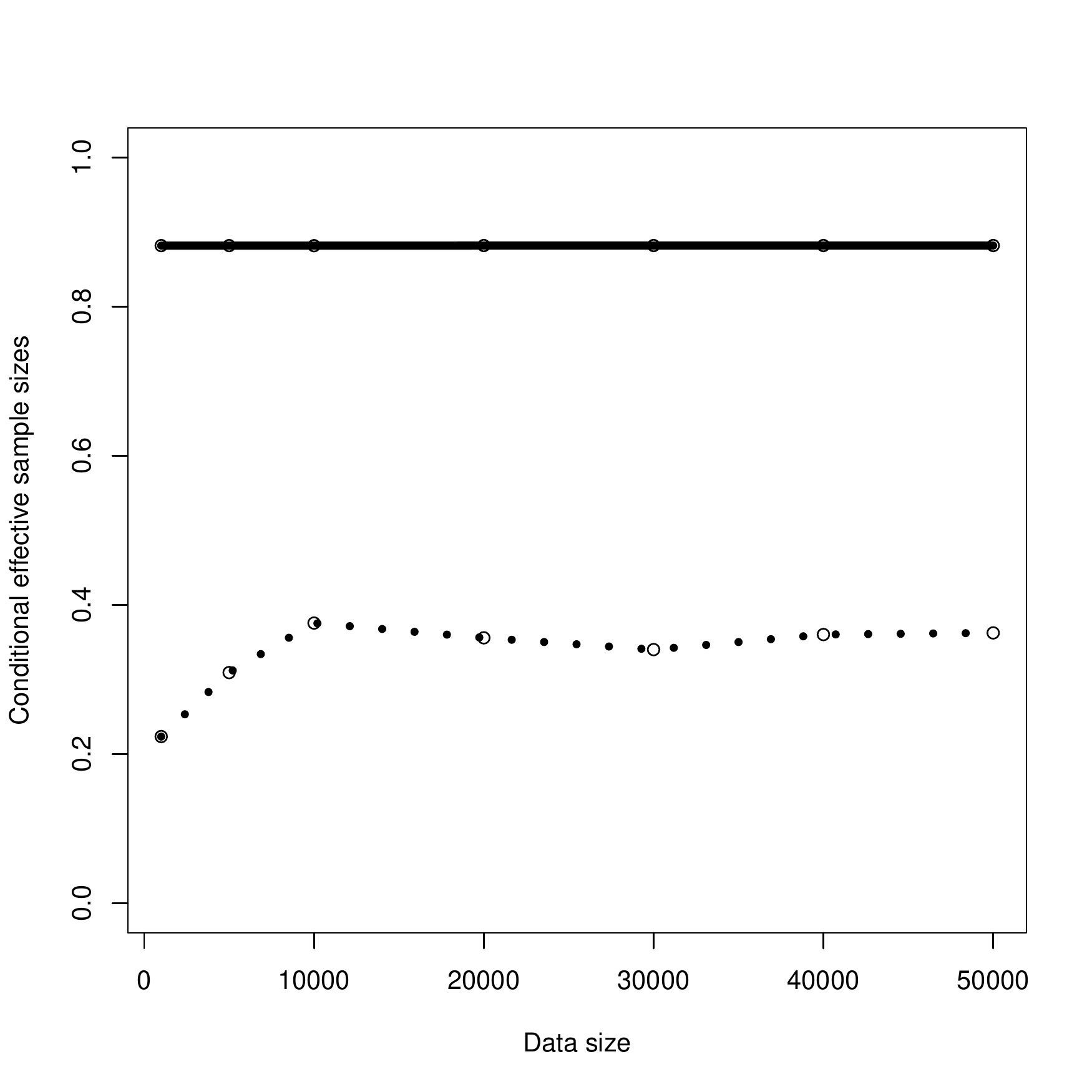}}
\hfill
\subfigure[Recommended scaling of $T$ and $n$. \label{subfig:shl:nscaled}]{\includegraphics[width=0.31\textwidth]{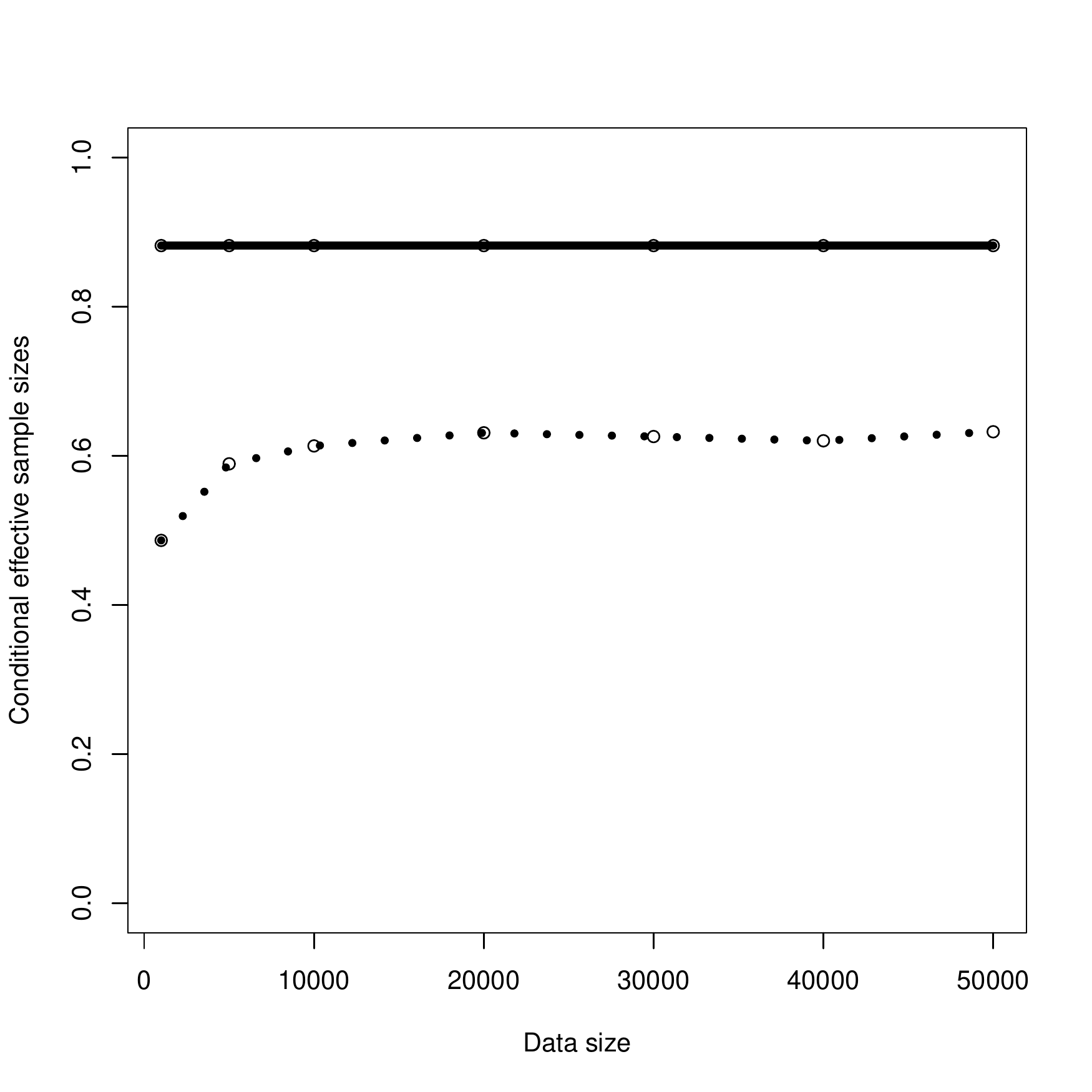}}
\caption{Conditional effective sample size of \algref{alg:bf} with increasing data size in \shl setting of \secref{sec:shl_case}. Solid lines denote initial conditional effective sample size (\ssj{0}, following \algstref{alg:bf}{st:initial})). Dotted lines denote averaged conditional effective sample size in subsequent iterations of \algref{alg:bf} ($(\sum^{n}_{j=1} \ssj{j})/n$, following \algstref{alg:bf}{st:iterate})).} \label{fig:shl}
\end{center}
\end{figure}

In line with the development of our guidance in \secref{sec:guidance}, we consider \ssj{0} and \ssj{j} ($j\in\{1,\dots,n\}$) with increasing data size by first considering fixed choices for $T$ and $n$ ($T=0.005$ and $n=\num{5}$), then choosing a robust scaling of $T$ but with fixed $n$ (as in \secref{sec:Tguide}), and then robustly scaling $T$ and $n$ (as in \secref{sec:nPguide}). This is presented in Figures \ref{subfig:shl:fixed}--\ref{subfig:shl:nscaled} respectively. 

Considering the results of fixing $T$ and $n$ in \figref{subfig:shl:fixed}, it is clear in this regime that \algref{alg:bf} would lack robustness with increasing data size. Although \ssj{0} improves with increasing data size as expected with increasingly similar sub-posteriors from \eqref{eq:rho0} of \thmref{thm:bf}, this comes with drastically decreasing \ssj{j} (as suggested by \thmref{thm:cessj}), which in totality would render the methodology impractical. Scaling $T$ following the guidance in \secref{sec:Tguide} immediately stabilises both \ssj{0} and \ssj{j} in the \shl setting, making \algref{alg:bf} robust to increasing data size (as shown in \figref{subfig:shl:Tscaled}). Additionally scaling $n$ substantively improves \ssj{j} for all data sizes. In both Figure \ref{subfig:shl:Tscaled} and \ref{subfig:shl:nscaled} the slightly decreased \ssj{j} for small data sizes can be explained by random variation in the simulation of the sub-posterior, which leads to slight mis-matching. 

\subsection{Sub-posteriors with dissimilar mean} \label{sec:ssh_case}

Now we examine the guidance for $T$ and $n$ in Bayesian Fusion under the \ssh setting of \conref{cond:ssh}. Recall this would be an extreme setting in which sub-posterior heterogeneity does not decay with data size, $m$. To investigate this setting we consider recovering a target distribution $f$, which is Gaussian with mean $\mu=0$ and variance $\sigma^2 = m^{-1}$, by using \algref{alg:bf} to unify $C=2$ sub-posteriors with mean $\mu_c=\pm 0.25$ and variance $\sigma^2_c= 2\sigma^2$. In this scenario as data size increases the sub-posteriors have increasingly diminishing common support, although our measure of heterogeneity is fixed with $\sigma_{\bolda}^2  = 0.0625$. In this example we consider a range of data sizes from $m=\num{250}$ to $m=\num{2500}$, and use a particle set of size $N=\num{10000}$. We again use \gpeb{} (\conref{con:rho_isb}) of \apxref{apx:ue} for simulating the unbiased estimator in \stepref{st:ue} when implementing \algref{alg:bf}.

\begin{figure}[ht]
\begin{center}
\subfigure[Fixed user-specified tuning parameters $T$ and $n$. 
\label{subfig:ssh:fixed}]{\includegraphics[width=0.31\textwidth]{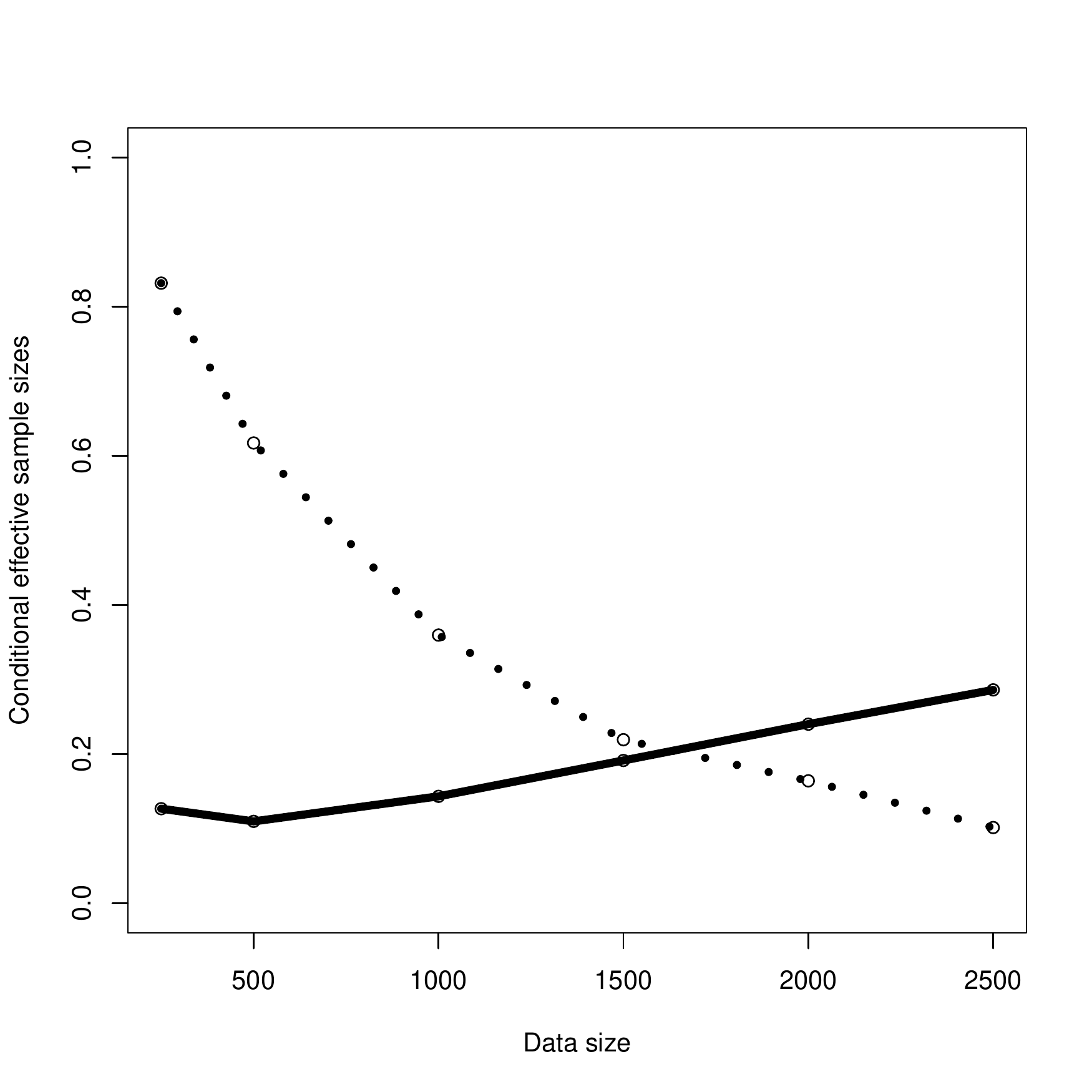}}
\hfill
\subfigure[Recommended scaling of $T$, fixed $n$. 
\label{subfig:ssh:Tscaled}]{\includegraphics[width=0.31\textwidth]{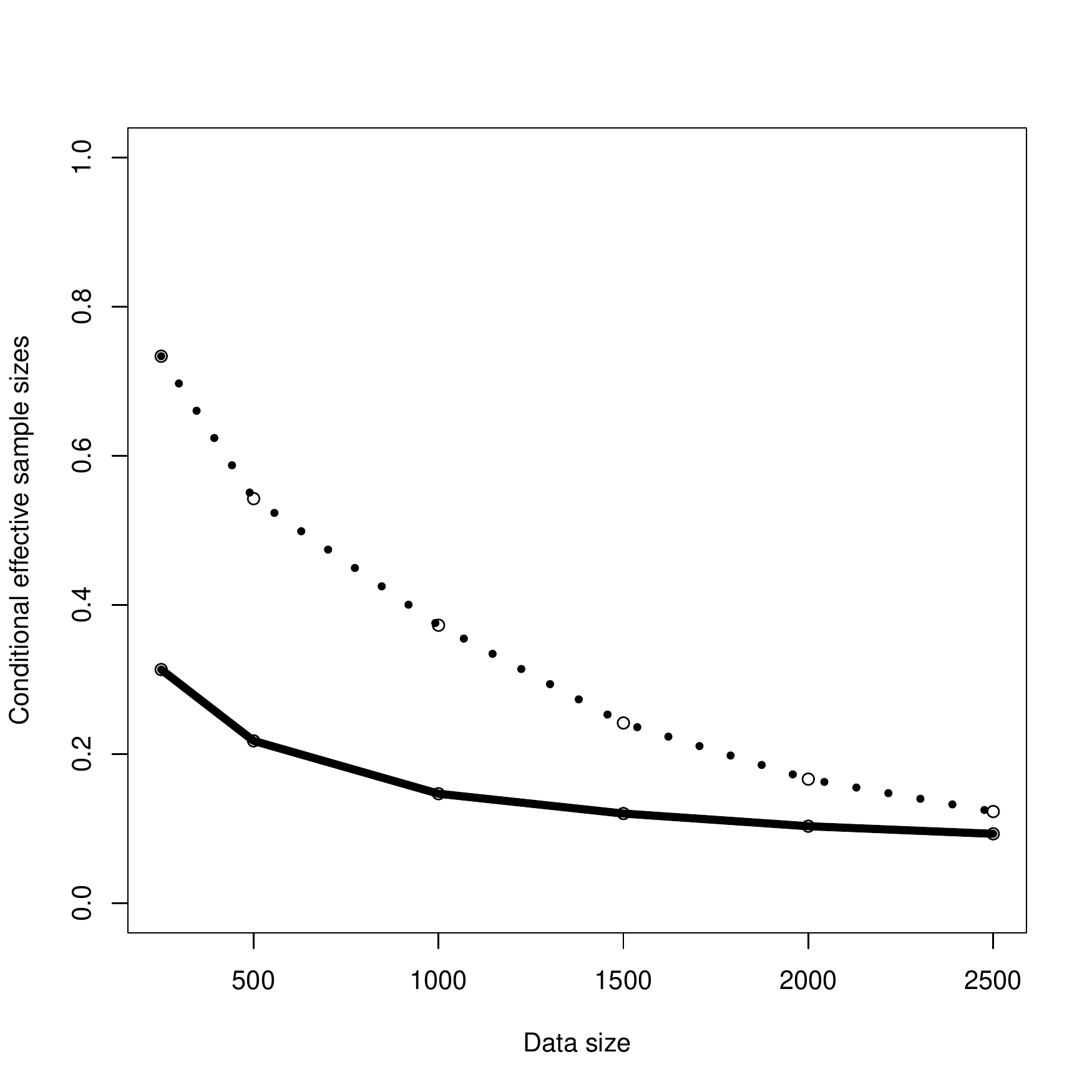}}
\hfill
\subfigure[Recommended scaling of $T$ and $n$. 
\label{subfig:ssh:nscaled}]{\includegraphics[width=0.31\textwidth]{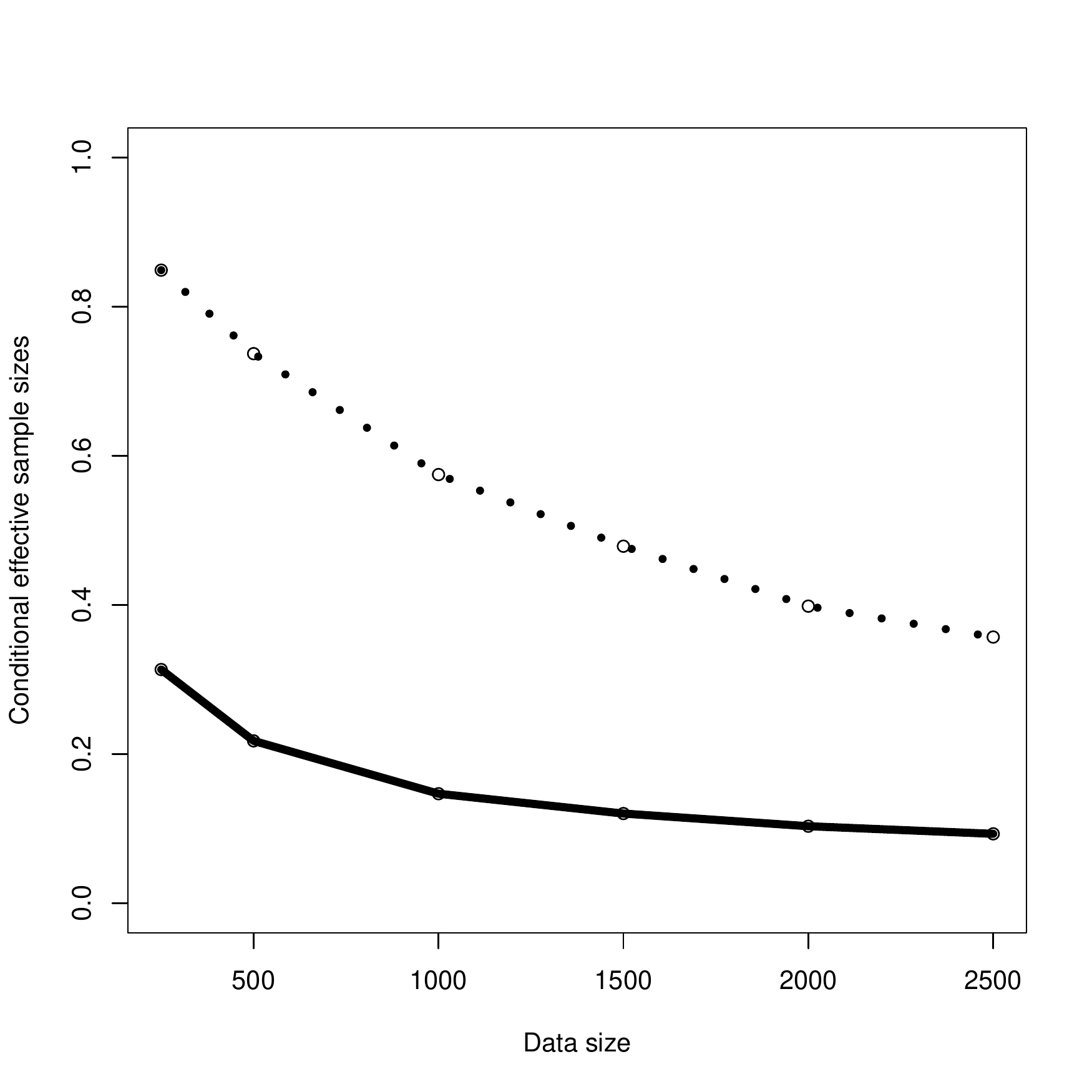}}
\caption{Conditional effective sample size of \algref{alg:bf} with increasing data size in \ssh setting of \secref{sec:ssh_case}. Solid lines denote initial conditional effective sample size (\ssj{0}, following \algstref{alg:bf}{st:initial})). Dotted lines denote averaged conditional effective sample size in subsequent iterations of \algref{alg:bf} ($(\sum^{n}_{j=1} \ssj{j})/n$, following \algstref{alg:bf}{st:iterate})).} \label{fig:ssh}
\end{center}
\end{figure}

As in the \shl setting of \secref{sec:shl_case}, for this \ssh setting we consider \ssj{0} and \ssj{j} ($j\in\{1,\dots,n\}$) with increasing data size with fixed choices for $T$ and $n$ ($T=0.01$ and $n=5$), then choose a robust scaling of $T$ but with fixed $n$ (as in \secref{sec:Tguide}), and then robustly scale both $T$ and $n$ (as in \secref{sec:nPguide}). This is presented in Figures \ref{subfig:ssh:fixed}--\ref{subfig:ssh:nscaled} respectively. 

It is clear looking at the results for the \ssh setting in \figref{fig:ssh}, and contrasting them with the \shl setting of \figref{fig:shl}, that the \ssh setting is considerably more challenging. This is to be expected as the sub-posteriors become increasingly mismatched as data size increases. However, the effect of including scaling $T$ and $n$ does substantively improve \algref{alg:bf} as it did in \secref{sec:shl_case}. Considering the results of fixing $T$ and $n$ in \figref{subfig:ssh:fixed}, it is clear in this regime that \algref{alg:bf} is degenerate. Incorporating scaling of $T$ in \figref{subfig:ssh:Tscaled} stabilises \ssj{0} and leads to a slower degradation with data size of \ssj{j}. However, incorporating scaling of $T$ and $n$ following our guidance earlier in \secref{sec:guidance} retains the stabilised \ssj{0} and substantively improves \ssj{j} to a level where it could lead to a practical algorithm. 

\subsection{Temporal regularity of partition} \label{sec:intervals}

In \secref{sec:nPguide} in order to simplify the guidance for selecting the partition $\partition$, we imposed a regular mesh. This allowed us to use the minimal guidance for the temporal distance between points in the partition we developed in \thmref{thm:cessj}, which in conjunction with the guidance already established for choosing $T$ in \secref{sec:Tguide}, allowed us to indirectly specify $n$ and in turn $\partition$. As discussed in \secref{sec:nPguide}, there may be some advantage of using an irregular mesh (in which the temporal distance between points in the partition decreases as $T\!\uparrow\! n$). In this section we investigate the impact of using a regular mesh on \ssj{j} ($j\in\{1,\dots,n\}$) as a function of the iteration of \algref{alg:bf}. 

\begin{figure}[ht]
\begin{center}
\subfigure[\shl setting. $\target{c}{x} = \normal(0, C\sigma^2)$. 
\label{subfig:partition:shl}]{\includegraphics[width=0.425\textwidth]{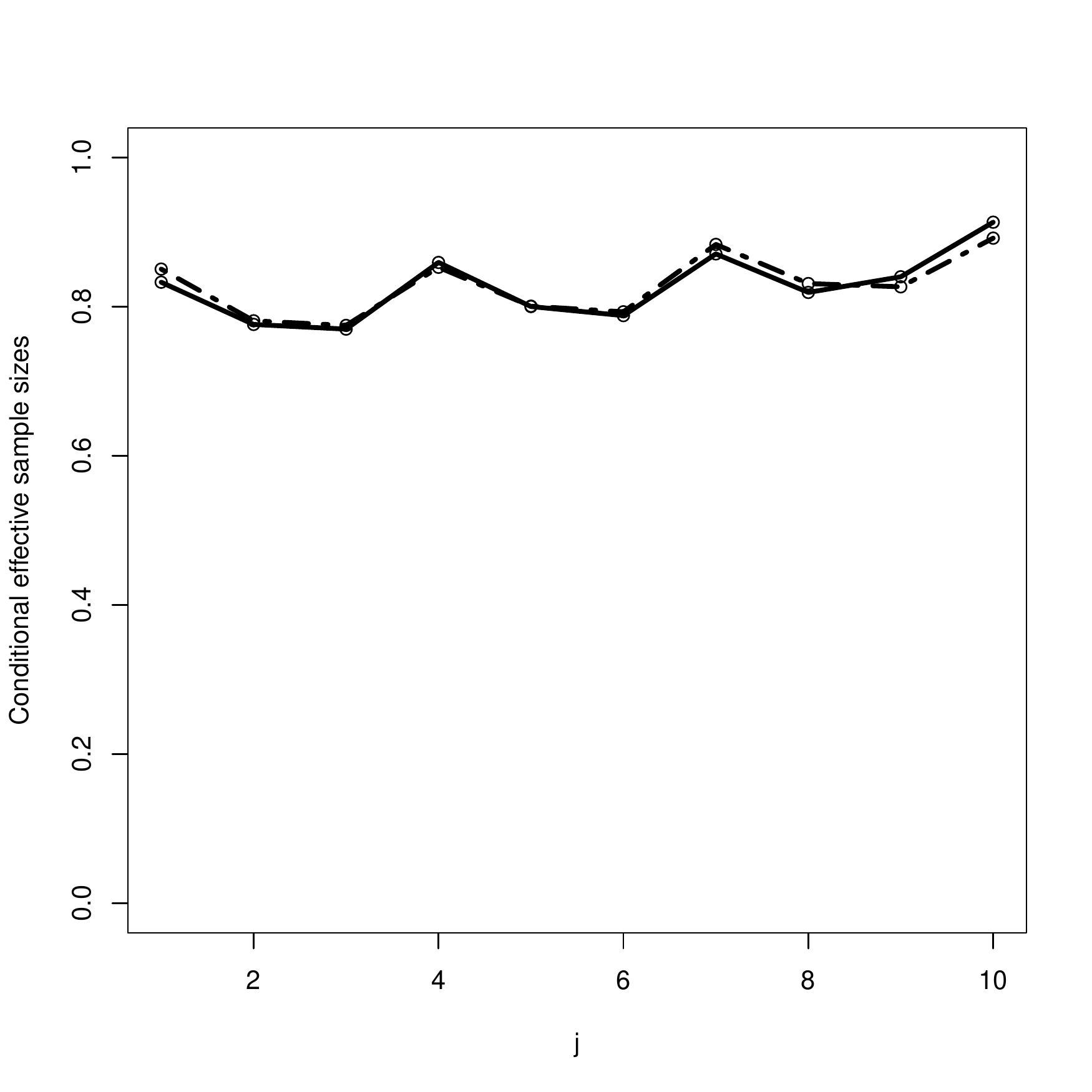}}
\hspace{1em}
\subfigure[\ssh setting. $\target{c}{x} = \normal(\mu_c, C\sigma^2)$, $\mu_c=\pm 0.25$. \label{subfig:partition:ssh}]{\includegraphics[width=0.425\textwidth]{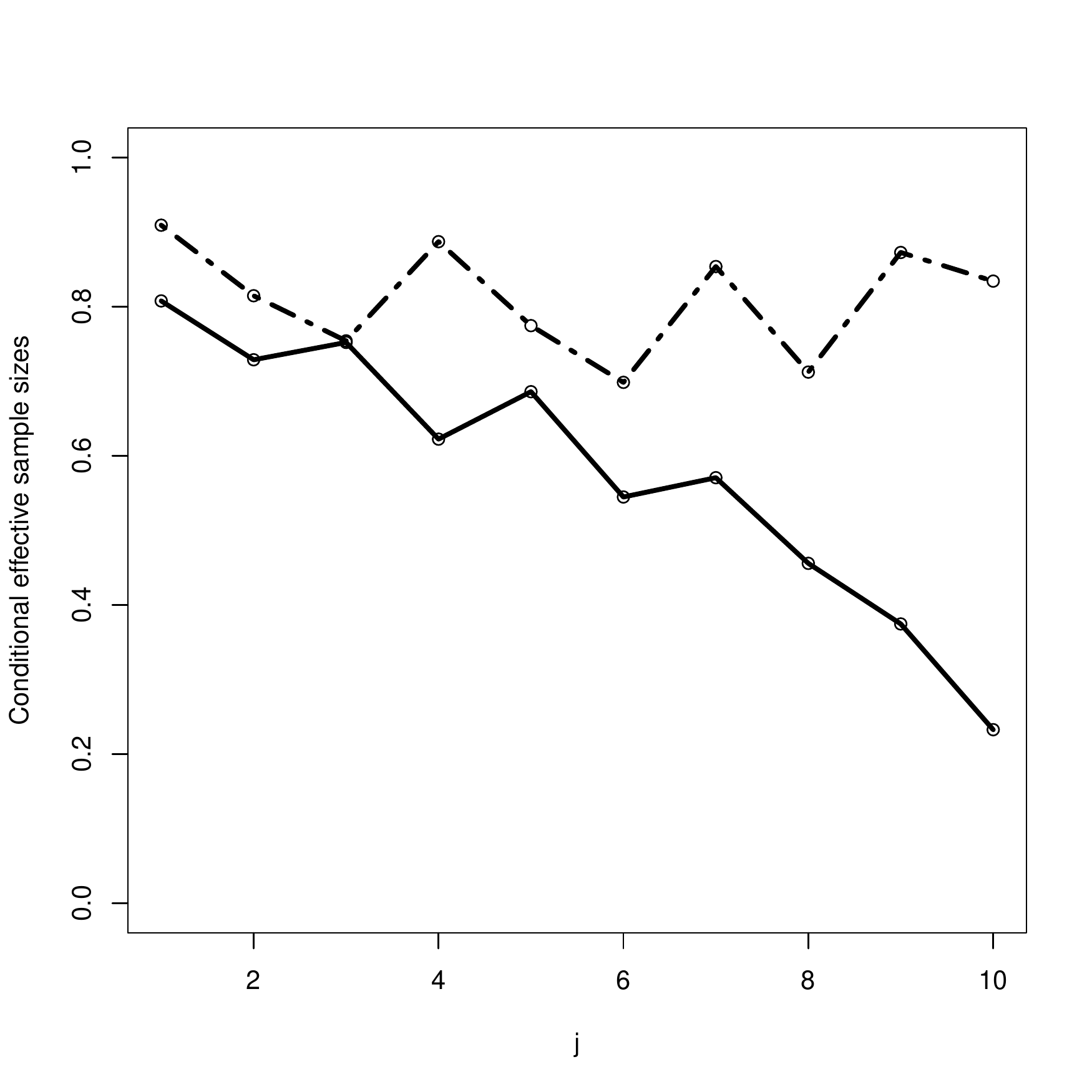}}
\caption{Conditional effective sample size at each iteration of \algref{alg:bf} ($j\in\{1,\dots,10\}$) under \shl and \ssh settings respectively. Solid lines denote results based upon selecting the unbiased estimator $\tilde{\rho}_j:=\gpeas{j}$. Dotted lines the unbiased estimator $\tilde{\rho}_j:=\gpebs{j}$.} \label{fig:partition}
\end{center}
\end{figure}

To investigate temporal regularity we revisit the idealised examples of the \shl and \ssh settings we introduced in Sections \ref{sec:shl_case} and \ref{sec:ssh_case} respectively. For both settings we consider a data size of $m=\num{1000}$ distributed across $C=\num{2}$ sub-posteriors, and specify a temporal horizon of $T=0.01$ and regular mesh of size $n=\num{10}$. In implementing Bayesian Fusion we use a particle set of size $N=\num{10000}$, and consider the use of two variants for the unbiased estimator in \stepref{st:ue} when implementing \algref{alg:bf} -- \gpea{} (\conref{con:rho_isa}) and \gpeb{} (\conref{con:rho_isb}) of \apxref{apx:ue} -- \gpea{} being a relatively straightforward construction, whereas \gpeb{} requiring slightly more specification but in general leading to a more robust estimator as defined by the variance of the estimator. The results are presented in \figref{fig:partition}. 

Considering the \shl setting of \figref{subfig:partition:shl} we find that \ssj{j} is stable across iterations of \algref{alg:bf}, which would suggest that there is little to be gained when heterogeneity is low in having a more flexible irregular mesh. The \ssh setting of \figref{subfig:partition:ssh} is slightly more complicated. The results here would suggest if using the \gpea{} in the \ssh setting there may be some advantage to using an irregular mesh to balance \ssj{j} across the iterations of \algref{alg:bf}. However, in both the \shl and \ssh settings when using the \gpeb{} unbiased estimator we find that \ssj{j} is stable. This would suggest that there is little to be gained from specifying an irregular mesh over the regular one we have imposed in \secref{sec:nPguide}. Choosing a good estimator for a regular mesh is far simpler than optimising an irregular mesh for a poor estimator, and so the more critical consideration is to ensure a suitable unbiased estimator is chosen -- a full discussion of which can be found in \apxref{apx:ue}. 

\subsection{Practical implementational considerations} \label{sec:practical}

As motivated in the introduction, the primary contribution of this paper is to develop a practical alternative to the \emph{Monte Carlo Fusion} approach of \citep{jap:dpr19} for inference in the \emph{fusion problem} (simulating from \eqref{eq:prod}). The methodological development of \secref{sec:theory}, and the practical guidance of Sections \ref{sec:Tguide} and \ref{sec:nPguide}, have been developed to this end. However, in some particular settings where this methodology is applied it is likely there will be a number of additional specific constraints that necessitate careful implementation, or some modification, of \algref{alg:bf}. For instance, \emph{latency} in communication between cores may be of particular concern, or in applications where there is a large amount of data on each individual sub-posterior the computational efficiency of some quantities in \algref{alg:bf} may need consideration. In this section we highlight some aspects and minor (non-standard) modifications of the methodology we have developed which may be useful for practitioners. 

For the purposes of clarity for the primary contributions of this paper, the methodology and examples given elsewhere in the paper do not exploit the modifications we present below. We discuss other more substantial possible directions for the practical development of the Bayesian Fusion methodology in the conclusions. We consider the possible modifications to Bayesian Fusion grouped into the constituent elements of \algref{alg:bf}: Initialisation; Propagation of the particle set; Computing importance weights; and, normalisation and resampling of the particle set. This is presented in Sections \ref{sec:effinit}--\ref{sec:effresamp} respectively. 

Note that Sequential Monte Carlo methods (upon which Bayesian Fusion is based) are in principle well-suited to parallel implementation in distributed environments (see for instance, \citet[Sec.\ 7.5.3]{bk:hgm:dl18} and \citet{aisp:cmr18}). A considerable literature has been developed on distributed resampling methodologies \citep{jcgs:lygdh10,jcgs:mlj16,sadm:lw16}, and methodological adaptations such as \emph{distributed particle filters} \citep{ieee:bdh05,spa:hw17}, and the \emph{island particle filter} \citep{sc:vdmm15}. The guidance provided in this subsection may be of interest in 
developing a truly parallel implementation of Bayesian Fusion, although note the particularities of the fusion problem make this a challenging problem outwith the scope of this paper. In particular, in the fusion setting the sub-posteriors (and accompanying data) are distributed across the available cores, and a natural implementation of \algref{alg:bf} would have the particle set common among all cores -- this is at odds with the setting typically addressed by the distributed SMC literature. We defer further discussion on this to the conclusions. 


\subsubsection{Initialising the particle set} \label{sec:effinit}

The initialisation of Bayesian Fusion as presented in \secref{sec:theory} utilises the fact that we have access to independent draws from the $C$ sub-posteriors. In particular, we propose $\vecX{0} := \X{0}{1:C}$ where for $\Crange$, $\X{0}{c} \sim f_c$ (as in \algstref{alg:bf}{st:init:draw}). Composing $\vecX{0}$ requires communication between the cores, and furthermore $\vecX{0}$ requires communication back to the cores for the computation of the proposal importance weight, $\rho_0(\vecX{0})$ (as in \algstref{alg:bf}{st:init:reweight}). Although $\rho_0(\vecX{0})$ can be trivially decomposed into a product of $C$ terms corresponding to the contribution from each core separately \eqref{eq:rho0}, computing $\rho_0(\vecX{0})$ still requires a third communication between cores during initialisation. Such a level of communication between the cores is undesirable, particular as \emph{latency} can make this communication expensive. In this setting, one could attempt to improve the quality of the proposals made on each core (in isolation, noting we do not wish to introduce additional communication), and reduce the level of communication. 

Consider choosing some $\tilde{\boldtheta}\in\Rd$ (for instance, by performing a single pre-processing step and choosing $\tilde{\boldtheta}$ to be the weighted average of the approximate modes of each sub-posterior), we can modify the proposal distribution for the initial draw from each core to be,
\begin{align}
\tilde{f}_c\big(\X{0}{c}\big) 
& \propto \exp \bigg\{- \frac{ \|\X{0}{c} - \tilde {\boldtheta}\|^2 }{2T} \bigg\}  \cdot  {f}_c\big(\X{0}{c}\big), \label{eq:tildef}
\end{align}
compensating for this modification by replacing $\rho_0$ within \algref{alg:bf} with
\begin{align}
\tilde{\varrho}_0(\vecX{0}) 
&:= \exp \bigg\{ \frac{ \|\barX{0} - \tilde {\boldtheta}\|^2 }{2T/C} \bigg\}, \qquad\text{where } \barX{0} = C^{-1} \sum\cores \X{0}{c}. \label{eq:varrho0}
\end{align}
The validity of these modifications can be established by noting that,
\begin{align*}
\tilde{\varrho}_0(\vecX{0}) \cdot \prod\cores \tilde{f}_c\big(\X{0}{c}\big)& \propto \rho_0(\vecX{0}) \cdot \prod\cores f_c\big(\X{0}{c}\big) ,
\end{align*}
and recalling that re-normalisation within \algref{alg:bf} removes the need to compute the constant of proportionality for $\tilde{\varrho}_0$. 

Noting that it is possible to sample from \eqref{eq:tildef} on each core in isolation by rejection sampling (using $f_c$ as a proposal), then this can be done by each core in parallel in advance of initialising the algorithm, and will lead to improved proposal quality. Furthermore, note that computation of the proposal importance weight, $\tilde{\varrho}_0(\vecX{0}) $ in \eqref{eq:varrho0}, \emph{does not} require further communication by the cores. In particular, we have removed two of the three communications required in the original formulation of the initialisation of Bayesian Fusion. This simple modification to the Bayesian Fusion algorithm is presented in \algref{alg:effinit}. 

\begin{algorithm}[ht]
    \caption{Modified Initialisation (in place of \algstref{alg:bf}{st:partinit}))} \label{alg:effinit} 
	\begin{enumerate}
	\item[(aii)] For $i$ in $1$ to $N$, \vspace{-0.25cm}
	\begin{enumerate}
	    \item[A.] \textbf{$\vecX{0,i}$:} For $c$ in $1$ to $C$, simulate $\X{0, i}{c} \sim \tilde{f}_c$. Set $\vecX{0,i}: = \X{0,i}{1:C}$.
	    \item[B.] \textbf{${w'}_{\!\!0,i}$:} Compute un-normalised weight ${w'}_{\!\!0,i}=\tilde \varrho_0(\vecX{0,i})$, as per \eqref{eq:varrho0}.
	\end{enumerate}
	\end{enumerate}
\end{algorithm}


\subsubsection{Propagation of the particle set} \label{sec:effprop}

Considering the iterative propagation of the particle set in \algstref{alg:bf}{st:covar}, note that for each particle we need to compute $\cvecM{j}$ and $\cV{j}$, from \eqref{eq:bfM} and \eqref{eq:bfS2}. In particular, communication between the cores is required as the computation of $\cvecM{j}$ and $\cV{j}$ requires the temporal position of every trajectory over all cores, which then needs communicated back to each core. Upon propagation further communication is required in order to compute the updated importance weight of the particle in \algstref{alg:bf}{st:ue}. This is clearly inefficient, and consequently we wish to minimise the number and size of communications. We would instead like to propagate $\vecX{j-1}$ to $\vecX{j}$ by considering the separate propagation of each of the $C$ parallel processes which compose $\vecX{j-1}$, namely $\X{j-1}{c}$ $\Crange$. To enable this we exploit the following corollary. 

\begin{corollary} \label{cor:parallel1}
Simulating $\vecX{j} \sim$ $\normal\left(\vecX{j-1}; \cvecM{j}, \cV{j} \right)$, the required transition from $\vecX{j-1}$ to $\vecX{j}$ in \algref{alg:bf}, can be expressed as
\begin{align}
\X{j}{c} 
& = \left(\frac{\Delta_j^2}{C(T-t_{j-1})} \right)^{1/2} \boldxi_{j} + \left(\frac{T-t_j}{T-t_{j-1}}\Delta_j\right)^{1/2} \boldeta_{j}^{(c)} + \cM{j}{c}, \label{eq:trans_new}
\end{align}
where $\boldxi_{j}$ and $\boldeta_{j}^{(c)}$ are standard Gaussian vectors,
and $\cM{j}{c}$ is the sub-vector of $\cvecM{j}$ corresponding to the $c$th component. 
\proof See \apxref{apx:parallel}. \qed
\end{corollary}

\corref{cor:parallel1} allows us to propagate the $c$th trajectory from $\vecX{j-1}$ to $\vecX{j}$ in relative isolation, noting that the interaction with the other trajectories solely appears in the mean of the trajectories at the previous iteration ($\barX{j-1}$). Computation of $\barX{j-1}$ can be conducted at the previous iteration of \algref{alg:bf} at the same time as the trajectories are communicated for composition and use in computing the importance weight --- thus removing an unnecessary communication. As we already compute $\barX{0,i}$, as required in the computation of  ${\rho}_0$ in \algstref{alg:bf}{st:init:reweight} (or alternatively as required by $\tilde{\varrho}_0$ in \secref{sec:effinit}), incorporating this into Bayesian Fusion requires only a minor modification of \algref{alg:bf}, as presented in \algref{alg:effprop}. 

\begin{algorithm}[ht]
\caption{Modified Propagation (in place of \algstref{alg:bf}{st:covar}).} \label{alg:effprop} 
\begin{enumerate}
{\setlength\itemindent{1cm} \item[b(ii)A.1.] For $c$ in $1$ to $C$, simulate $\X{j,i}{c}\big|\big(\barX{j-1,i}, \X{j-1,i}{c}\big)$ as per \eqref{eq:trans_new}.}
{\setlength\itemindent{1cm} \item[b(ii)A.2.] Set $\vecX{j,i} := \X{j,i}{1:C}$, and compute $\barX{j,i}:=  \sum\cores \X{j,i}{c}/C$.}
\end{enumerate}
\end{algorithm}


\subsubsection{Updating the particle set weights} \label{sec:effweight}

In many settings it may not be practical to compute the required functionals of each sub-posterior ($f_c$, $\Crange$), and so rendering the evaluation of $\phi_c$, and in turn $\tilde{\rho}_j$ in \algstref{alg:bf}{st:ue}, unfeasible. This may be due to a form of intractability of the sub-posteriors, (such as the settings considered by \citet{as:ar09}), or simply that their evaluation is computationally too expensive (such as in the large data settings considered by \citet{jrssb:pfjr20}). 

This particular issue can be circumvented by noting it is possible to construct an unbiased estimator of $\tilde{\rho}_j$ as follows. 

\begin{corollary} \label{cor:scale_pe}
The estimator
\begin{align*}
\tilde{\varrho}_j := \prod\cores \frac{\Delta_j^{\kappa_c}\cdot e^{-\bar{U}_{j}^{(c)}\Delta_j}}{\kappa_c! \cdot p(\kappa_c|R_c)} \prod^{\kappa_c}_{k_c=1}\left(\bar{U}_{j}^{(c)}- \phihatfn{c}{\X{\chi_{c,k}}{c}}\right),
\end{align*}
where $\kappa_c$, $p$, $R_{c}$ and $\chi_{\cdot}$ are as defined in \thmref{thm:pe}, $\hat{\phi}_c$ is an unbiased estimator of $\phi_c$, and $\bar{U}^{(c)}_j$ is a constant such that $\phihatfn{c}{\X{t}{c}} \leq \bar{U}^{(c)}_j$ for all $\X{t}{c} \sim \brown{j,c}|R_{c}$, is an unbiased estimator of $\tilde{\rho}_j$.
\proof Follows directly from the proof of \thmref{thm:pe} in \apxref{apx:ue}. \qed
\end{corollary}

The estimator $\tilde{\varrho}_j$ in \corref{cor:scale_pe} can be used as an immediate replacement for $\tilde{\rho}_j$ in \algstref{alg:bf}{st:ue}, and simulated by direct modification of \algref{alg:ue}. To take advantage of \corref{cor:scale_pe} one simply has to find a \emph{suitable} unbiased estimator of ${\phi}_c$, which in many settings will be straightforward to construct as ${\phi}_c$ is linear in terms of $\nabla\log f_c(\x)$ and $\Delta\log f_c(\x)$. To find a suitable unbiased estimator to use in place of $\tilde{\rho}_j$, it is important to recognise the penalty for its introduction. In particular, introducing the estimator $\tilde{\varrho}_j$ will (typically) increase the variance of the estimator, which will manifest itself in the variance of the particle set weights in \algref{alg:bf}. To control this we will (typically) require a heavier tailed choice of discrete distribution $p$ in \corref{cor:scale_pe}. An extensive discussion on finding low variance estimators of the type in \thmref{thm:bf} can be found in \apxref{apx:ue}, and can be adapted directly to the setting in \corref{cor:scale_pe}. A concrete application of \corref{cor:scale_pe} can be found in \apxref{apx:scale}, where we consider a simple large data setting. 


\subsubsection{Normalisation and resampling of the particle set} \label{sec:effresamp}

For simplicity in the presentation of Bayesian Fusion, and in our examples, we have employed \emph{multinomial} resampling of the particle set \citep{ieeeprsp:gss93}. It is common within the Sequential Monte Carlo (SMC) literature for alternative resampling schemes to be employed which minimise the introduction of additional variance, and can be used in place of multinomial resampling within Bayesian Fusion (typically with better performance  \citep{tr:dcm05}). These include \emph{systematic} resampling \citep{jcgs:k96}, \emph{stratified} resampling \citep{ieeprsn:ccf99} and \emph{residual} resampling \citep{jcgs:h97,jasa:lc98}. Further detail on resampling schemes can be found in \citet{bk:smcinp}, which includes a review of more advanced methodologies.


\FloatBarrier
\section{Illustrative comparisons with competing methodologies} \label{sec:comparisons}

In this section we contrast Bayesian Fusion with competing methodologies. In \secref{sec:compmcf} we compare it to Monte Carlo Fusion \citep{jap:dpr19}. In \secref{sec:comparisons:competitors} we consider a simple logistic regression model and the relative performance of Bayesian Fusion with the approximate \emph{Consensus Monte Carlo} \citep{ijmsem:setal16} and \emph{Weierstrass Refinement Sampler} \citep{arxiv:wd13} methodologies. Note that in both subsections we are considering pedagogical and illustrative examples in order to illustrate the strengths and weaknesses of each methodology, before returning to more substantive examples in \secref{sec:examples}.


\subsection{Comparison with Monte Carlo Fusion} \label{sec:compmcf}

As discussed in the introduction and \secref{sec:methodology}, the primary motivation for developing Bayesian Fusion is to address the scalability of the (otherwise exact) \emph{Monte Carlo Fusion} approach of \citet{jap:dpr19}. Recall that Monte Carlo Fusion is a rejection sampling based approach, and as a consequence to be computationally practical requires acceptance probabilities which are sufficiently large. However, when contrasting Bayesian Fusion with competing methodologies in \secref{sec:comparisons:competitors} in more realistic (albeit idealised) settings, and when considering the practical application of Bayesian Fusion in \secref{sec:examples}, Monte Carlo Fusion proves to be impractical. As Monte Carlo Fusion
is the progenitor of this methodological approach to the fusion problem of \eqref{eq:prod}, we explicitly contrast the scalability of Monte Carlo Fusion and Bayesian Fusion in idealised settings well suited to Monte Carlo Fusion.

\begin{figure}[ht]
\begin{center}
\subfigure[Increasing $C$ in the \shl setting. 
Here $f(x) \propto \normal(0,1/2m)$, $f_c(x) \propto f^{1/m}(x)$, and with $m\in\{1,\dots,6\}$ 
\label{fig:mcf_comp_incC}]{\includegraphics[width=0.35\textwidth]{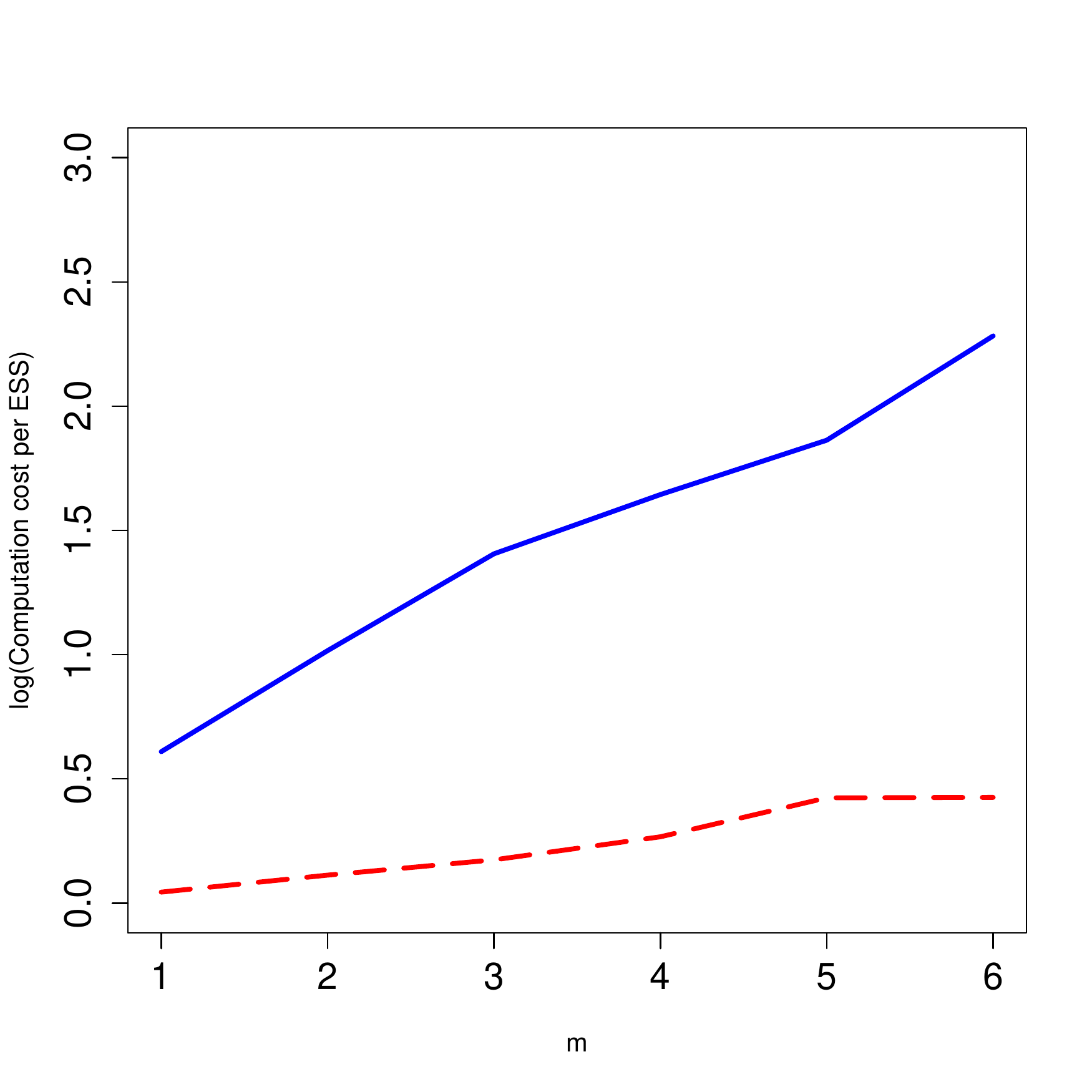}}
\hspace{1em}
\subfigure[Increasing sub-posterior heterogeneity setting: 
Here $f(x)\propto \normal(0,1/2)$ with component densities $\propto \normal(\pm\mu/2, C/2)$, $T=1$
\label{fig:mcf_comp_fixedC}]{\includegraphics[width=0.35\textwidth]{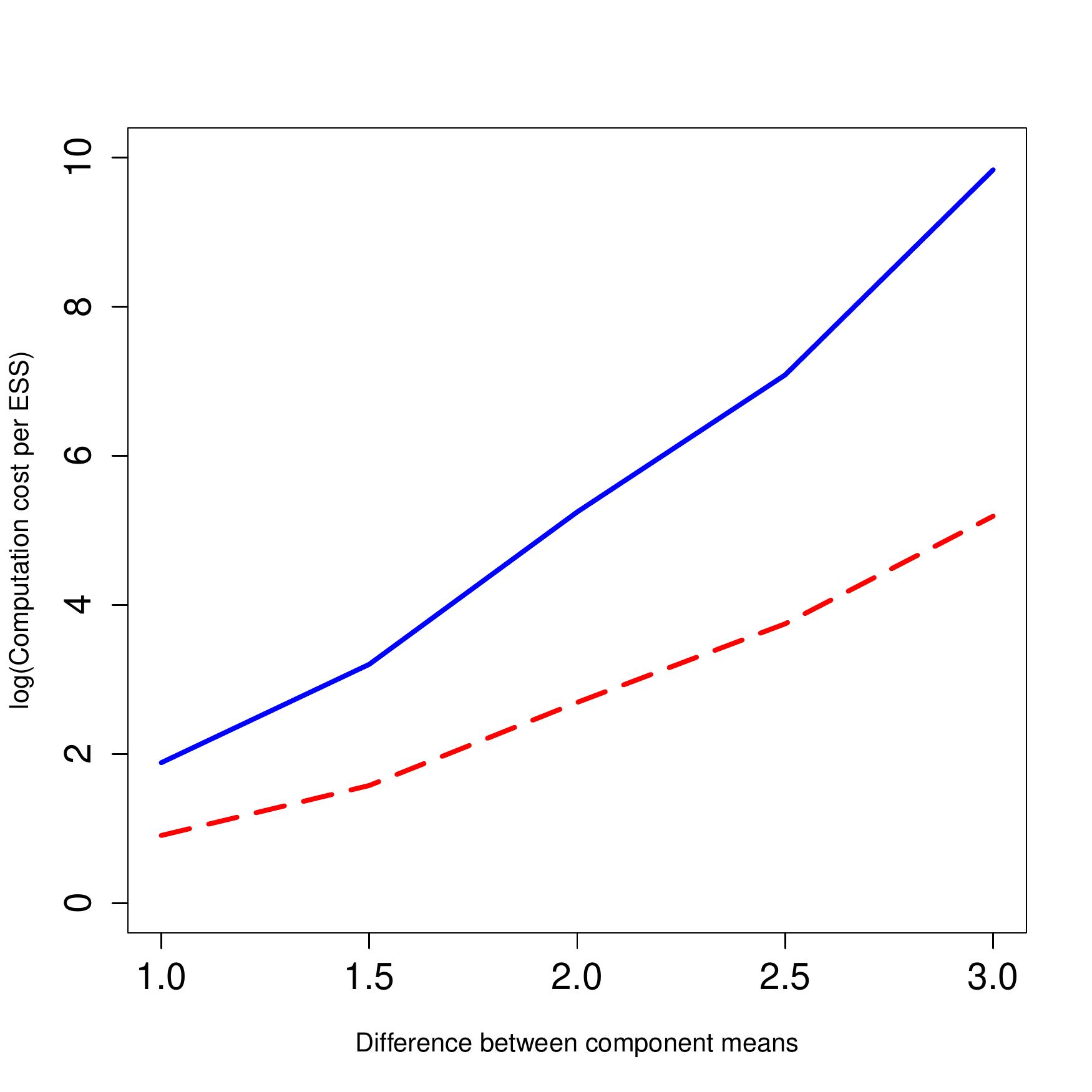}}
\caption{Log computational cost comparison of Bayesian Fusion (red dashed line) and Monte Carlo Fusion (blue solid line) in idealised scaling scenarios.} \label{fig:mcf_comp}
\end{center} 
\end{figure}

In our first scenario, illustrated in \figref{fig:mcf_comp_incC}, we consider the fusion of an increasing number of identical sub-posteriors. The challenge in this setting is that despite the sub-posterior homogeneity, the (fusion) target we want to recover is becoming increasingly concentrated 
relative to the sub-posteriors. In \figref{fig:mcf_comp_fixedC}, we consider the fusion of two Gaussian sub-posteriors with the same variance but with different means, and consider the computational cost of each methodology to achieve a fixed ESS, while fixing $T$, as the means of the sub-posteriors increase in distance from one another. This corresponds (loosely) to the increasing sub-posterior heterogeneity scenario of \secref{sec:guidance}. 

It is clear in both scenarios in \figref{fig:mcf_comp} that even without employing the optimized guidance on the implementation of Bayesian Fusion of \secref{sec:guidance} (in particular, $T$ in \figref{fig:mcf_comp_fixedC}), Bayesian Fusion still has far better scaling properties and offers considerable advantage over Monte Carlo Fusion, even in idealised settings well suited to Monte Carlo Fusion.

\subsection{Comparison with approximate methodologies} \label{sec:comparisons:competitors}

In this section we study the performance of Bayesian Fusion against other competing (approximate) methodologies for simulating from \eqref{eq:prod}. A synthetic data set of size $m=\num{1000}$ was simulated from the following logistic regression model, 
\begin{align}
\label{eq:logistic}
y_i = \left\lbrace \begin{array}{ll} 
1 & \quad\text{with probability } \frac{\exp\{\boldz^T_i \boldbeta\}}{1+\exp\{\boldz^T_i \boldbeta\}},
\\*[10pt] 0 & \quad\text{otherwise}.  \end{array}  \right.
\end{align}
The true $\boldbeta:=(-4,-2)$ (where the first co-ordinate corresponds to the intercept). Each record contained a single covariate in addition to an intercept, which was independently simulated from a Gaussian distribution with mean $0.7$ and variance $1$. The BayesLogit R package was used to fit logistic regression on the entire data set, using a Gaussian prior distribution for both $\beta_1$ and $\beta_2$ with mean $0$ and variance $10$. We term this the \emph{benchmark} posterior distribution, and use it to compare methodologies in this section. 

\begin{figure}[ht]
\begin{center}
\subfigure[$\beta_1$ \label{subfig:subpost:beta1}]{\includegraphics[width=0.35\textwidth]{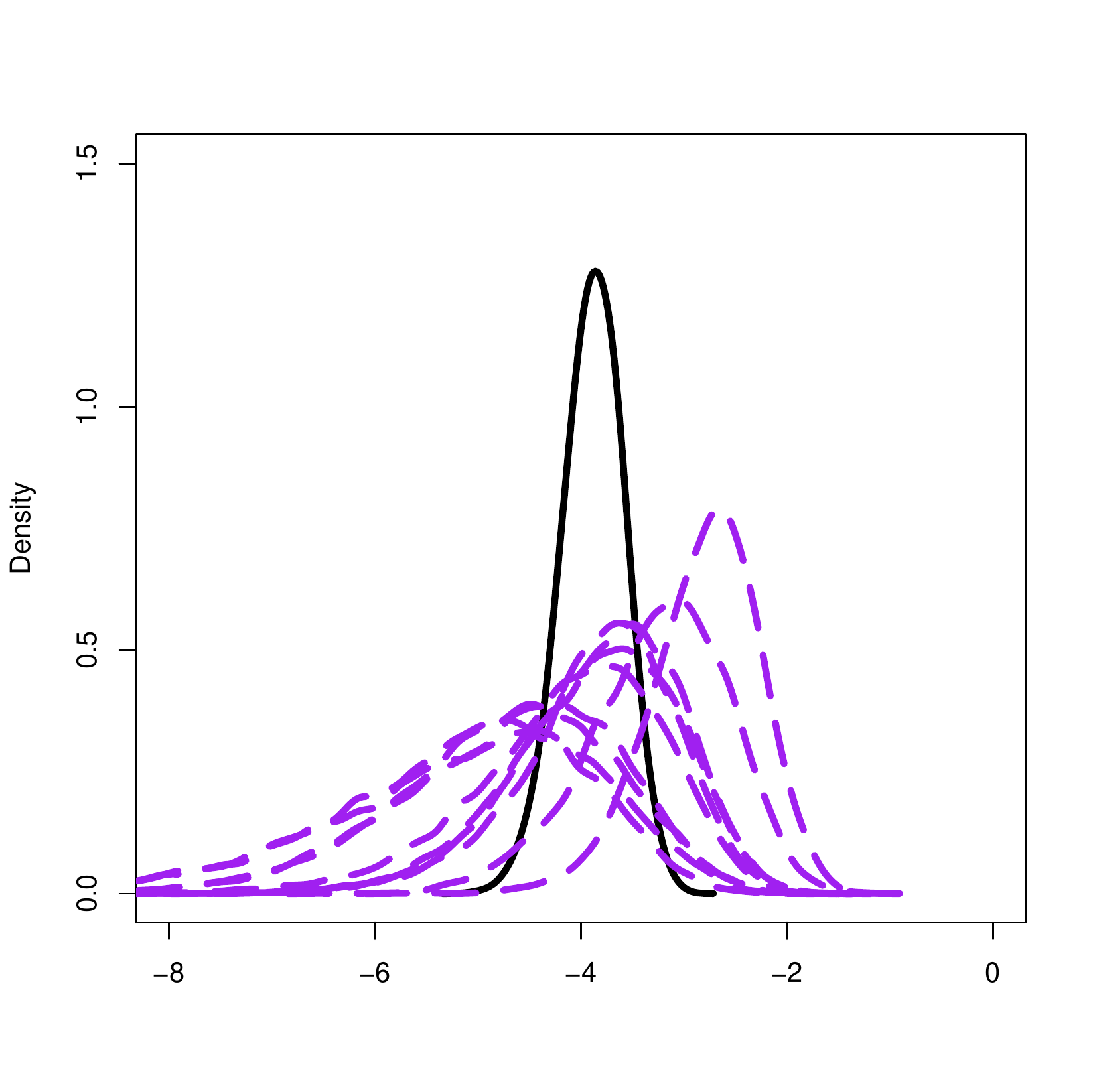}}
\hspace{1em}
\subfigure[$\beta_2$ \label{subfig:subpost:beta2}]{\includegraphics[width=0.35\textwidth]{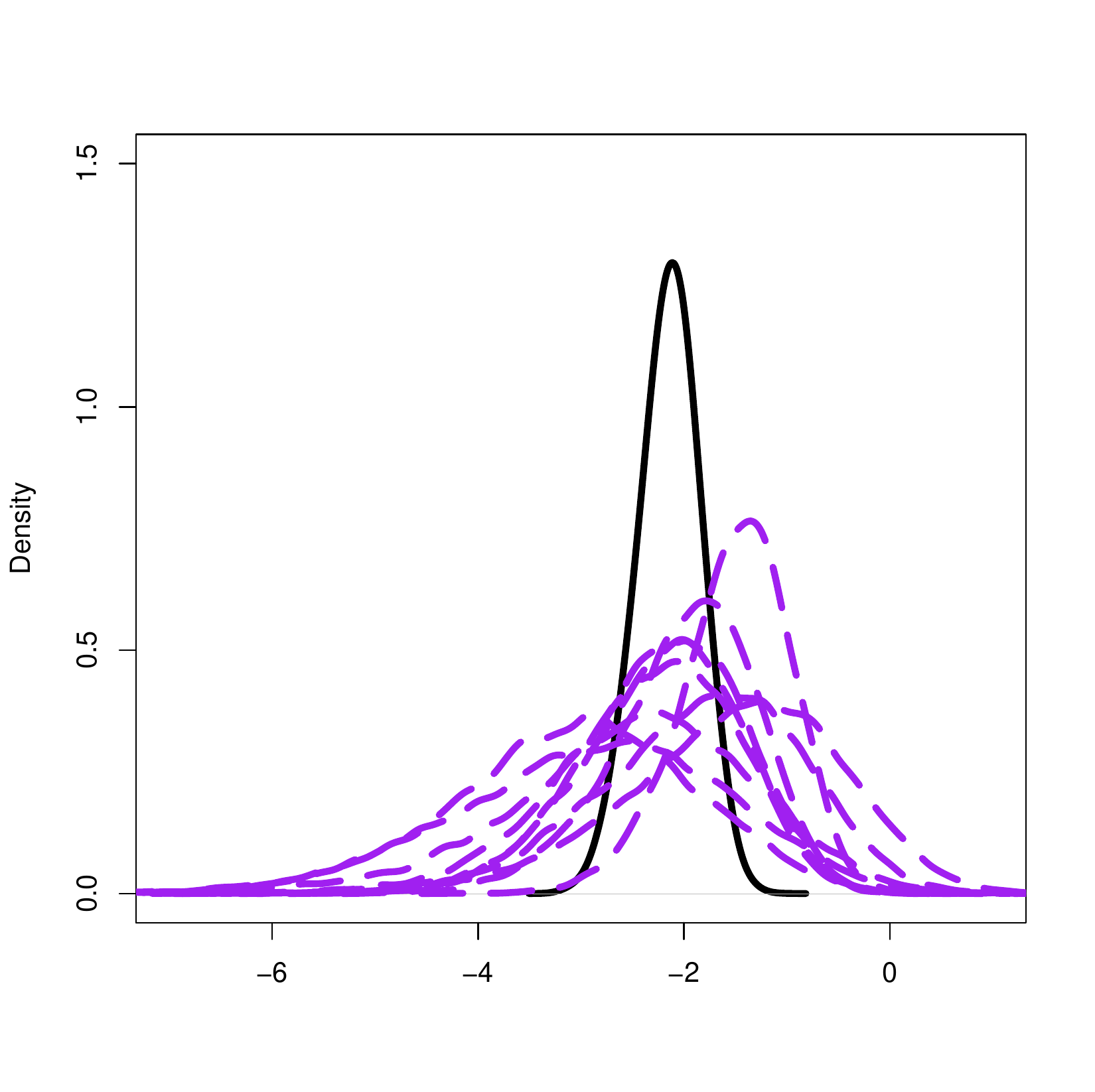}}
\caption{Sub-posterior marginals for logistic regression problem of \secref{sec:comparisons:competitors} with $C=\num{10}$. Black solid line denotes benchmark posterior distribution. Purple dashed lines denote sub-posterior distributions.} \label{fig:subpos}
\end{center} 
\end{figure}

For this data set we consider recovering the posterior distribution by unifying sub-posteriors across an increasing number of cores $C\in\{5,10,20,40\}$. To obtain our $C$ sub-posteriors we evenly distributed the data among the $C$ cores ($m_c=m/C$), and for each core we specified a prior distribution by raising the prior distribution specified for the entire data set to the power $1/C$. We then fit logistic regression using the BayesLogit R package. The specification (data size, parameterisation, and number of cores) we have chosen above is particularly challenging when considering the fusion problem in \eqref{eq:prod} for all methodologies. The lack of data on each core (particularly in the case where $C=40$), and the scarcity of positive responses in the entire data set (we had $\sum_i y_i = 30$), results in sub-posteriors which are both irregular and exhibit a high degree of dissimilarity, and are consequently difficult to unify. To illustrate this the sub-posterior marginals are shown in \figref{fig:subpos} for the case where $C=10$. 

We contrasted Bayesian Fusion (\algref{alg:bf}) with the approximate \emph{Consensus Monte Carlo} (CMC) method of \citet{ijmsem:setal16}, and the approximate \emph{Weierstrass Refinement Sampler} (WRS) of \citet{arxiv:wd13}. Recall that unlike the  approximate schemes in the literature, Bayesian Fusion is an asymptotically consistent methodology, and so increased posterior accuracy can be obtained over these competing methodologies given sufficient computational budget. We also attempted to implement \emph{Monte Carlo Fusion} \citep{jap:dpr19} and the \emph{Weierstrass Rejection Sampler} \citep{arxiv:wd13}, but the small acceptance probabilities resulting from the dis-similarity in the sub-posteriors, together with the number of sub-posteriors, rendered applying these methodologies computationally infeasible.  Bayesian Fusion was implemented following the guidance in \secref{sec:guidance} with a particle set of size $N=\num{10000}$, and the other methodologies were implemented following the guidance suggested by the authors and tuned to this particular data set.

\begin{figure}[hp]
\begin{center}
\subfigure[$\beta_1$ \label{subfig:comparisons:beta1}]{\includegraphics[width=0.35\textwidth]{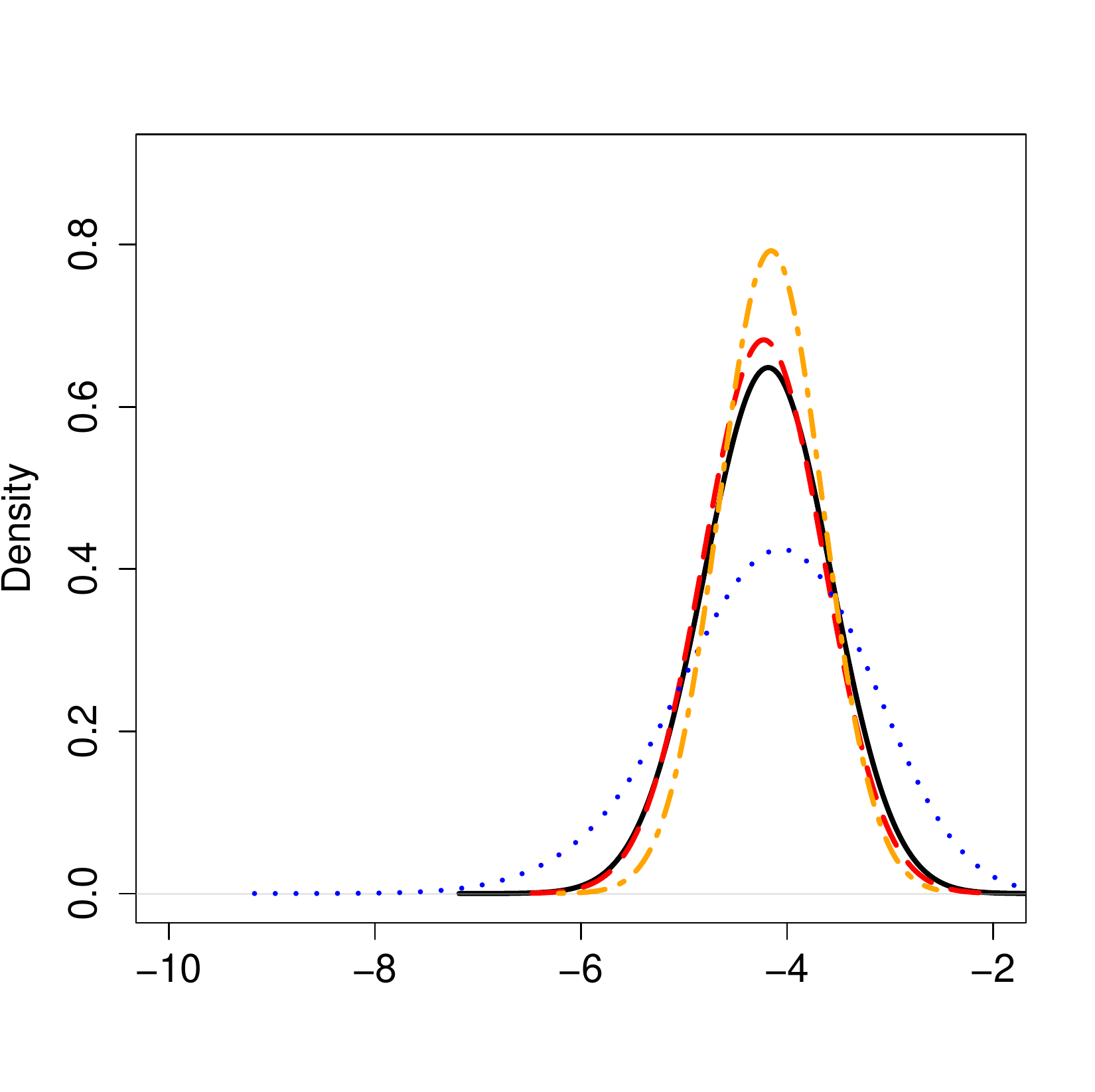}}
\hspace{1em}
\subfigure[$\beta_2$  \label{subfig:comparisons:beta2}]{\includegraphics[width=0.35\textwidth]{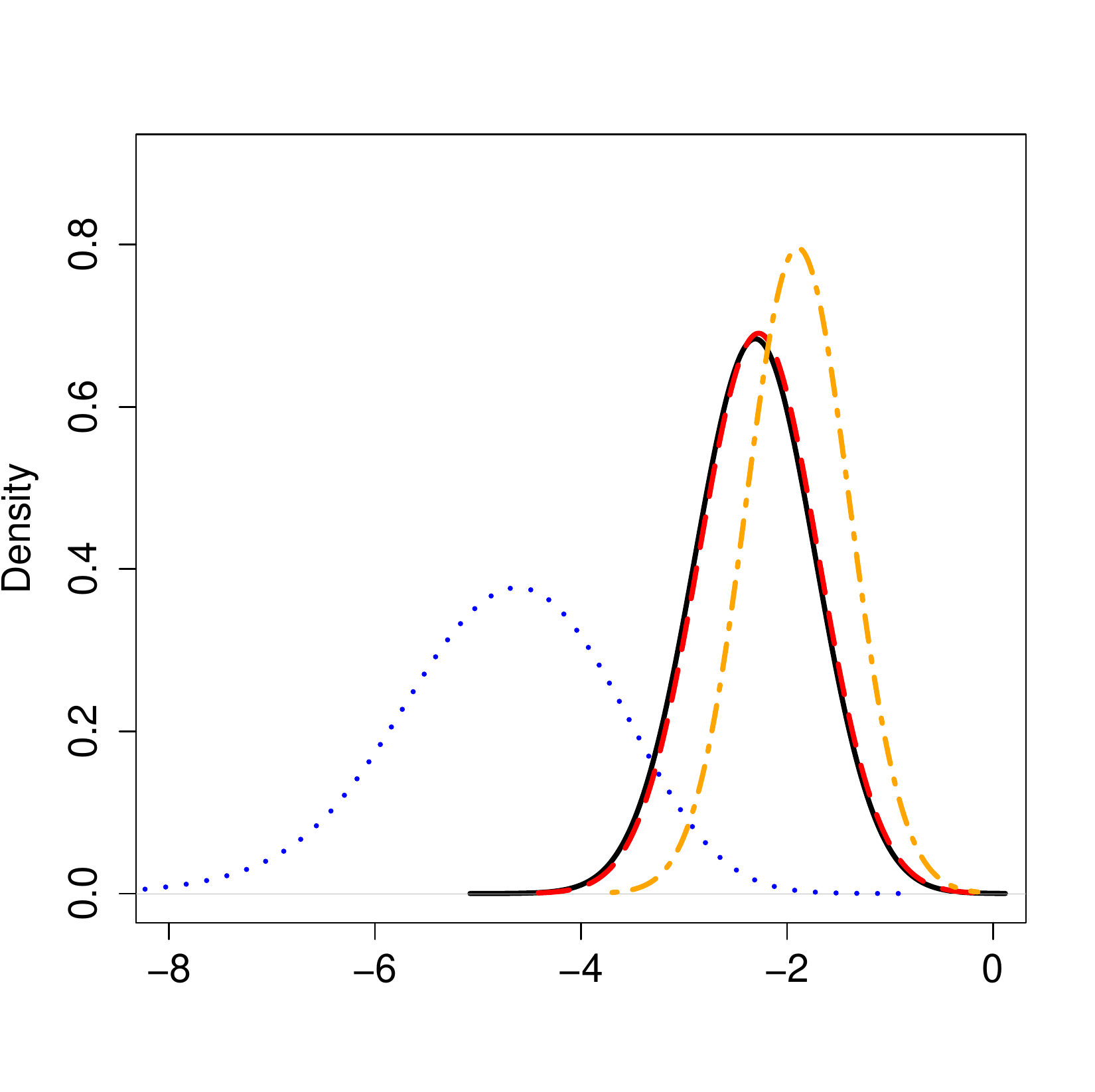}}
\hspace{1em}
\subfigure[IAD with respect to benchmark\label{subfig:comparisons:iad}]{\includegraphics[width=0.35\textwidth]{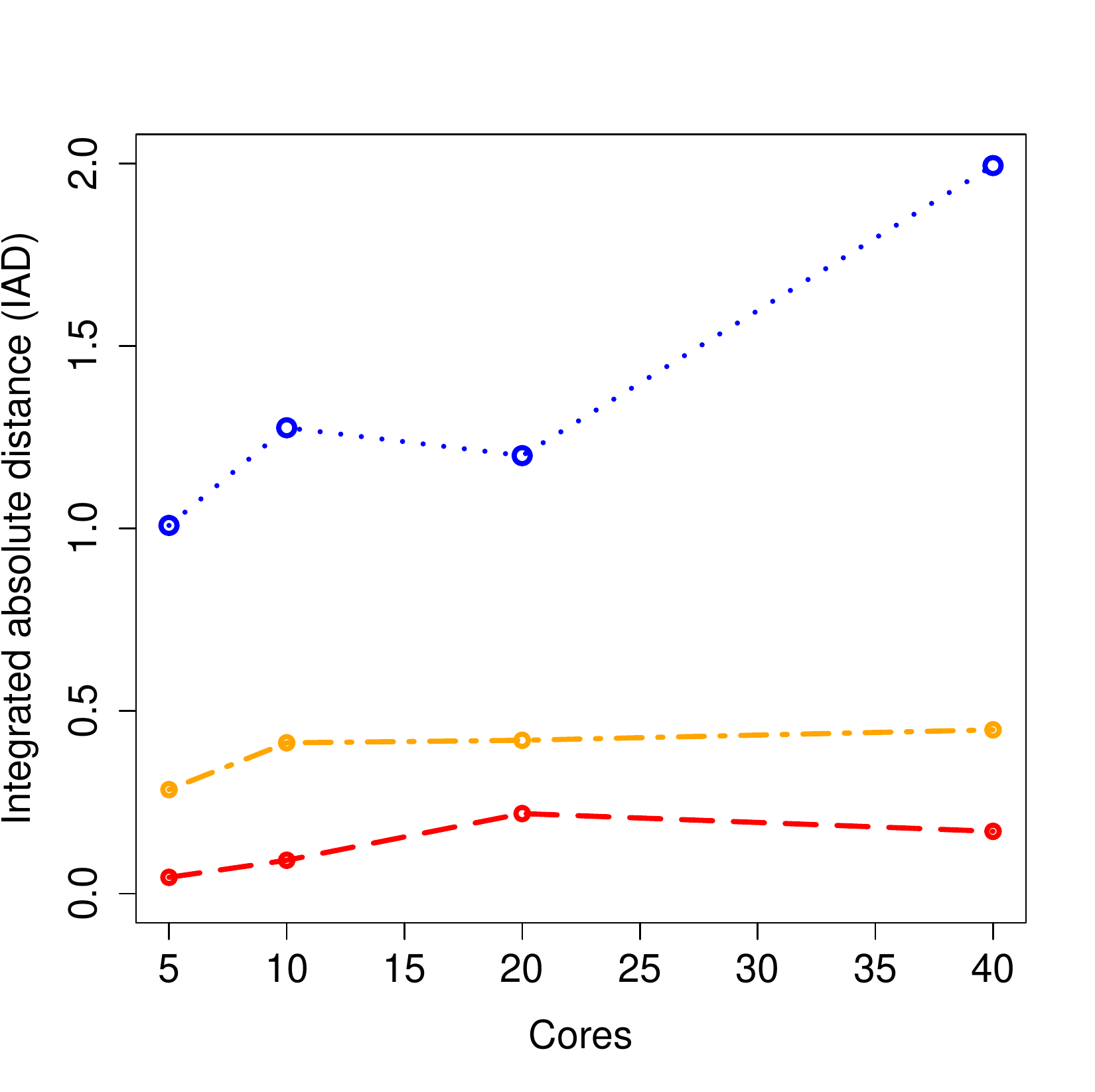}}
\hspace{1em}
\subfigure[Log running times \label{subfig:comparisons:compcost}]{\includegraphics[width=0.35\textwidth]{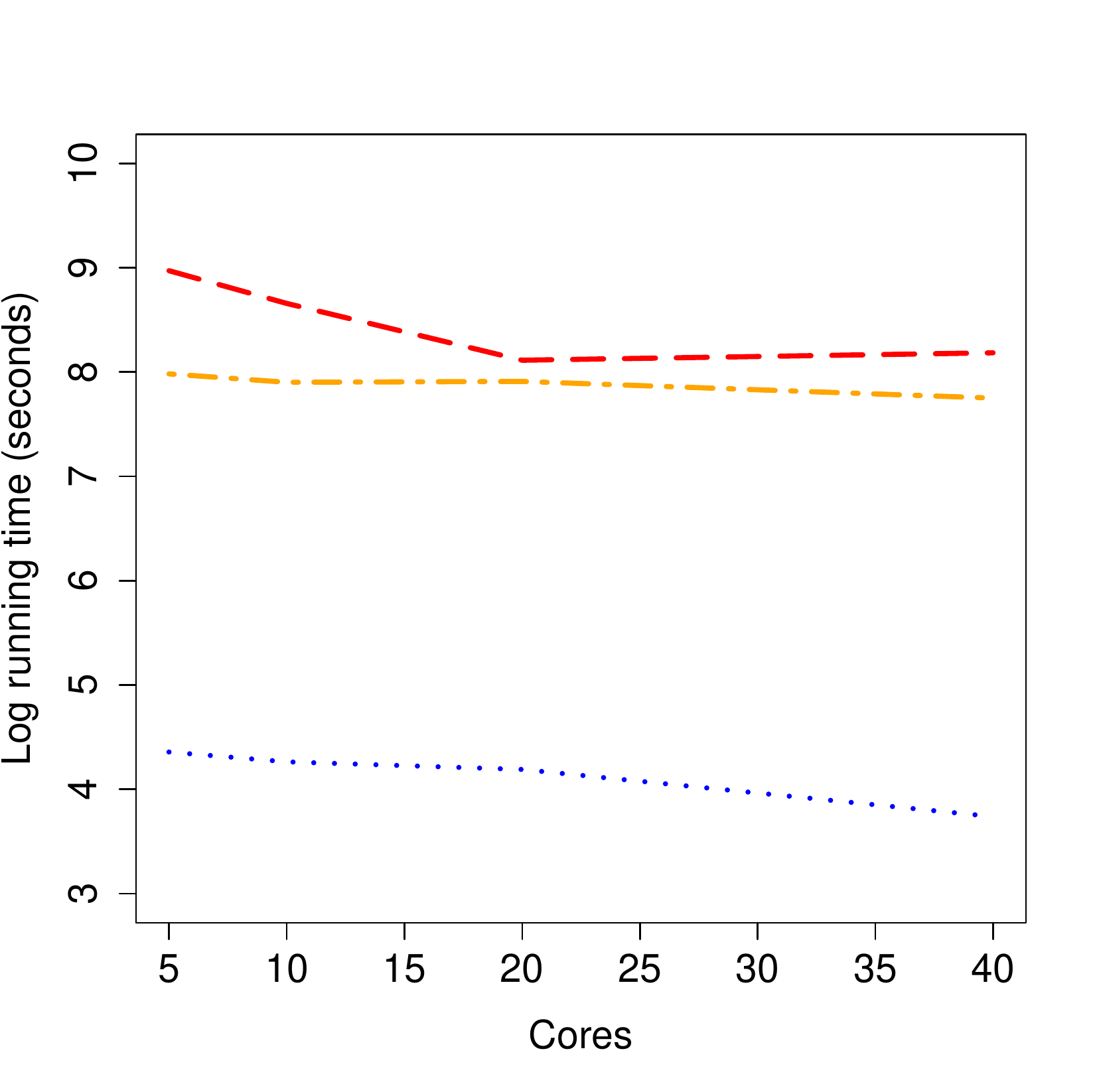}}
\caption{Comparison of competing algorithms to Bayesian Fusion applied to the logistic regression problem of \secref{sec:comparisons:competitors}. Upper marginal densities are of the $C=\num{40}$ case. Black solid lines denote the benchmark fitted target distribution. Red dashed lines denote Bayesian Fusion. Blue dotted lines denote Consensus Monte Carlo (CMC). Orange dotted and dashed lines denote the Weierstrass Refinement Sampler (WRS).} \label{fig:comparison}
\end{center} 
\end{figure}

The marginal densities for each methodology were obtained together with their running times, and are presented in \figref{fig:comparison}. As the competing methodologies are approximate it is important to determine the accuracy of each method for the given computational budget. To do so we define and compute the \emph{Integrated Absolute Distance} (IAD) for each method with respect to the benchmark distribution we obtained earlier. We obtain IAD by simply considering the difference between the marginal for the methodology and the benchmark for each dimension. In particular,
\begin{align}
\text{IAD} := \frac{1}{d} \sum_{j=1}^d \int \Big|\hat f(\theta_j) - f(\theta_j)\Big| \ud \theta_j \in [0,2], \label{eq:iad}
\end{align}
where $f$ is the benchmark distribution and $\hat{f}$ is the distribution obtained from the methodology employed, both computed using a kernel density estimate as necessary. It is important to note that as Bayesian Fusion is asymptotically consistent, and so when considering IAD and the associated running time (\figref{subfig:comparisons:iad} and \figref{subfig:comparisons:compcost} respectively) this is one possible combination -- IAD can be improved to a user-specified accuracy for Bayesian Fusion given sufficient computational budget (which is not true for other schemes in the literature). 

Considering solely the computational cost of each scheme (\figref{subfig:comparisons:compcost}), it is clear Consensus Monte Carlo (CMC) is substantially faster than both Bayesian Fusion and the Weierstrass Refinement Sampler (WRS). This is largely due to the desirable lack of communication between cores CMC achieves. However, scrutinising the marginals in \figref{subfig:comparisons:beta1} and \figref{subfig:comparisons:beta2}, it is apparent that even for this standard two dimensional logistic regression problem, CMC incorrectly estimates both the location of the modes (in particular $\beta_2$) and tail structure of the benchmark distribution. This is indeed summarised by the IAD in \figref{subfig:comparisons:iad}, further suggesting the methodology is not robust to unifying the target distribution with increasing numbers of cores, $C$. The WRS substantially improves upon CMC, appears to better capture both the mode and tail structure of the benchmark distribution, and seems to be more robust to increasing numbers of cores. However, Bayesian Fusion recovers the benchmark distribution for only a modest increase in computational budget over the WRS. Indeed, the IAD obtained for Bayesian Fusion in \figref{subfig:comparisons:iad} is driven by Monte Carlo error (and not approximation error), and so could be further improved if necessary by increasing the computational budget. In truly large data or distributed network settings, Bayesian Fusion could be further optimised when the extensions discussed in \secref{sec:practical} are incorporated.

\section{Examples} \label{sec:examples}


\subsection{U.S. Census Bureau population surveys} \label{sec:ex1}

In this example we applied Bayesian Fusion to the 1994 and 1995 U.S. Census Bureau population surveys, obtained from \citet{uci:bl13}, and of size $m=\num{199523}$. For the purposes of this example we investigated the effect of education on gross income. We took \emph{gross income} as our observed data, treating it as a binary taking a value of one if income was greater than \$\num{50000}. An income in excess of \$\num{50000} is moderately rare with only \num{12382} individuals exceeding this threshold (which represents approximately \num{6}\% of the data). In addition to the intercept, we extracted three further \emph{education} covariates indicating educational stages attained by the individual (each of which were binary). We then fitted the logistic regression model of \eqref{eq:logistic}, with prior distribution $\normal(0,10\identity{4})$, to the data set to obtain a \emph{benchmark} posterior distribution to assess the quality of Bayesian Fusion. The data size for this example exceeded the capabilities of the BayesLogit R package used in \secref{sec:comparisons:competitors}, and so we instead obtained our benchmark by applying Markov chain Monte Carlo to the full data set. 
 
For this data set we considered recovering the benchmark distribution by unifying sub-posteriors across an increasing number of cores $C\in\{10,20,40\}$. We again contrasted Bayesian Fusion with Consensus Monte Carlo (CMC) and the Weierstrass Refinement Sampler (WRS). To construct sub-posteriors we distributed the data among the available $C$ cores, and fit the logistic regression model of \eqref{eq:logistic} to each using a prior obtained by raising the prior distribution specified for the entire data set to the power $1/C$. In contrast with \secref{sec:comparisons:competitors}, in constructing the sub-posteriors we did not allocate the data randomly (or evenly) among the $C$ cores. This is more representative of a typical application, and introduces dis-similarity in the sub-posteriors, particularly when observations or covariates are rare. For instance, in the case where $C=40$, three of the cores contained data comprising in excess of $\num{99}$\% of the individuals earning in excess of \$\num{50000}. 

\begin{figure}[hp]
\begin{center}
\subfigure[$\beta_1$ \label{subfig:ex1marg:beta1}]{\includegraphics[width=0.35\textwidth]{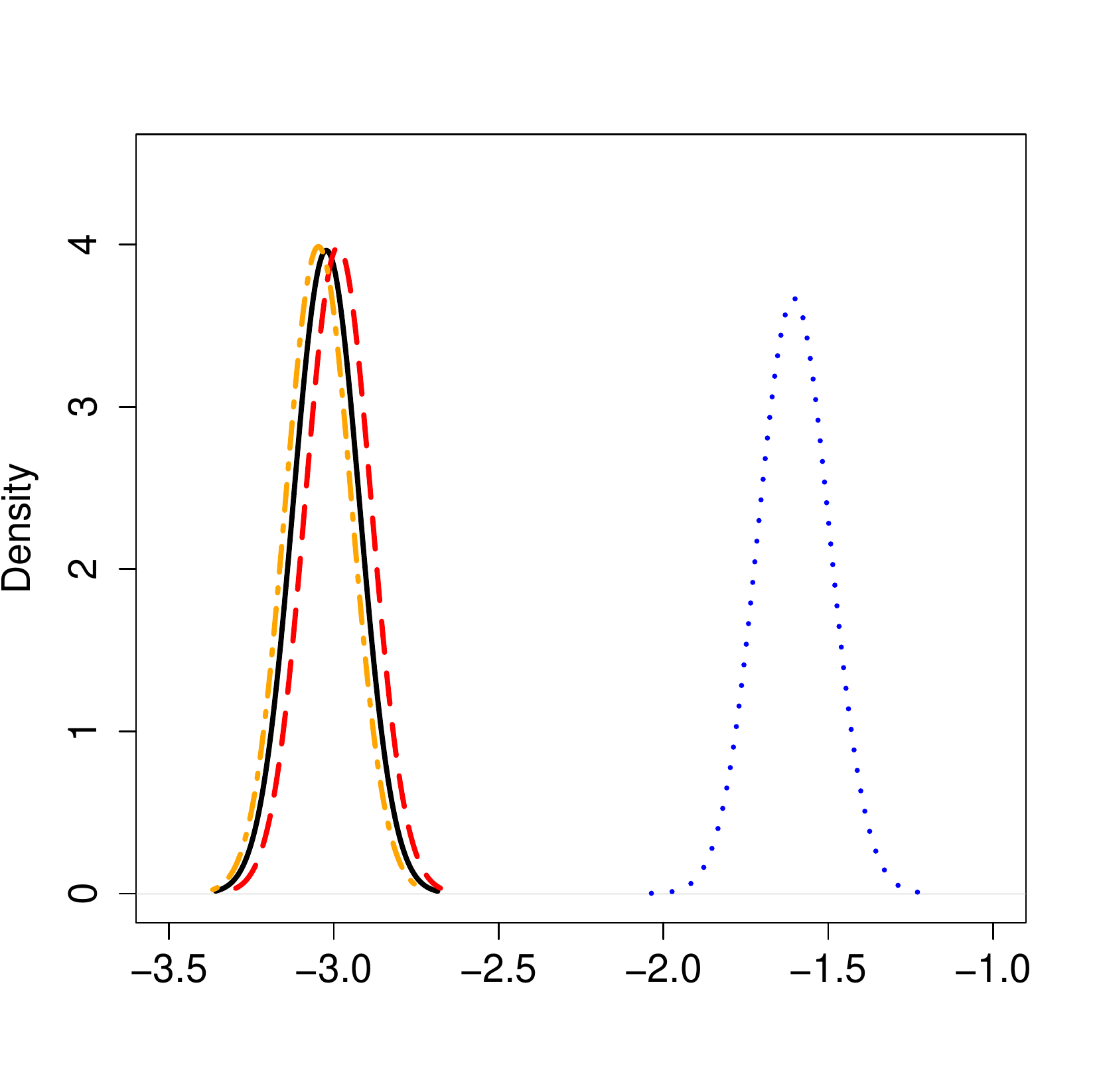}}
\hspace{1em}
\subfigure[$\beta_2$ \label{subfig:ex1marg:beta2}]{\includegraphics[width=0.35\textwidth]{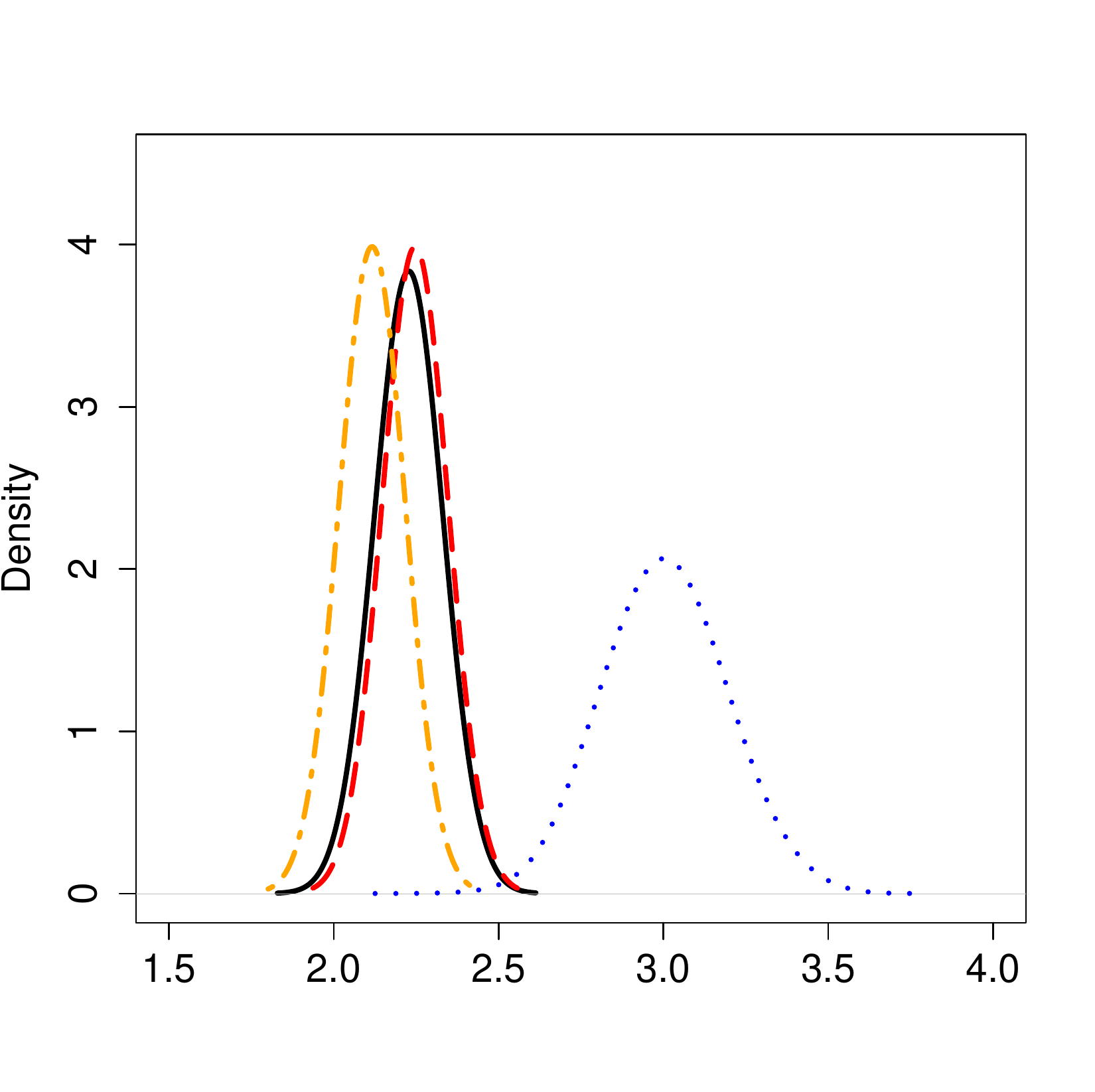}}
\hspace{1em}
\subfigure[$\beta_3$ \label{subfig:ex1marg:beta3}]{\includegraphics[width=0.35\textwidth]{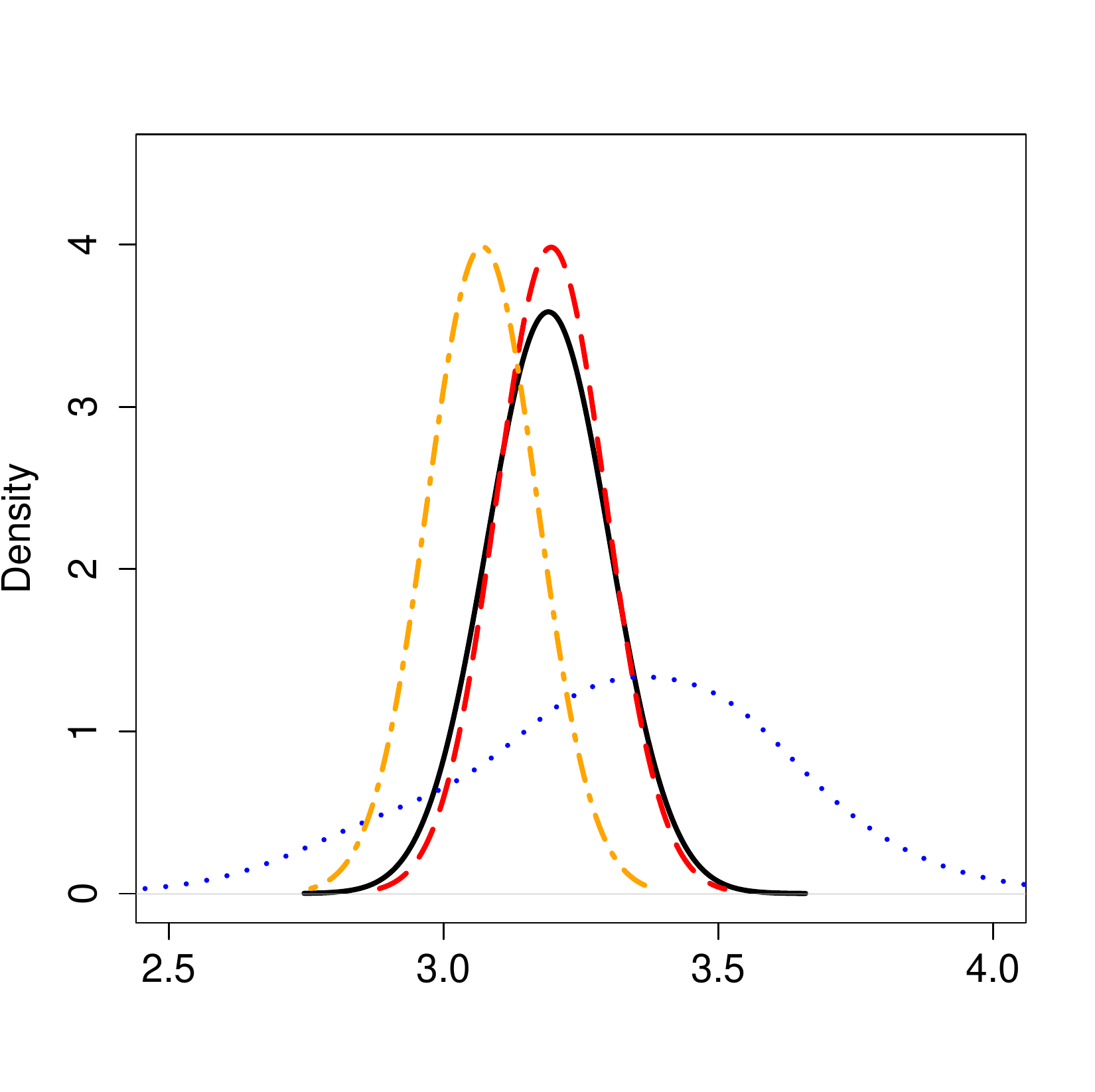}}
\hspace{1em}
\subfigure[$\beta_4$ \label{subfig:ex1marg:beta4}]{\includegraphics[width=0.35\textwidth]{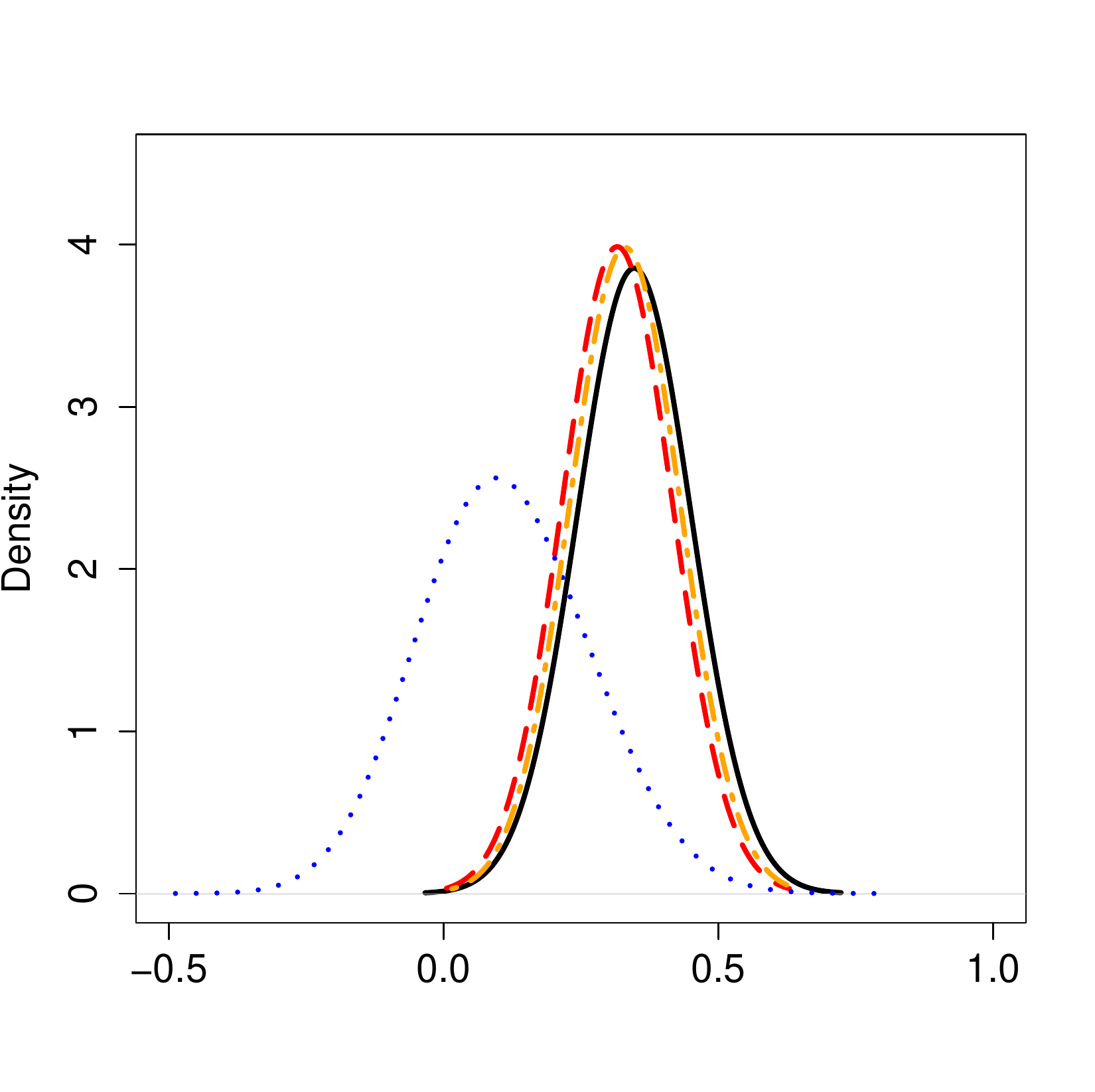}}
\caption{Marginal density estimates of Bayesian Fusion and competing algorithms applied to the U.S. Census Bureau population survey data set of \secref{sec:ex1}. Black solid lines denote the benchmark fitted target distribution. Red dashed lines denote Bayesian Fusion. Blue dotted lines denote Consensus Monte Carlo (CMC). Orange dotted and dashed lines denote the Weierstrass rejection sampler (WRS).} \label{fig:ex1marg}
\end{center} 
\end{figure}

Bayesian Fusion was implemented with a particle set of size $N=\num{30000}$, and following the guidance of \secref{sec:guidance}. CMC and WRS were again implemented as fairly as possible, following the guidance suggested by the authors. The marginal densities are presented in \figref{fig:ex1marg} for the $C=40$ setting, and in \figref{fig:ex1perf} we again present the Integrated Absolute Distance (IAD, see \eqref{eq:iad}) of each methodology with respect to the benchmark distribution, together with their computational costs for the range of cores considered. 

For this data set CMC performs extremely poorly, capturing neither the marginals of the benchmark distribution (particularly, $\beta_1$ and $\beta_2$) or showing any robustness with respect to the numbers of cores. Considering the marginals in \figref{fig:ex1marg}, the  WRS substantially improves upon CMC (only having difficulty capturing the benchmark for $\beta_2$ and $\beta_3$). However, for slightly more computational expenditure (\figref{subfig:ex1perf:compcost}), Bayesian Fusion substantially improves upon IAD over the WRS (\figref{subfig:ex1perf:iad}), and also appears to show robustness with increasing $C$. 
\begin{figure}[hp]
\begin{center}
\subfigure[IAD with respect to benchmark \label{subfig:ex1perf:iad}]{\includegraphics[width=0.35\textwidth]{./Figures/
bigdata_diff_C40_weier_without5cores.pdf}}
\hspace{1em}
\subfigure[Log running times \label{subfig:ex1perf:compcost}]{\includegraphics[width=0.35\textwidth]{./Figures/
dataanalysisruntime_without5cores.pdf}}
\caption{Performance of Bayesian Fusion and competing algorithms applied to the U.S. Census Bureau population survey data set of \secref{sec:ex1}. Black solid lines denote the benchmark fitted target distribution. Red dashed lines denote Bayesian Fusion. Blue dotted lines denote Consensus Monte Carlo (CMC). Orange dotted and dashed lines denote the Weierstrass rejection sampler (WRS).} \label{fig:ex1perf}
\end{center} 
\end{figure}


\subsection{U.K. road accidents} \label{sec:ex2}

In this example we considered the `Road Safety Data' data set published by the Department for Transport of the U.K. government \citep{gov:road_safety_data}. It comprises road accident data set from 2011--2018, and in total is of size $m=\num{1111320}$. We treated our observation for each record to be binary taking a value of one if a \emph{severe} accident was recorded. In total in the full data set there were \num{13358} such severe accidents. We selected a number of covariates to investigate what effect (if any) they have on accident severity. In particular, and in addition to an intercept, we considered \emph{road speed limit}, \emph{lighting condition} (which we treated as binary taking a value of one if lighting was \emph{good}, and zero if lighting was \emph{poor}), and \emph{weather condition} (binary, taking one if \emph{good} and zero if \emph{poor}). The logistic regression model of \eqref{eq:logistic} was fit to the data set, together with a $\normal(0,10\identity{4})$ prior distribution. Our \emph{benchmark} posterior distribution was obtained by applying Markov chain Monte Carlo to the full data set. 

We again considered recovering the benchmark distribution by unifying sub-posteriors across an increasing number of cores $C\in\{10,20,40\}$. The sub-posteriors were obtained following the same approach as \secref{sec:ex1} --- with the allocation of data to each core being in temporal order. We contrasted Bayesian Fusion with a particle set of size $N=\num{30000}$, with fair implementations of Consensus Monte Carlo (CMC) and the Weierstrass Refinement Sampler (WRS). Marginal densities for the $C=40$ setting are presented in \figref{fig:ex1marg}, and IAD \eqref{eq:iad} with respect to the benchmark together with their computational costs for the range of cores considered. The results are in keeping with those of \secref{sec:ex1}. CMC performs extremely poorly, and for a modest increase in computational budget Bayesian Fusion obtains substantially better results than the WRS. 
\begin{figure}[hp]
\begin{center}
\subfigure[IAD with respect to benchmark \label{subfig:ex2perf:iad}]{\includegraphics[width=0.35\textwidth]{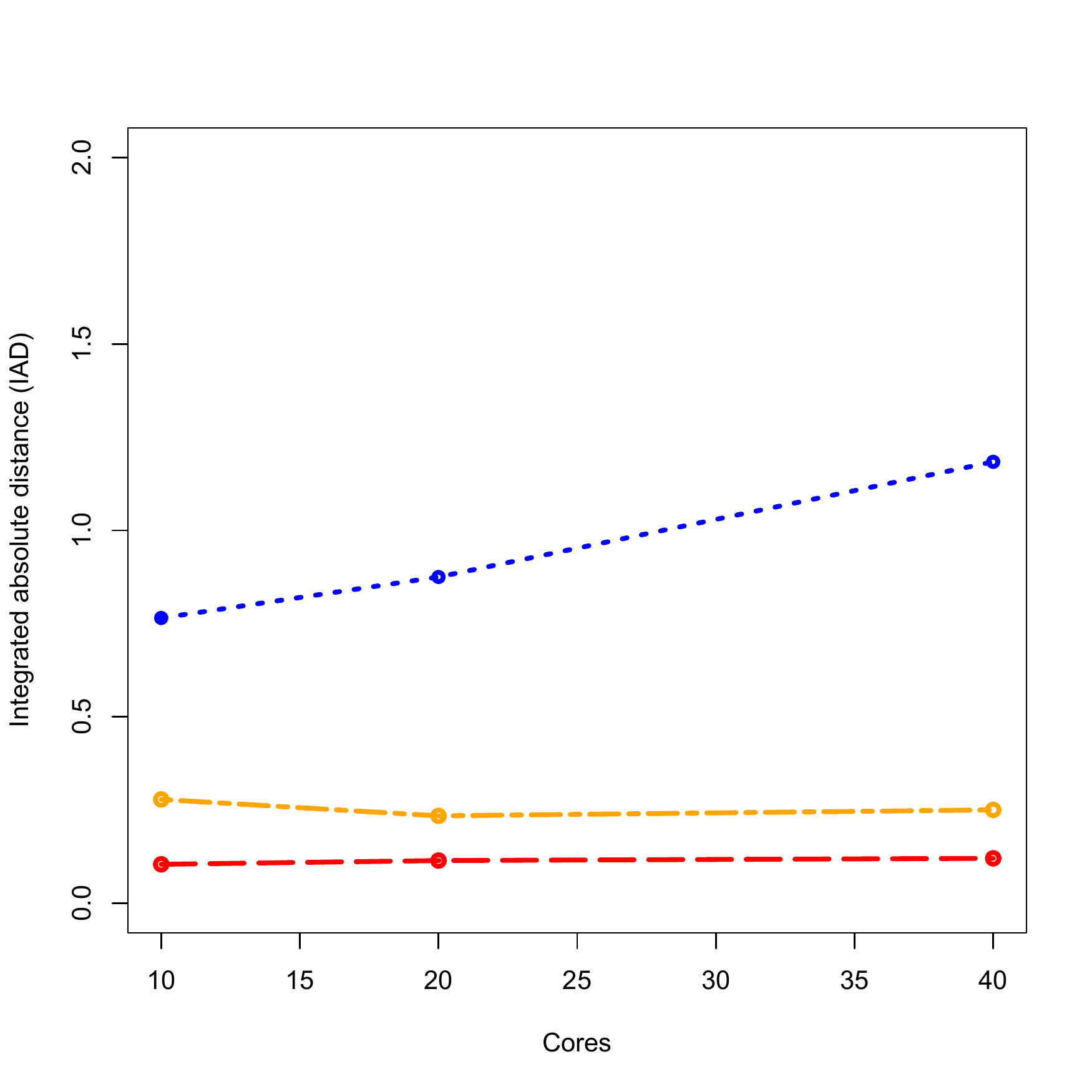}}
\hspace{1em}
\subfigure[Log running times \label{subfig:ex2perf:compcost}]{\includegraphics[width=0.35\textwidth]{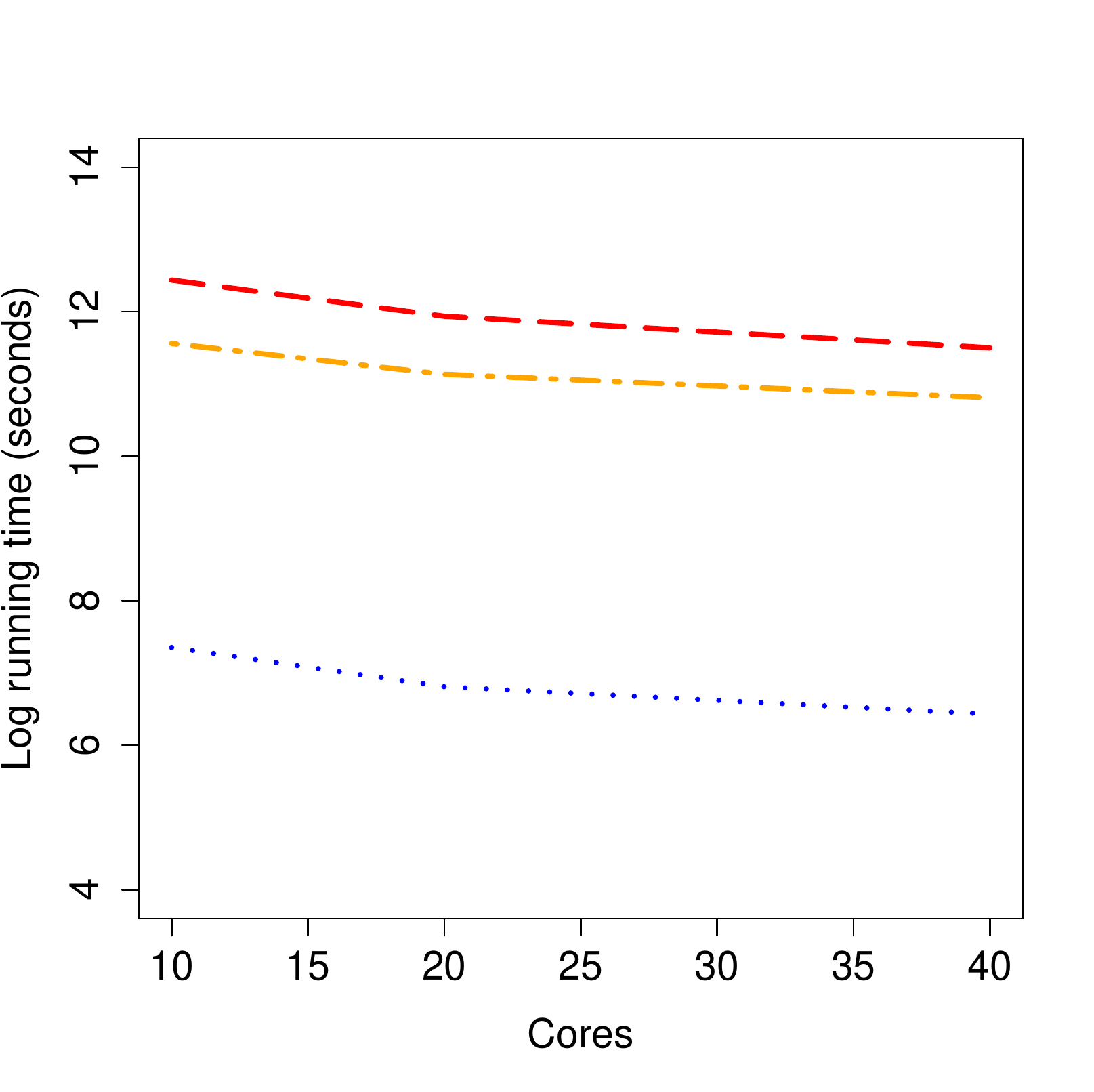}}
\caption{Performance of Bayesian Fusion and competing algorithms applied to the U.K. road accident data set of \secref{sec:ex2}. Black solid lines denote the benchmark fitted target distribution. Red dashed lines denote Bayesian Fusion. Blue dotted lines denote Consensus Monte Carlo (CMC). Orange dotted and dashed lines denote the Weierstrass rejection sampler (WRS).} \label{fig:ex2perf}
\end{center} 
\end{figure}
\begin{figure}[hp]
\begin{center}
\subfigure[$\beta_1$ \label{subfig:ex2marg:beta1}]{\includegraphics[width=0.35\textwidth]{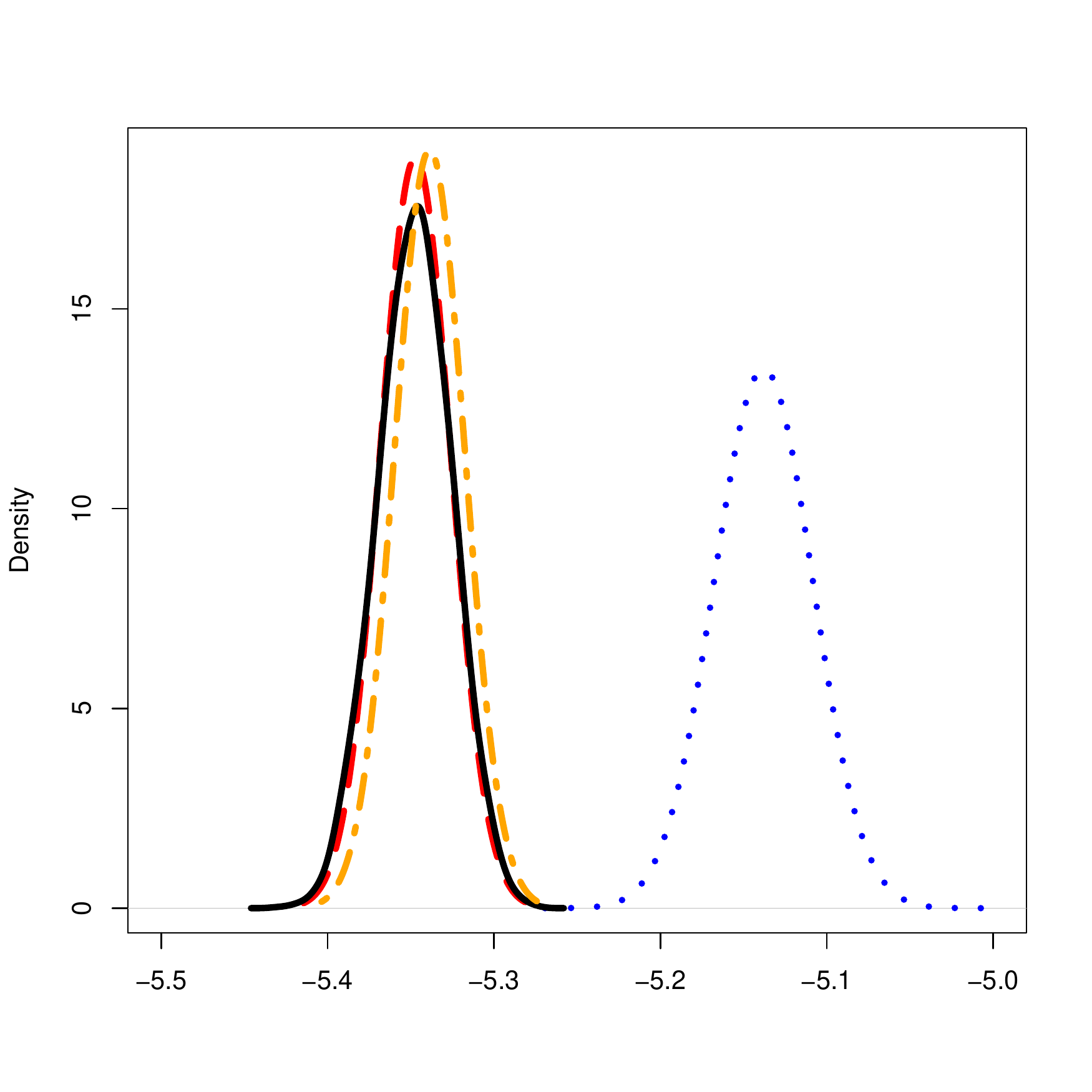}}
\hspace{1em}
\subfigure[$\beta_2$ \label{subfig:ex2marg:beta2}]{\includegraphics[width=0.35\textwidth]{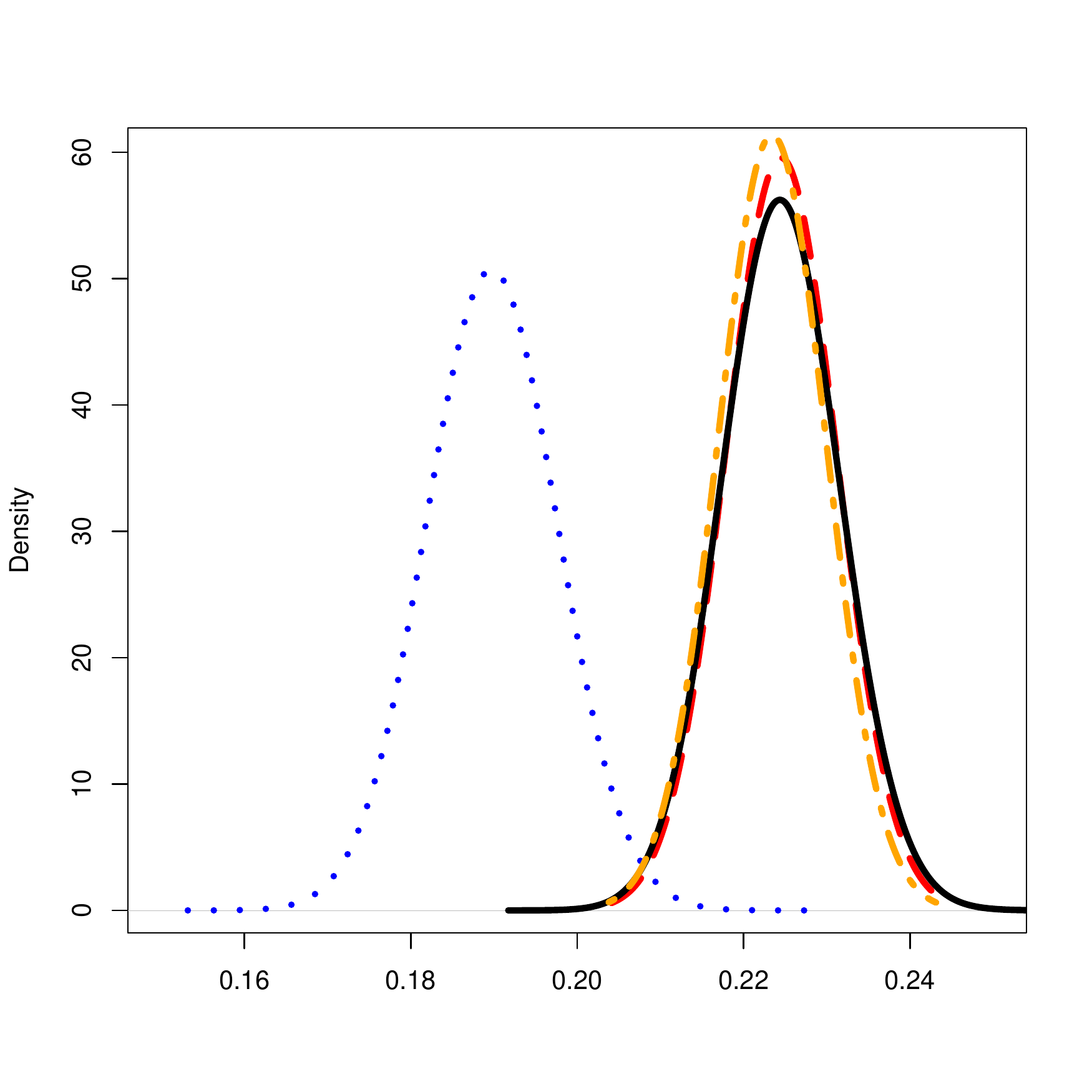}}
\hspace{1em}
\subfigure[$\beta_3$ \label{subfig:ex2marg:beta3}]{\includegraphics[width=0.35\textwidth]{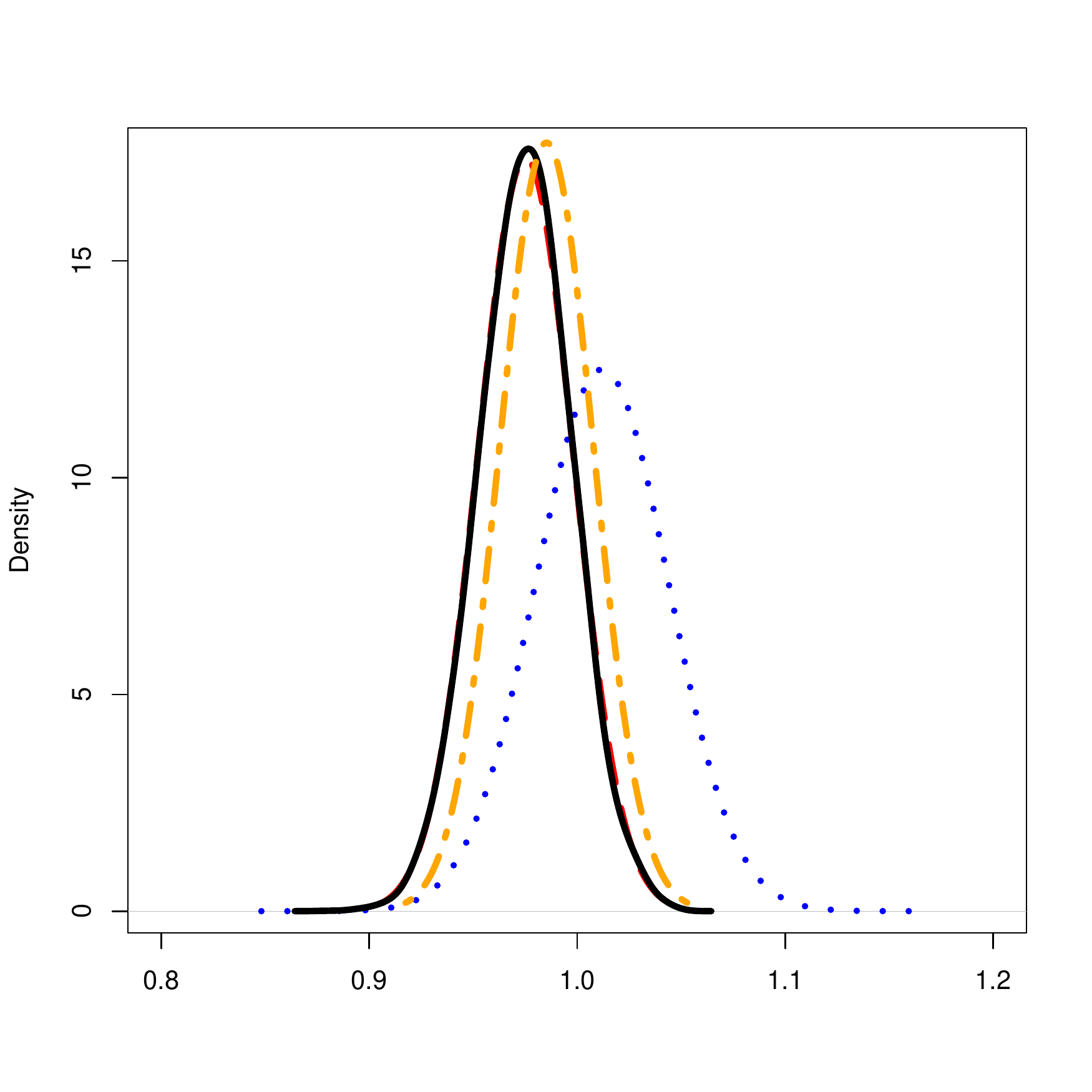}}
\hspace{1em}
\subfigure[$\beta_4$ \label{subfig:ex2marg:beta4}]{\includegraphics[width=0.35\textwidth]{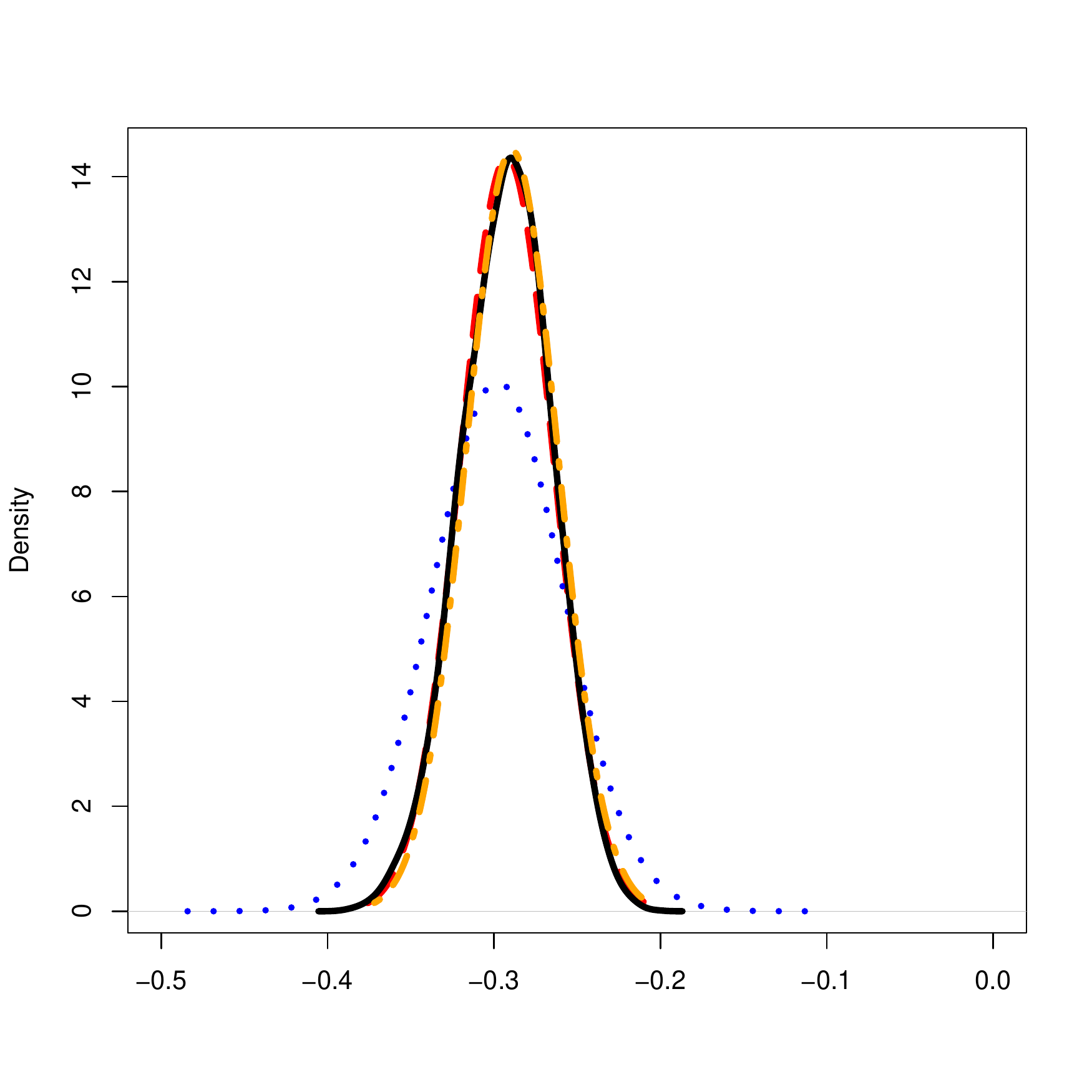}}
\caption{Marginal density estimates of Bayesian Fusion and competing algorithms applied to the U.K. road accident data set of \secref{sec:ex2}. Black solid lines denote the benchmark fitted target distribution. Red dashed lines denote Bayesian Fusion. Blue dotted lines denote Consensus Monte Carlo (CMC). Orange dotted and dashed lines denote the Weierstrass rejection sampler (WRS).} \label{fig:ex2marg}
\end{center} 
\end{figure}


\section{Conclusions} \label{sec:conclusions}

In this paper we have developed a theoretical framework, and scalable sequential Monte Carlo (SMC) methodology, for unifying distributed statistical analyses on shared parameters from multiple sources (which we term \emph{sub-posteriors}) into a single coherent inference. The work significantly extends the theoretical underpinning, and addresses the practical limitations, of the \emph{exact Monte Carlo Fusion} approach of \citet{jap:dpr19}. Monte Carlo Fusion is a rejection-sampling based approach which is the first methodology in general settings in which the product pooled posterior distribution of the \emph{fusion problem} in \eqref{eq:prod} is recovered without approximation. However, it lacked scalability with respect to the number of sub-posteriors to be unified, and robustness with sub-posterior dis-similarity. This is addressed by the \emph{Bayesian Fusion} approach introduced in this paper, resulting in a methodology which both recovers the correct target distribution and is computationally competitive with leading approximate schemes. Fundamental to the methodology introduced is the construction of the fusion measure via an SMC procedure driven by the SDE in \eqref{eq:SDE}, and leading to \algref{alg:bf}.

In addition to the theoretical and methodological development of Bayesian Fusion presented in \secref{sec:theory}, in \secref{sec:guidance} we provide concrete theory and guidance on how to choose the free parameters of \algref{alg:bf} to ensure robustness with increasing numbers of sub-posteriors, and robustness with sub-posterior dis-similarity. This includes in \secref{sec:practical} providing extensive practical guidance on how Bayesian Fusion may be implemented by practitioners, including in distributed network and big data settings.

\secref{sec:comparisons} provides an extensive comparison of the performance of Bayesian Fusion, and competing approximate methodologies, in an idealised synthetic data setting. In \secref{sec:examples} we apply Bayesian Fusion to real data sets, including the \emph{`U.S.\ Census Bureau population surveys'} data set in \secref{sec:ex1}, and a \emph{`U.K.\ road accidents'} data set in \secref{sec:ex2}, together with the competing approximate methodologies. In all settings our implementation of Bayesian Fusion performs extremely well, demonstrating appreciable scope for its broader application.  

One of the key advantages of Bayesian Fusion is that it is underpinned methodologically by sequential Monte Carlo (SMC), which allows us to leverage many of the existing theoretical results and methodology found in that literature. As is typical within SMC it is desirable to attempt to minimise the discrepancy between the sequence of proposal and target distributions. In our setting this entails ensuring the propagated temporal marginal of $g$ in \eqref{eq:bf} (say $g^{N}_{j-1}$), is well-matched with the following temporal marginal of $g$ (say $g^{N}_j$). Although not emphasised within the main text, there is clear scope to improve Bayesian Fusion in this sense by modifying the diffusion theory presented in its development (\apxref{apx:bfproof}), to one which better incorporates information about each sub-posterior (for instance, this could be knowledge of the volume of data on each core). One approach explored in \citet[Sec.\ 4]{jap:dpr19} is to consider an underlying Ornstein-Uhlenbeck proposal measure (appropriately parameterised), which could well-approximate posterior distributions which are approximately Gaussian, and thus lead to better propagation of temporal marginals if incorporated within Bayesian Fusion.  Another feasible direction is to estimate the covariance structure of each sub-posterior and transform the $C$ spaces accordingly, which would lead to the Brownian proposals being more attuned to the target distribution \citep{chan:poll:rob:21}. This would be equivalent to modifying the \emph{Fusion measure} in \eqref{eq:fusion_reg}, in which the transition densities for each sub-posterior are that of a Langevin diffusion with unit volatility, to one with volatility which matches the covariance structure of its respective sub-posterior. The theory remains valid provided the transition densities of the chosen diffusion in \eqref{eq:fusion_reg} have the same invariant distribution, and the proposal chosen has matching volatility.

We have provided considerable practical guidance in \secref{sec:practical} to render many aspects of Bayesian Fusion which are non-standard due to the particularities of the fusion problem into standard SMC structures. A truly parallel implementation of Bayesian Fusion is a very attractive prospect for future development. As discussed in \secref{sec:effresamp}, although SMC is inherently well-suited to parallel implementation in distributed environments \citep{bk:hgm:dl18}, in the fusion setting the natural direct interpretation of Bayesian Fusion would be to consider the sub-posteriors (and associated data) as being distributed across cores, but the particle set to be shared across all cores. This is not the setting typically addressed by distributed SMC literature, and raises interesting challenges which require further innovation to be resolved. For instance, developing theory to support methodology in which the particles are not shared by all cores. 

A number of other methodological directions for Bayesian Fusion are possible. As presented in \secref{sec:theory} and \secref{sec:guidance}, the $C$ sub-posteriors are unified together in a \emph{`fork-and-join'} manner. An alternative would be to unify the sub-posteriors in stages gradually by constructing a tree to perform the operation hierarchically, for instance by exploiting \emph{`divide-and-conquer'} SMC theory and methodologies such as that of \citet{jcgs:dc17}. Another direction would be to consider how approximations could be used within the methodology. Many approximate approaches tackling the fusion problem are highly computationally efficient, albeit at the expense of introducing an approximation error which can be difficult to quantify and on occasion significant. The work of \citet{aap:wprs20} constructs an explicit Monte Carlo scheme in which approximations can be readily used to develop exact Monte Carlo schemes. There is tangible theory linking this paper with \citet{jrssb:pfjr20} and \citet{aap:wprs20}, and so finding a similar approach to embedding approximations may be viable. 

There is considerable scope for application of Bayesian Fusion, as inference in the setting of \eqref{eq:prod} arises directly and indirectly in many interesting practical settings. Many of these applications were discussed in \secref{sec:intro}. One interesting direction considers the use of Fusion methodologies within the \emph{Markov melding} framework of \citet{ba:gpldw19}, in which a modular approach is taken to statistical inference where separate sub-models are fit to data sources in isolation (often of varying dimensionality), and then joined. This type of application would necessitate theoretical developments to the Fusion methodologies to support sub-posteriors on mismatched dimensions. However, such a theoretical development when combined with \emph{`divide-and-conquer'} SMC theory such as that developed in \citet{jcgs:dc17} may also make Fusion methodologies more robust to increasing dimensionality. 

A number of future directions for the Bayesian Fusion methodology are currently being pursued by the authors. One interesting avenue of research is to apply Fusion methodologies within statistical cryptography. In the simplest setting a number of trusted parties who wish to securely share their distributional information on a common parameter space and model, but would prefer not to reveal their individual level distributions, could do so by means of applying cryptography techniques and exploiting the exactness and linear contributions to computations of individual sub-posteriors within the Fusion approach. In a further example, the authors are investigating the application of Bayesian fusion for purely algorithmic reasons. One motivation for this (rather like the motivation for tempering MCMC approaches) is that the simulation of a multimodal target density could be prohibitively difficult, whereas the target density might be readily written as a product of densities with less pronounced multi-modal behaviour, thus making it far more amenable to Monte Carlo sampling (see \citet{chan:poll:rob:21}).


\section{Acknowledgements}
We would like to thank Louis Aslett, Ryan Chan, Paul Jenkins, Yuxi Jiang and Adam Johansen for helpful discussions on aspects of the paper. This work was supported by the Engineering and Physical Sciences Research Council (EPSRC) grant numbers EP/K014463/1, EP/N031938/1, EP/R018561/1, EP/R034710/1, and the Alan Turing Institute's Defence \& Security Programme. The authors would like to thank the Isaac Newton Institute for Mathematical Sciences for support and hospitality during the programme ``Scalable inference; statistical, algorithmic, computational aspects (SIN)'' where aspects of the work in this paper were undertaken. 


\FloatBarrier
\appendix

\section{Background notation, and proof of \texorpdfstring{\thmref{thm:key}}{Theorem \ref{thm:key}} and \texorpdfstring{\thmref{thm:bf}}{Theorem \ref{thm:bf}}} \label{apx:bfproof}

We begin by more formally considering the $C$ correlated continuous-time Markov processes in $[0,T]$ introduced in \secref{sec:theory}, which are initialised separately but coalesce to a single point $\y$ at time $T$. A typical realisation of the object $\tX = \{\vecX{t},t\in [0,T]\} = \{ \X{t}{c},\,c\in\{1,\dots,C\},\,\Trange\}$ is given in \figref{fig:partition_object}, and is defined on the following space $\boldOmega_\boldzero$:
\begin{apx_definition}[$\boldOmega_\boldzero$] \label{defn:omega0}
\begin{align*}
\boldOmega_\boldzero &:= \left\{ \tX: \X{}{c} \in C^{d}[0,T] , c\in\{1,\dots, C\}, \X{T}{1} = \dots = \X{T}{C} = \y\right\},
\end{align*}
where $C^{d}[0,T]$ denotes the $d$-dimensional continuous function space with domain $[0,T]$.
\end{apx_definition}

In proving the results presented in this appendix, we impose the following regularity assumptions. 
\begin{apx_assumption} \label{ass:cont_diff}
$\nabla\log \target{c}{\x}$ is at least once continuously differentiable, where  $\nabla$ is the gradient operator. 
\end{apx_assumption}
\begin{apx_assumption} \label{ass:lower_bound}
$\phifn{c}{\x}$ is bounded below by some $\Phifn{c} \leq \inf\{\phifn{c}{\x}:\x\in\Rd\} \in \Rone$.
\end{apx_assumption}

Both Assumptions \ref{ass:cont_diff} and \ref{ass:lower_bound} are easily verified in practice and will typically be satisfied for many statistical applications. \assref{ass:cont_diff} is required in order to establish the \rnd in the proof of \thmref{thm:bf} below, although in principle could be weakened to consider discontinuous drifts following the approach of \citet{aap:prt16} (at the expense of adapting \thmref{thm:bf} and complicating the resulting methodology). Within the context of the sequential Monte Carlo methodology developed in \secref{sec:methodology}, \assref{ass:lower_bound} can be weakened, but as discussed in \apxref{apx:ue} ensures the estimator presented in \thmref{thm:pe} has finite variance (and so ensures robustness of \algref{alg:bf}). 

We can now proceed to the proof of \thmref{thm:key}, and show that if $\tX\sim\fusion$, then the marginal $\y\,(:=\vecX{T})\sim f$ as desired. 

\begin{proof}[Proof (\thmref{thm:key})]

We begin by marginalising $\fusion$ onto the values of $\vecX{T}$. Since all densities are written with respect to $\proposal$ we first take an expectation 
with respect to $\fusion$ 
of each of the $C$ coalescing diffusion paths ($\{\X{t}{c}, \Trange\}\cores$) and condition on their respective endpoints (for the $c$th path this is $\X{0}{c}$ and $\X{T}{c}$ respectively). Note that by construction these paths are independent Brownian bridges. The calculation for the remaining expectation (for $\vecX{0}$) appears in \citet[Prop. 2]{jap:dpr19}. Therefore the marginal distribution of the common endpoint $\y:=\X{T}{1}=\dots=\X{T}{C}$ has density $f$.

To show that the law of $C$ independent Brownian motions initialised from their respective distributions ($\vecX{0}=\{\X{0}{c}\}\cores$ where $\X{0}{1}\sim f_1,\dots,\X{0}{C}\sim f_C$) and conditioned to coalesce at time $T$ satisfies \eqref{eq:SDE}, we use Doob $h$-transforms (see for instance, \citet[Section IV.6.39]{rogers2000diffusions}). As such, we introduce the space-time harmonic function
\begin{align*}
{\mathrm h}(t, \vecX{t}) 
& = \int \prod\cores \frac{1}{\sqrt{2\pi (T-t)}} \exp\left\{- \frac{\|\y- \X{t}{c}\|^2}{2(T-t)} \right\} \ud \y
\end{align*}
which represents the integrated density of coalescence at time $T$ given the current state $\vecX{t}$. As a consequence we have that the $C$ conditioned processes satisfy a SDE of the form,
\begin{align*}
\ud \cvecX{t} = \ud \cvecW{t} + \nabla \log ({\mathrm h}(t, \cvecX{t})) \ud t,
\end{align*}
where $\nabla \log ({\mathrm h}(t, \vecX{t})) = (\boldv^{(1)}_t ,\dots,\boldv^{(C)}_t)$ is the concatenation of $C$ $d$-dimensional vectors (which we denote $\{\boldv^{(c)}_t\}\cores$). Considering the $c$th term we have
\begin{align*}
\boldv^{(c)}_t 
& = \frac{\int \frac{\y - \X{t}{c}}{T-t}\prod\cores \frac{1}{\sqrt{2\pi (T-t)}}\exp\left\{- \frac{\|\y- \X{t}{c} \|^2}{2(T-t)}\right\} \ud \y
    }{
    \int \prod\cores \frac{1}{\sqrt{2\pi (T-t)}}\exp \left\{-\frac{\|\y- \X{t}{c} \|^2}{ 2(T-t)} \right\} \ud \y} \\
& = \frac{\int \frac{\y}{T-t} \frac{1}{ \sqrt{2\pi (T-t)}} \exp  \left\{- \frac{C\|\barX{t}- \X{t}{c} \|^2 }{2(T-t)} \right\} \ud \y 
    }{ \int \frac{1}{\sqrt{2\pi (T-t)}}\exp  \left\{- \frac{C\|\barX{t}- \X{t}{c} \|^2}{2(T-t)} \right\} \ud \y} - \frac{\X{t}{c}}{T-t} \\
& = \frac{\barX{t}-\X{t}{c}}{T-t}.
\end{align*}
As a consequence we have
\begin{align*}
\nabla \log ({\mathrm h}(t, \vecX{t})) 
& = \left(\frac{\barX{t} - \X{t}{1}}{T-t}, \frac{\barX{t} - \X{t}{2}}{T-t}, \dots , \frac{\barX{t} - \X{t}{C}}{T-t}\right),
\end{align*}
and \eqref{eq:SDE} holds as required.
\end{proof}

Simulating from $\proposal$ without discretisation error relies on having explicit access to the finite-dimensional distributions of the process given by the SDE in \eqref{eq:SDE}. This is established by \thmref{thm:bf}. 

\begin{proof}[Proof (\thmref{thm:bf})]
\begin{enumerate}[leftmargin=*]
\item From \thmref{thm:key} we have established that $\tX\sim\fusion$ is Markov (in time), and so without loss of generality we only need to consider its incremental distribution. 
    
For $\tX\sim \proposal$ we have that for all $\Crange$ $\{\X{t}{c}, t\in [0,T]\}$ is the realisation of a $d$-dimensional Brownian bridge conditioned on starting at $\X{0}{c}$ and ending at $\y= \X{T}{c}$. Furthermore, under $\proposal$ we have that conditional on $\vecX{0}$ then $\y$ is distributed according to a Gaussian distribution with mean $\barX{0}$ and covariance matrix $TC^{-1}\identity{d\times d}$.
    
To derive the joint density of $\cvecX{t}$ conditional on $\cvecX{s}$ (for $0\leq s < t < T$), we begin by considering the joint $d(C+1)$-dimensional density of $\cvecX{t}$ and $\y$ conditional on $\cvecX{s}$, which we denote by $p_1$.
\begin{align*}
& -2 \log p_1 
 = \frac{C \| \y - \barX{s}  \|^2}{T-s} +\sum\cores \frac{T-s}{(t-s)\cdot (T-t)} \left\|\X{t}{c} - \frac{t-s}{T-s}\y - \frac{T-t}{T-s} \X{s}{c} \right\|^2 \\
& = \frac{C}{T-t} \left[ \|\y \|^2 - 2\y ' \barX{t} \right] + \sum\cores \frac{T-s}{(t-s)\cdot (T-t)} \left\| \X{t}{c} - \frac{T-t}{T-s} \X{s}{c} \right\|^2 + k_5 \\
& = \frac{C}{T-t}\left\|\y -   \barX{t}\right\|^2 - \frac{C }{T-t}\|\barX{t} \|^2 +\frac{T-s}{(t-s)\cdot (T-t)}\sum\cores \left\|\X{t}{c} - \frac{T-t}{T-s}\X{s}{c} \right\|^2\ + k_5,
\end{align*}
where $k_5$ is a constant. Now, integrating out $\y$ we obtain the $dC$-dimensional density of $\cvecX{t}$ conditional on $\cvecX{s}$, which we denote $p_2$,
\begin{align*}
-2 \log p_2 
& = - \frac{1}{T-t} \frac{\|\sum\cores \X{t}{c}\|^2}{C}+\frac{T-s}{(t-s)\cdot(T-t)}\sum\cores \left\|\X{t}{c}- \frac{T-t}{T-s}\X{s}{c} \right\|^2 + k_6 \\
& = \vecX{t}'\cV{s,t}^{-1}\vecX{t}-2\vecX{t}'\cV{s,t}^{-1} \cvecM{s,t} + k_7,
\end{align*}
where $k_6$ and $k_7$ are constants, and $\cV{s,t}$ and $\cvecM{s,t}$ are terms we will now derive. We have $\cV{s,t}^{-1}={\bSigma}^{-1} \otimes \identity{C\times C}$ with 
\begin{align*}
\bSigma^{-1}
& = \frac{T-s}{(t-s)\cdot (T-t)} \identity{C\times C} - \frac{1}{C(T-t)}\onemat{C\times C},
\end{align*}
where $\onemat{C\times C}$ is the $C\times C$ matrix containing all elements $1$. Inverting $\bSigma^{-1}$ we have 
\begin{align*}
\Sigma_{ii}
= \frac{(t-s)\cdot (T-t)}{T-s} + \frac{(t-s)^2}{C(T-s)}, \qquad 
\Sigma_{ij} = \frac{(t-s)^2}{C(T-s)}.
\end{align*}
$\cvecM{s,t}$ can be written as $(\cM{s,t}{1}, \dots, \cM{s,t}{C})$ with
\begin{align*}
\cM{s,t}{c}
& ={\frac{T-t}{T-s}\X{s}{c} +  \frac{t-s}{T-s}\barX{s} },
\end{align*}
as required by the statement of the theorem. 
\item From \citet{jap:dpr19} we have that for $\Crange$ the law of $\{\X{t}{c},\, t\in(0,T)\}$ conditional on the endpoints $\X{0}{c}$ and $\y$ is that of a Brownian bridge. As a consequence, the result holds from standard properties of Brownian bridges. 
\end{enumerate}
\end{proof}


\FloatBarrier
\section{Proof of \texorpdfstring{\thmref{thm:pe}}{Theorem \ref{thm:pe}}, and unbiased estimation of \texorpdfstring{$\rho_j$}{the importance weights}} \label{apx:ue}

In this appendix we provide a proof of \thmref{thm:pe}, together with practical guidance on how to simulate a low variance, positive, unbiased estimator $\tilde{\rho}_j$. This is accompanied with pseudo-code which is presented in \algref{alg:ue}. The approach we take is a variant of \citet{jrssb:bprf06}, \citet[Section 4]{jrssb:fpr08}, and \citet[Chapter 7.1]{phd:p13}, applied to our particular setting.

To construct such an estimator we rely on the property that the function $\phi_c$ for $\Crange$ is bounded on compact sets, which follows directly from \assref{ass:cont_diff} \citep{mcap:bpr08}. In particular, suppose there exists some compact region $R_c \subset \Rd$ such that $\X{}{c}(\omega) \in R_c$, then there exists some $L^{(c)}_j := L\big(\X{}{c}(\omega)\big)\in\Rone$ and $U^{(c)}_j := U\big(\X{}{c}(\omega)\big)\in\Rone$ such that $\phifn{c}{\X{}{c}(\omega)}\in\big[L^{(c)}_j ,U^{(c)}_j\big]$. 

We can exploit this property of $\phi_c$ by simulating as required $\X{}{c}$ (with law $\brown{j,c}$) in two steps: (i) partitioning the path-space of $\brown{j,c}$ into disjoint \emph{layers} and simulating to which $\X{}{c}$ belongs (denoting  $R_c:=R_c(\X{}{c})\sim\layer_c$); (ii) simulating the path at time marginals as required conditional on the simulated layer (i.e. $\X{t}{c} \sim \brown{j,c}|R_{c}$). This two step procedure then allows us to identity $L^{(c)}_j$ and $U^{(c)}_j$ for use when constructing our estimator. Full detail on step (i) can be found in \citet[Section 8.1 and Algorithm 16]{b:pjr16}, and on step (ii) can be found in \citet[Section 8.5 and Algorithm 22]{b:pjr16}, but both are omitted from this paper for brevity.

We can now proceed to the proof of \thmref{thm:pe}. 

\begin{proof}[Proof (\thmref{thm:pe})]
Recalling $R_{c}:=R_{c}(\X{}{c})$ is a function of the Brownian bridge sample path $\X{}{c}\sim\brown{j,c}$ which determines a compact subset of $\Rd$ for which $\X{}{c}$ is constrained, further denote $\layer$ as the law of $R_{1},\dots,R_{C}$, and $\brownbar$ as the law of the $C$ Brownian bridges $\X{}{1},\dots,\X{}{C}$. Let $\lawK$ denote the law of $\kappa_1,\dots,\kappa_C$, and $\lawU$ denote the law of $\chi_{1,1}$, $\dots$, $\chi_{1,\kappa_1}$, $\dots$, $\chi_{C,1}$, $\dots$, $\chi_{C,\kappa_C} {\sim}\uniform[t_{j-1},t_j]$. Then for $j\in\{1,\dots,n\}$ we have,
\begin{align*}
& \expect_{\layer}\expect_{\brownbar|\layer}\expect_{\lawK}\expect_{\lawU}[\hat{\rho}_j] = \expect_{\layer}\expect_{\brownbar|\layer}\expect_{\lawK}\expect_{\lawU}\left[\prod\cores \frac{\Delta_j^{\kappa_c}\cdot e^{- (U_{j}^{(c)}-\Phifn{c})\Delta_j}}{\kappa_c! \cdot p(\kappa_c|R_c)} \prod^{\kappa_c}_{k_c=1}\left(U_{j}^{(c)}- \phifn{c}{\X{\chi_{c,k}}{c}}\right)\right] \\
& \qquad 
= \expect_{\layer}\expect_{\brownbar|\layer}\expect_{\lawK}\left[\prod\cores \frac{\Delta^{\kappa_c}_j\cdot e^{- (U_{j}^{(c)}-\Phifn{c})\Delta_j}}{\kappa_c! \cdot p(\kappa_c|R_c)} \left[\int^{t_j}_{t_{j-1}} \frac{U_{j}^{(c)}- \phifn{c}{\X{t}{c}}}{\Delta_j} \ud t\right]^{\kappa_c}\right] \\
& \qquad = \expect_{\layer}\expect_{\brownbar|\layer} \left[\prod\cores \sum^\infty_{k_c=0} \left(\frac{\Delta^{k_c}_j\cdot e^{- (U_{j}^{(c)}-\Phifn{c})\Delta_j}}{k_c!} \left[\int^{t_j}_{t_{j-1}} \frac{U_{j}^{(c)}- \phifn{c}{\X{t}{c}}}{\Delta_j} \ud t\right]^{k_c}\right)\right] \nonumber \\
& \qquad = \expect_{\layer}\expect_{\brownbar|\layer} \left[\prod\cores  e^{- (U_{j}^{(c)}-\Phifn{c})\Delta_j} \cdot \exp\left\{\int^{t_j}_{t_{j-1}} \left(U_{j}^{(c)}- \phifn{c}{\X{t}{c}}\right) \ud t\right\} \right] \\
& \qquad = \prod\cores{\expect}_{\brown{j,c}} \left[ \exp \left\{ -   \int_{t_{j-1}}^{t_j} \left( \phifn{c}{\X{t}{c}}  - \Phifn{c} \right) \ud t \right\} \right]  =: \rho_{j}.
\end{align*}
\end{proof}

\thmref{thm:pe} allows for significant flexibility in choosing the law $\lawK$. As we are embedding the estimator within a sequential Monte Carlo framework, we want to choose the law $\lawK$ to minimise the variance of the estimator (or equivalently in our case, the second moment). 

\begin{apx_lemma} \label{lem:rho}
The second moment of the estimator $\hat{\rho}_j$ is minimised when  $p(\kappa_1|R_1)$, $\dots$, $p(\kappa_c|R_c)$ are chosen to be Poisson distributed with intensities,
\begin{align}
\lambda_c 
& := \left[\Delta_j\int^{t_j}_{t_{j-1}} \left(U^{(c)}_j - \phifn{c}{\X{t}{c}}\right)^2 \ud t\right]^{1/2}, \qquad \Crange. \label{eq:rho_opt}
\end{align}
\proof
We have,
\begin{align}
& \expect_{\layer}\expect_{\brownbar|\layer}\expect_{\lawK}\expect_{\lawU}[\hat{\rho}^2_j] \nonumber \\
& \, =  \expect_{\layer}\expect_{\brownbar|\layer}\expect_{\lawK}\left[\prod\cores\left(\frac{\Delta_j^{2\kappa_c}\cdot e^{-2 (U_{j}^{(c)}-\Phifn{c})\Delta_j}}{(\kappa_c!)^2 \cdot p^2(\kappa_c|R_c)} \left[\int^{t_j}_{t_{j-1}} \frac{\big(U_{j}^{(c)}- \phifn{c}{\X{\chi_{c,k}}{c}}\big)^2}{\Delta_j} \ud t\right]^{\kappa_c}\right)\right]  \nonumber \\
& \, =   \expect_{\layer}\expect_{\brownbar|\layer}\Bigg[\prod\cores e^{-2(U_{j}^{(c)}-\Phifn{c})\Delta_j} \cdot \sum^\infty_{k=0}\underbrace{\frac{\left.\big[\Delta_j \cdot \int^{t_j}_{t_{j-1}} \big(U_{j}^{(c)}- \phifn{c}{\X{\chi_{c,k}}{c}}\big)^2 \ud t\big]^{k}/k!^2\right.}{p(k|R_c)}}_{=: f_k/p_k}\Bigg]. \label{eq:rho_second}
\end{align}
Recalling we have the flexibility to choose the discrete probability distributions given by $p(k|R_c)$ for $\Crange$, we want to make our selection to minimise \eqref{eq:rho_second}. To do so we consider each sub-posterior separately and use Lagrange multipliers to optimise $\sum_k f_k/p_k + \lambda(\sum_k p_k -1)$, finding that $-f_k/p^2_k + \lambda =0$. As $p_k\in(0,1)$ we have $p_k=\sqrt{f_k/\lambda}$, and further noting $\sum_k p_k = 1$ then $\lambda= (\sum_k \sqrt{f_k})^2$. Hence we find the optimal distribution of $p(k|R_c)$ to be Poisson,
\begin{align*}
    p(k|R_c) = \frac{\lambda_c^{k}}{k!} e^{ -\lambda_c}, 
\end{align*}
with $\lambda_c$ as given in \eqref{eq:rho_opt}. Substituting this selection into \eqref{eq:rho_second}, and recalling from \thmref{thm:bf} that $\Phifn{c}$ is a constant such that $\inf_{\x} \phifn{c}{\x}\geq \Phifn{c} > -\infty$, we have $\lambda_c \leq \Delta_j\cdot(U_{j}^{(c)}-\Phifn{c})$, then we show finiteness as required:
\begin{align*}
\expect_{\layer}\expect_{\brownbar|\layer}\expect_{\lawK}\expect_{\lawU}[\hat{\rho}^2_j]
& =  \expect_{\layer}\expect_{\brownbar|\layer}\Bigg[\prod\cores e^{-2(U_{j}^{(c)}-\Phifn{c})\Delta_j} \cdot \sum^\infty_{k=0}\frac{\lambda_c^k\cdot e^{\lambda_c}}{k!}\Bigg] \nonumber\\
& = \expect_{\layer}\expect_{\brownbar|\layer}\left[\exp\left\{2\sum\cores\left(\lambda_c-\big(U_{j}^{(c)}-\Phifn{c}\big)\Delta_j\right) \right\} \right]  \leq 1 < \infty.
\end{align*} \qed
\end{apx_lemma}

As noted with \secref{sec:methodology}, normalisation within \algref{alg:bf} permits us to use the estimator $\tilde{\rho}$ (given by \eqref{eq:rhotilde}) in place of the estimator $\hat{\rho}$, thus avoiding the need to compute the constants $\Phifn{1},\dots,\Phifn{C}$.

\begin{apx_corollary} \label{cor:rho}
The second moment of the estimator $\tilde{\rho}_j$ is minimised when  $p(\kappa_1|R_1)$, $\dots$, $p(\kappa_c|R_c)$ are chosen as in \lemref{lem:rho}, and is finite. 
\proof
Follows directly from \eqref{eq:rhotilde} and \lemref{lem:rho}. \qed
\end{apx_corollary}

Although \lemref{lem:rho} and \corref{cor:rho} suggest an optimal distribution and parameterisation for the simulation of the law $\lawK$ in \thmref{thm:pe}, the integral in \eqref{eq:rho_opt} precludes this choice. In this paper we consider the following two possible choices for $\lawK$ which attempt to mimic the optimal parameterisation in \eqref{eq:rho_opt} (but erring on having heavier tails for robustness): (i) a Poisson distribution with a higher intensity; (ii) a Poisson distribution with a random mean approximating $\lambda_c$ given by a Gamma distribution (which leads to the negative binomial distribution). We term these \emph{Unbiased Estimator A} and \emph{B} respectively (\gpea\ and \gpeb) respectively (and are based upon GPE-$1$ and GPE-$2$ within \citep{jrssb:fpr08} applied to our setting). 

\begin{apx_condition}[\gpeas{}] \label{con:rho_isa}
Choosing $p(\kappa_1|R_1)$, $\dots$, $p(\kappa_c|R_c)$ to be Poisson distributed with intensity
\begin{align*}
\Lambda_c := \Delta_j \big( U^{(c)}_j - L^{(c)}_j\big), 
\end{align*}
leads to the estimator
\begin{align*}
\gpeas{j} :=  \prod\cores \left(\frac{e^{- L_{j}^{(c)}\Delta_j}}{\big(U_{j}^{(c)}- L_{j}^{(c)}\big)^{\kappa_c}}\prod^{\kappa_c}_{k_c=1}\left(U_{j}^{(c)}- \phifn{c}{\X{\chi_{c,k}}{c}}\right)\right).
\end{align*}
\end{apx_condition}

\begin{apx_condition}[\gpebs{}] \label{con:rho_isb}
Choosing $p(\kappa_1|R_1)$, $\dots$, $p(\kappa_c|R_c)$ to be Negative Binomial distributed with mean parameter
\begin{align}
m_c := \Delta_j U^{(c)}_j - \int^{t_j}_{t_{j-1}} \phifn{c}{\frac{\X{j-1}{c}(t_j-s) + \X{j}{c}s}{\Delta_j}} \ud s \label{eq:mean_gpeab}
\end{align}
and dispersion parameter $r_c$, leads to the estimator,
\begin{align*}
\gpebs{j} :=  \prod\cores \left(\Delta_j^{\kappa_c} e^{- U_{j}^{(c)}\Delta_j}\cdot \frac{\Gamma(r_c)\cdot (m_c+r_c)^{r_c+\kappa_c}}{\Gamma(r_c+\kappa_c)\cdot r_c^{r_c} m_c^{\kappa_c}} \prod^{\kappa_c}_{k_c=1}\left(U_{j}^{(c)}- \phifn{c}{\X{\chi_{c,k}}{c}}\right)\right).
\end{align*}
\end{apx_condition}

Although \gpeas{j}\ of \conref{con:rho_isa} is a more natural estimator (it is trivially bounded by $\exp\{-\Delta_j\sum\cores \Phifn{c}\}$, and consequently has finite variance), \citet{jrssb:fpr08} recommend using an estimator of the form \gpebs{j}\ as given in \conref{con:rho_isb} as it is more robust in practice. The mean parameterisations suggested in \eqref{eq:mean_gpeab} of the Negative Binomial are typically tractable, but if it is too unavailable then a crude estimate of the integral for each sub-posterior can be used (for instance by taking the mean of $\phi$ evaluated at $\X{j-1}{c}$ and $\X{j}{c}$) and this does not introduce bias to the estimator (it simply inflates the variance). The dispersion parameterisations in \eqref{eq:mean_gpeab} of the Negative Binomial can be chosen to approximately match the tail thickness of the optimal Poisson distribution. If the dispersion parameters are chosen to be constant for every sub-posterior ($r_1=\dots=r_c$) then a further normalising constant can be removed from the estimator (noting the normalising constant will be common for all particles in \algref{alg:bf}, and so will be lost when the particle weights are re-normalised). As the distribution for $\lawK$ has heavier tails under the choice in \conref{con:rho_isb} than \conref{con:rho_isa}, a variation of \citet{jrssb:fpr08} shows that the variance of \gpebs{j}\ is finite too. 

An algorithmic summary of the construction of the unbiased estimator $\tilde{\rho}_{j}$ for use in \algstref{alg:bf}{st:ue} is given in \algref{alg:ue}. In practice we have found using the slightly more complicated \gpeb\ to be more robust than \gpea\ within \algstref{alg:ue}{st:choice}), particularly when the individual sub-posterior trajectories are out-with the domain of attraction of their corresponding sub-posterior (for instance with increasing sub-posterior heterogeneity, and as $j\to n$).

\begin{algorithm}[ht]
	\caption{Simulating the unbiased estimator $\tilde{\rho}_j$ (\algstref{alg:bf}{st:ue}).} \label{alg:ue}
    \begin{enumerate}
	\item For $c$ in $1$ to $C$,
	\begin{enumerate}
	\item \textbf{${R}_{c}$:} Simulate $R_{c} \sim \layer_c$ as per \citet[Alg. 16]{b:pjr16}.
	\item \textbf{$p_c$:} Choose $p(\cdot|R_c)$ (e.g. following guidance in \conref{con:rho_isa} or \conref{con:rho_isb}). \label{st:choice}
	\item \textbf{$\kappa_c$:} Simulate $\kappa_c \sim p(\cdot|R_c)$.
	\item \textbf{$\chi_{\cdot}$:} Simulate $\chi_{c,1},\dots,\chi_{c,\kappa_c} \sim \uniform[t_{j-1},t_j]$.
	\item \textbf{$\X{\cdot}{c}$:}  Simulate $\X{\chi_{c,1}}{c},\dots,\X{\chi_{c,\kappa_c}}{c} \sim \brown{j,c}|R_{c}$ as per \citet[Alg. 22]{b:pjr16}.
	\end{enumerate}
	\item \textbf{Output:} $\tilde{\rho}_j := \prod\cores \Big(\frac{\Delta_j^{\kappa_c}\cdot e^{- U_{j}^{(c)}\Delta_j}}{\kappa_c! \cdot p(\kappa_c|R_c)} \prod^{\kappa_c}_{k_c=1}\big(U_{j}^{(c)}- \phifn{c}{\X{\chi_{c,k}}{c}}\big)\Big)$ (e.g. following guidance in \conref{con:rho_isa} or \conref{con:rho_isb}).
	\end{enumerate}
\end{algorithm}

\FloatBarrier
\section{Proofs of Theorems \ref{thm:cess0} and \ref{thm:cessj}, and Corollaries \ref{cor:cess0} and \ref{cor:parallel1}} \label{apx:guidanceproofs}

To prove \thmref{thm:cess0} and \corref{cor:cess0} of \secref{sec:Tguide}, we introduce the following lemma,

\begin{apx_lemma}\label{lem:sigma_mgf}
The moment generating function (mgf) for $ \sigma^2 := \frac{1}{C} \sum\cores \|\X{0}{c} - \barX{0}  \|^2$, where $\X{0}{c}, c=1,\dots, C$ are independent with $\X{0}{c} \sim \normal(\bolda_c, m^{-1}Cb\identity{} )$, is given by
\begin{align*}
M_{\sigma^2}(s)  
&= \exp\left\{\frac{m \sigma_{\bolda}^2 s}{m-2sb} \right\} \cdot \left(1-2s\frac{b}{m}\right)^{-\frac{(C-1)d}{2}} , \qquad \text{where, } \frac{sb}{m} < \frac{1}{2}.
\end{align*}
\proof We have
\begin{align*}
\frac{1}{C} \sum\cores \| \X{0}{c} - \bolda\|^2 
& = \sigma^2 + \frac{1}{C} \sum\cores \left \| \bolda - \barX{0}  \right \|^2
\end{align*}
where the right hand side is a sum of two independent variables. If we denote 
\begin{align*}
\lambda 
& = \sum\cores \left\| \frac{\bolda_c - \bolda }{\sqrt{m^{-1}Cb} }\right\|^2 = \frac{\sigma_{\bolda}^2 m}{b}
\end{align*}
then $\frac{m}{Cb} \sum\cores \| \X{0}{c} - \bolda\|^2$ is a non-central $\chi^2(Cd, \lambda)$, with moment generating function (mgf) 
\begin{align*}
M_1(s) 
& = \frac{\exp\left\{ \frac{\lambda s}{1-2s} \right\}}{ (1-2s)^{\frac{Cd}{2}} }.
\end{align*}
Furthermore, $\frac{m}{b} \left \| \bolda - \barX{0}  \right \|^2$ is a $\chi^2(d)$ random variable with mgf $M_2(s) = (1-2s)^{-\frac{d}{2}}$. Therefore, the mgf of $\sigma^2$ is 
\begin{align*}
M_{\sigma^2}(s) 
& = \frac{M_1(sb/m)}{M_2(sb/m)} = \left. \frac{\exp\left\{ \frac{\sigma_{\bolda}^2 s}{1-2s\frac{b}{m}} \right\}}{ \left(1-2s\frac{b}{m}\right)^{\frac{Cd}{2}} } \right / 
\left(1-2s\frac{b}{m}\right)^{-\frac{d}{2}},
\end{align*} 
and the statement of \lemref{lem:sigma_mgf} follows directly. \qed
\end{apx_lemma}

We can now present the proofs of \thmref{thm:cess0} and \corref{cor:cess0}.

\begin{proof}[Proof (\thmref{thm:cess0})]
The conditional effective sample size \ssj{0} for particles with weight $\rho_{0,i}$, $i=1,\dots, N$ is such that, as $N\rightarrow \infty$,
\begin{align*}
N^{-1} \ssj{0} 
& =  N^{-1} \left[ \sum_i \frac{(\rho_{0,i})^2}{(\sum_j \rho_{0,j})^2 }\right]^{-1} 
\rightarrow  \frac{(\expect \rho_{0,i})^2}{\expect (\rho_{0,i}^2) } 
=  \frac{ \left[\expect \left( e^{-\frac{C\sigma^2}{2T}} \right) \right]^2}{\expect \left( e^{-\frac{C\sigma^2}{T}} \right) }  = \frac{ \left( M_{\sigma^2}\left(- \frac{C}{2T} \right) \right)^2 }{M_{\sigma^2}\left(-\frac{C}{T}\right)}.
\end{align*}

From \lemref{lem:sigma_mgf}, we have
\begin{align}
\frac{ \left[ M_{\sigma^2}\left( -\frac{C}{2T} \right) \right]^2 }{M_{\sigma^2}\left(- \frac{C}{T}\right) } 
& = \frac{ \left[ \exp\left\{-\frac{m \sigma_{\bolda}^2 \frac{C}{2T}}{m + 2\frac{C}{2T }b} \right\} \cdot \left(1+2\frac{C}{2T} \frac{b}{m}\right)^{-\frac{(C-1)d}{2}}  \right]^2 }{ \exp\left\{-\frac{m \sigma_{\bolda}^2 \frac{C}{T}}{m + 2\frac{C}{T}b} \right\}\cdot \left(1 + 2\frac{C}{T}\frac{b}{ m}\right)^{-\frac{(C-1)d}{2}}} \nonumber \\
& = \exp\left\{ - \frac{ \frac{\sigma_{\bolda}^2 b}{m }}{\left(\frac{T}{C} + \frac{b}{m}\right)
\cdot\left(\frac{T}{C} + \frac{2b}{m}\right)} \right\} \cdot \left [1 + \frac{\left( \frac{Cb}{T m}\right)^2 }{ 1+\frac{2Cb}{T m}  } \right]^{-\frac{(C-1)d}{2}}, \label{eq:bfproof}
\end{align}
and so \thmref{thm:cess0} immediately follows.
\end{proof}

\begin{proof}[Proof (\corref{cor:cess0})]
First considering the proof of part (a): In the \shl setting (\conref{cond:sh1}) we have that $\sigma^2_{\bolda}\leq bC\lambda/m$. For the first term in \eqref{eq:bfproof} we have
\begin{align*}
\exp\left\{ - \frac{\frac{\sigma_{\bolda}^2 b}{m}}{\left(\frac{T}{C} + \frac{b}{m}\right)
\cdot \left(\frac{T}{C} + \frac{2b}{m}\right)} \right\}
& \ge \exp\left\{ - 
{\frac{\sigma_{\bolda}^2 b C^2}{T^2m}} \right\} 
\ge \exp\left\{ - 
\frac{b^2 C^3 \lambda}{T^2m^2} \right\} \ge \exp\left\{ -\frac{\lambda}{k_1^2} \right\},
\end{align*}
and for the second term in \eqref{eq:bfproof} we have
\begin{align}
\left [1 + \frac{\left( \frac{Cb}{T m}\right)^2 }{ 1+\frac{2Cb}{T m}  } \right]^{-\frac{(C-1)d}{2}}
&\ge \exp\left\{ - \frac{\left( \frac{Cb}{T m}\right)^2\cdot (C-1) d}{2 (1 + \frac{2Cb}{T m} ) }\right\} 
\ge \exp\left\{ - \frac{d}{2 k_1^2}\right\}, \label{eq:corproofparta2}
\end{align}
which together prove part (a).

Now considering the proof of part (b): Using the assumed bounds in \eqref{eq:T} and \eqref{eq:ssh_guide}, we have 
\begin{align*}
\left(\frac{T}{C} + \frac{b}{m}\right) \cdot \left(\frac{T}{C} + \frac{2b}{m}\right)
& \ge 
\frac{T^2}{C^2} \ge \frac{bk_1 k_2}{m}.
\end{align*}
In the \ssh setting (\conref{cond:ssh}) we can deduce that
\begin{align*}
\exp\left\{ - \frac{\frac{\sigma_{\bolda}^2 b}{ m}}{\left(\frac{T}{C} + \frac{b}{m}\right) \cdot \left(\frac{T}{C} + \frac{2b}{m}\right)} \right\}
& \ge \exp\left\{ - \frac{b^2 \gamma /m}{bk_1 k_2 /m} \right\} 
= \exp\left\{ -\frac{b \gamma}{k_1 k_2} \right\},
\end{align*}
which when taken together with the bound in \eqref{eq:corproofparta2} prove part (b).
\end{proof}

In \thmref{thm:cessj} for simplicity we derive the conditional effective sample size \ssj{j} for particles with importance weight $\rho_{j,i} = \prod\cores \rho_{j,i}^{(c)}$, $i=1\dots, N$.

\begin{proof}[Proof (\thmref{thm:cessj})]
For large $N$, and ${\boldxi}_j$ as given in \eqref{eq:trans_new},
\begin{align*}
 N^{-1}\ssj{j} \approx  \frac{ \left[\expect \left( \exp\left\{- \sum\cores \int^{t_j}_{t_{j-1}} \phifn{c}{\X{t}{c}} \ud t \right\} \Big|\,{\boldxi}_j \right) \right]^2 }{\expect \left(\exp\left\{- 2 \sum\cores \int^{t_j}_{t_{j-1}} \phifn{c}{\X{t}{c}} \ud t \right\} \Big|\,{\boldxi}_j \right) }.
\end{align*}
For Gaussian sub-posteriors we have $\phifn{c}{\X{t}{c}} =  \frac{1}{2} (Cb)^{-2} m^2 \|\X{t}{c} - \bolda_c\|^2 - \frac{1}{2}  (Cb)^{-1}md$. If we consider very small intervals $(t_{j-1}, t_j)$, we have
\begin{align}
N^{-1}\ssj{j} 
& \approx  \frac{ \left[\expect \left( \exp\left\{- \frac{\Delta_j}{2} \sum\cores (Cb)^{-2} m^2 \|\X{j}{c} - \bolda_c\|^2 \right\} \Big|\,{\boldxi}_j \right) \right]^2 }{\expect \left( \exp\left\{-  \Delta_j  \sum\cores (Cb)^{-2} m^2 \|\X{j}{c} - \bolda_c\|^2 \right\} \Big|\,{\boldxi}_j \right)}. \label{eq:approx_cessj}
\end{align}
On the other hand, we have
$\left(\frac{T-t_j}{T-t_{j-1}}\Delta_j\right)^{-1} \sum\cores \|\X{j}{c} - \bolda_c\|^2$ is a non-central $\chi^2(Cd, \lambda_{j} ')$, where
\begin{align*}
\lambda_{j}' 
& = \sum\cores \left \| \frac{ {\expect \left(\X{j}{c}\Big|\,{\boldxi}_j \right) } - \bolda_c }{\sqrt{\frac{T-t_j}{T-t_{j-1}}\Delta_j}} \right\|^2.
\end{align*}
Using arguments similar to those in \thmref{thm:cess0} and \lemref{lem:sigma_mgf}, we can write \eqref{eq:approx_cessj} as 
\begin{align*}
N^{-1}\ssj{j}  
& \approx   \frac{ \left[ \exp\left\{ \frac{\lambda_{j}' s}{1-2s} \right\}\cdot \left(1-2s\right)^{-\frac{Cd}{2}}  \right]^2 }{ \exp\left\{ \frac{2 \lambda_{j}' s}{1-4s} \right\} \cdot \left(1-4s\right)^{-\frac{Cd}{2}}  } = \frac{  \exp\left\{ - \frac{2 \sigma_{t_j}^2 C^{-1}m^2 b^{-2} \Delta_j}{1-2s} \right\}\cdot \left(1-2s \right)^{-Cd}}{ \exp\left\{ - \frac{2 \sigma_{t_j}^2 C^{-1}m^2 b^{-2} \Delta_j}{1-4s} \right\} \cdot \left(1-4s\right)^{-\frac{Cd}{2}}  } ,
\end{align*}
where $s = - \frac{1}{2}  (Cb)^{-2} m^2 \frac{T - t_{j}}{T-t_{j-1}}\Delta_j^2$ and $\sigma_{t_j}^2 = C^{-1}\sum\cores\| { \expect \big(\X{j}{c}\big|\,{\boldxi}_j \big)} - \bolda_c \|^2$. This can be further simplified as
\begin{align}
N^{-1}\ssj{j}  
& \approx \exp\left\{ 2s \frac{2 \sigma_{t_j}^2 C^{-1}m^2 b^{-2} \Delta_j}{(1-2s)(1-4s)} \right\}
\cdot \frac{  \left(1-4s + 4s^2 \right)^{-\frac{Cd}{2}}  }{ \left(1-4s\right)^{-\frac{Cd}{2}}}. \label{eq:simplified}
\end{align}
If we take the limiting regime prescribed in \eqref{eq:cessj} this implies that $s\to 0$. Further bounding the right hand expression in \eqref{eq:simplified} as follows
\begin{align*}
\frac{\left(1-4s+4s^2\right)^{-Cd/2}}{\left(1-4s\right)^{-Cd/2}} & = \left(1+\frac{4s^2}{1-4s}\right)^{-Cd/2} \ge \exp \left\{-
\frac{2s^2 Cd}{1-2s}
\right\},
\end{align*}
and substituting in the bounds in \eqref{eq:cessj}, we arrive at the required result.
\end{proof}

\FloatBarrier
\section{Proof of Corollary \ref{cor:parallel1}} \label{apx:parallel}

\begin{proof}[Proof of \corref{cor:parallel1}]
From \algstref{alg:bf}{st:covar} we have $\vecX{j} \sim \normal\left(\vecX{j-1}; \cvecM{j}, \cV{j} \right)$, where $\cvecM{j}$ and $\cV{j}$ are as given in \thmref{thm:bf}. In addition we have,
\begin{align*}
\cV{j} 
& =  \left\{\frac{(t_j-t_{j-1})^2}{C(T-t_{j-1})} \identity{C\times C} + \frac{T-t_j}{T-t_{j-1}}(t_j-t_{j-1}) \identity{C\times C}\right\}\otimes \identity{d\times d}.
\end{align*}

From \eqref{eq:trans_new} we also have the mean and covariance matrix of $\vecX{j}$ given $\vecX{j-1}$ are given by $\cvecM{j}$ and the above $\cV{j}$, as required. 
\end{proof}


\section{Application of Corollary \ref{cor:scale_pe} to large data settings} \label{apx:scale}

An example of where \corref{cor:scale_pe} may be useful to practitioners is in a large data setting, in which each core contains a large volume of data ($m_c\gg 1$, $\Crange$), and where computing $\phi_c$ is consequently an (expensive) $\bigO(m_c)$ operation on each core. Consider a simple model which admits a structure with conditional independence, and hence the following factorisation 
\begin{align}
f_c(\x) \propto \prod_{i=0}^{m_c} \ell_{i,c}(\x), \label{eq:bigdatalike}
\end{align}
where $\ell_{0,c}$ and $\ell_{i,c}(\x)$, $i\in\{1,\dots,m_c\}$, are the prior and $m_c$ likelihood terms respectively corresponding to the $c$th sub-posterior. Recalling $\phi_c$ is linear in terms of $\nabla\log \ell_{i,c}(\x)$ and $\Delta\log \ell_{i,c}(\x)$, one such simple unbiased estimator for $\phi_c$ that could be used in \algref{alg:bf} would be
\begin{align}
\phihatfn{c}{\x} = (m_c+1) \cdot \left[(m_c+1)\cdot\left(\nabla\log \ell_{I,c}(\x\right))^{T}\left(\nabla\log \ell_{J,c}(\x\right)) + \Delta \log \ell_{I,c}(\x)\right]/2, \label{eq:phichat1}
\end{align}
where $I,J \overset{\iid}{\sim} \uniform\{0,\dots,m_c\}$. However, in this setting one would naturally be interested in the robustness of Bayesian Fusion as the volume of data on each sub-posterior increases ($m_c\to\infty$). As discussed above, the critical consideration when using \corref{cor:scale_pe} is to take note that the expected number of functional evaluations in \algref{alg:bf} will increase, from say $K$ to $K'$, and so the critical quantity to consider is the ratio $K'/K$ and its growth as $m_c\to\infty$. Unfortunately use of \eqref{eq:phichat1} would in a futile manner exchange the $\bigO(m_c)$ evaluations of the sub-posterior in the case of $\tilde{\rho}_j$, with an $\bigO(m_c)$ inflation in the in the expected number of functional evaluations for the case of $\tilde{\varrho}_j$ modified by \eqref{eq:phichat1}. 

Instead, to exploit \corref{cor:scale_pe} in the large data setting one could apply directly the approach of \citet{jrssb:pfjr20}, and develop an $\bigO(1)$ unbiased estimator $\hat{\phi}_c$, with $\bigO(1)$ scaling of the ratio $K'/K$, in place of $\phi_c$, and so in turn find a suitable $\tilde{\varrho}_j$. \citet[Sec.\ 4]{jrssb:pfjr20} propose using a small number of suitably chosen \emph{control variates} in the construction of such an estimator. In the Bayesian Fusion setting it is natural to choose a set of control variates for each sub-posterior: $\nabla\log f_c$ and $\Delta\log f_c$ computed at both a point \emph{close} to the mode of the sub-posterior (say $\hat{\x}_c$), and a point \emph{close} to the posterior mode (say $\hat{\x}$) -- where close in this sense is within  $\bigO(m_c^{-1/2})$ of the true respective modes. Such points can be found by applying an appropriate mode finding algorithm \citep{compstat:b10,bk:nesterov,arxiv:jnj17}, although note that these will involve full likelihood calculations and so are both likely to be one-time $\bigO(m_c)$ computations. With this, we instead recommend the following choice as an unbiased estimator for $\phi_c$,
\begin{align}
\phihatfn{c}{\x} = (\hat{\alpha}_{I,c}(\x))^{T}\left(2\nabla\log f_c(\x^*)+ \hat{\alpha}_{J,c}(\x)\right) + \text{div\,}\hat{\alpha}_{I,c}(\x))/2 + C \label{eq:phichat2}
\end{align}
where $\x^*$ is either $\hat{\x}_c$ or $\hat{\x}$ and is chosen to be that closest to $\x$, $I,J \overset{\iid}{\sim} \uniform\{0,\dots,m_c\}$, $C := [\| \nabla\log f_c(\x^*)\|^2
+  \Delta \log f_c(\x^*)]/2$ is a constant, and where
\begin{align*}
\hat{\alpha}_{I,c}(\x) & := (m_c+1)\cdot\left[\nabla\log \ell_{I,c}(\x)-\nabla\log \ell_{I,c}(\x^*)\right],\\
\text{div\,}\hat{\alpha}_{I,c}(\x) & := (m_c+1)\cdot\left[\Delta \log \ell_{I,c}(\x)-\Delta \log \ell_{I,c}(\x^*)\right].
\end{align*}
\citet[Thm.\ 3]{jrssb:pfjr20} show that under some mild technical assumptions, and where control variates as described above are used, then in the regular setting where the sub-posteriors contract at the rate ${m_c^{-1 /2 }}$, that $K'/K$ grows with data size like $\bigO(1)$. They also consider scaling under different contraction rates, and the effect if the control variates are not as per the guidance above. Note that although the unbiased estimator $\hat{\phi}_c$ indicated above uses only two draws from \eqref{eq:bigdatalike}, it may be worthwhile using multiple draws (sampled with replacement) as the variance of the estimator ultimately impacts the stability of the particle set weights in \algref{alg:bf} (as discussed in \apxref{apx:ue}). Further note that conveniently the constant $C$ in \eqref{eq:phichat2} does not need to be computed in \algref{alg:bf} as it forms part of the normalisation constant.  


\bibliographystyle{chicago}
\bibliography{./Sections/bibliography1}

\begin{thebibliography}{}

\bibitem[\protect\citeauthoryear{Agarwal and Duchi}{Agarwal and
  Duchi}{2011}]{neurips:ad11}
Agarwal, A. and J.~Duchi (2011).
\newblock {Distributed delayed stochastic optimization}.
\newblock In {\em Advances in Neural Information Processing Systems}, pp.\
  873--881.

\bibitem[\protect\citeauthoryear{Andrieu and Roberts}{Andrieu and
  Roberts}{2009}]{as:ar09}
Andrieu, C. and G.~Roberts (2009).
\newblock {The pseudo-marginal approach for efficient Monte Carlo
  computations}.
\newblock {\em Annals of Statistics\/}~{\em 37}, 697--725.

\bibitem[\protect\citeauthoryear{Bache and Lichman}{Bache and
  Lichman}{2013}]{uci:bl13}
Bache, K. and M.~Lichman (2013).
\newblock {\em {UCI Machine Learning Repository}}.
\newblock Irvine CA: University of California, School of Information and
  Computer Science.

\bibitem[\protect\citeauthoryear{Berger}{Berger}{1980}]{bk:b80}
Berger, J. (1980).
\newblock {\em {Statistical decision theory and Bayesian analysis}}.
\newblock Springer.

\bibitem[\protect\citeauthoryear{Beskos, Papaspiliopoulos, and Roberts}{Beskos
  et~al.}{2008}]{mcap:bpr08}
Beskos, A., O.~Papaspiliopoulos, and G.~Roberts (2008).
\newblock {A factorisation of diffusion measure and finite sample path
  constructions}.
\newblock {\em Methodology and Computing in Applied Probability\/}~{\em 10},
  85--104.

\bibitem[\protect\citeauthoryear{Beskos, Papaspiliopoulos, Roberts, and
  Fearnhead}{Beskos et~al.}{2006}]{jrssb:bprf06}
Beskos, A., O.~Papaspiliopoulos, G.~Roberts, and P.~Fearnhead (2006).
\newblock {Exact and computationally efficient likelihood-based estimation for
  discretely observed diffusion processes (with discussion)}.
\newblock {\em Journal of the Royal Statistical Society, Series B (Statistical
  Methodology)\/}~{\em 68\/}(3), 333--382.

\bibitem[\protect\citeauthoryear{Bolic, Djuric, and Hong}{Bolic
  et~al.}{2005}]{ieee:bdh05}
Bolic, M., P.~Djuric, and S.~Hong (2005).
\newblock {Resampling algorithms and architectures for distributed particle
  filters}.
\newblock {\em IEEE Transactions on Signal Processing\/}~{\em 53\/}(7),
  2442--2450.

\bibitem[\protect\citeauthoryear{Bottou}{Bottou}{2010}]{compstat:b10}
Bottou, L. (2010).
\newblock {Large-scale machine learning with stochastic gradient descent}.
\newblock In {\em Proceedings of COMPSTAT'2010}, pp.\  177--186. Springer.

\bibitem[\protect\citeauthoryear{Buchholz, Ahfock, and Richardson}{Buchholz
  et~al.}{2019}]{arxiv:bar20}
Buchholz, A., D.~Ahfock, and S.~Richardson (2019).
\newblock {Distributed Computation for Marginal Likelihood based Model Choice}.
\newblock arXiv e-prints, arXiv:1910.04672.

\bibitem[\protect\citeauthoryear{Carpenter, Clifford, and Fearnhead}{Carpenter
  et~al.}{1999}]{ieeprsn:ccf99}
Carpenter, J., P.~Clifford, and P.~Fearnhead (1999).
\newblock {An Improved Particle Filter for Non-linear Problems}.
\newblock {\em IEEE Proceedings - Radar, Sonar and Navigation\/}~{\em 146},
  2--7.

\bibitem[\protect\citeauthoryear{Chan, Johansen, Pollock, and Roberts}{Chan
  et~al.}{2021}]{chan:poll:rob:21}
Chan, R., A.~Johansen, M.~Pollock, and G.~Roberts (2021).
\newblock {Hierarchical Monte Carlo Fusion}.
\newblock In preparation.

\bibitem[\protect\citeauthoryear{Crisan, M{\'\i}guez, and
  R{\'\i}os-Mu{\~n}oz}{Crisan et~al.}{2018}]{aisp:cmr18}
Crisan, D., J.~M{\'\i}guez, and G.~R{\'\i}os-Mu{\~n}oz (2018).
\newblock {On the performance of parallelisation schemes for particle
  filtering}.
\newblock {\em EURASIP Journal on Advances in Signal Processing\/}~{\em
  2018\/}(1), 31.

\bibitem[\protect\citeauthoryear{Dai, Pollock, and Roberts}{Dai
  et~al.}{2019}]{jap:dpr19}
Dai, H., M.~Pollock, and G.~Roberts (2019).
\newblock {Monte Carlo Fusion}.
\newblock {\em Journal of Applied Probability\/}~{\em 56}, 174--191.

\bibitem[\protect\citeauthoryear{Douc, Capp\'{e}, and Moulines}{Douc
  et~al.}{2005}]{tr:dcm05}
Douc, R., O.~Capp\'{e}, and E.~Moulines (2005, September).
\newblock {Comparison of resampling schemes for particle filtering}.
\newblock In {\em 4th International Symposium on Image and Signal Processing
  and Analysis (ISPA)}, Zagreb, Croatia.

\bibitem[\protect\citeauthoryear{Doucet, de~Freitas, and Gordon}{Doucet
  et~al.}{2001}]{bk:smcinp}
Doucet, A., N.~de~Freitas, and N.~Gordon (2001).
\newblock {\em {Sequential Monte Carlo Methods in Practice}\/} (1st ed.).
\newblock Springer.

\bibitem[\protect\citeauthoryear{Doucet and Lee}{Doucet and
  Lee}{2018}]{bk:hgm:dl18}
Doucet, A. and A.~Lee (2018).
\newblock {Sequential Monte Carlo Methods}.
\newblock In M.~Maathuis, M.~Drton, S.~Lauritzen, and M.~Wainwright (Eds.),
  {\em {Handbook of Graphical Models}}, Chapter~7, pp.\  165--189. CRC Press.

\bibitem[\protect\citeauthoryear{Fearnhead, Papaspiliopoulos, and
  Roberts}{Fearnhead et~al.}{2008}]{jrssb:fpr08}
Fearnhead, P., O.~Papaspiliopoulos, and G.~Roberts (2008).
\newblock {Particle filters for partially-observed diffusions}.
\newblock {\em Journal of the Royal Statistical Society, Series B (Statistical
  Methodology)\/}~{\em 70\/}(4), 755--777.

\bibitem[\protect\citeauthoryear{Fleiss}{Fleiss}{1993}]{smmr:f93}
Fleiss, J. (1993).
\newblock {Review papers: The statistical basis of meta-analysis}.
\newblock {\em Statistical methods in medical research\/}~{\em 2\/}(2),
  121--145.

\bibitem[\protect\citeauthoryear{Genest and Zidek}{Genest and
  Zidek}{1986}]{ss:gz86}
Genest, C. and J.~Zidek (1986).
\newblock {Combining probability distributions: A critique and an annotated
  bibliography}.
\newblock {\em Statistical Science\/}~{\em 1\/}(1), 114--135.

\bibitem[\protect\citeauthoryear{Gordon, Salmond, and Smith}{Gordon
  et~al.}{1993}]{ieeeprsp:gss93}
Gordon, N., J.~Salmond, and A.~Smith (1993).
\newblock {A novel approach to nonlinear/non-Gaussian Bayesian state
  estimation}.
\newblock {\em IEEE Proceedings on Radar and Signal Processing\/}~{\em 140},
  107--113.

\bibitem[\protect\citeauthoryear{Goudie, Presanis, Lunn, De~Angelis, and
  Wernisch}{Goudie et~al.}{2019}]{ba:gpldw19}
Goudie, R., A.~Presanis, D.~Lunn, D.~De~Angelis, and L.~Wernisch (2019).
\newblock {Joining and splitting models with Markov melding}.
\newblock {\em Bayesian analysis\/}~{\em 14\/}(1), 81.

\bibitem[\protect\citeauthoryear{gov.uk}{gov.uk}{2019}]{gov:road_safety_data}
gov.uk (2019).
\newblock {`Road Safety Data' dataset, Department for Transport, U.K.
  Government}.
\newblock
  \url{https://data.gov.uk/dataset/cb7ae6f0-4be6-4935-9277-47e5ce24a11f/road-safety-data}.
\newblock Update Version: 2019-12-17. Accessed: 2020-09-17.

\bibitem[\protect\citeauthoryear{Heine and Whiteley}{Heine and
  Whiteley}{2017}]{spa:hw17}
Heine, K. and N.~Whiteley (2017).
\newblock {Fluctuations, stability and instability of a distributed particle
  filter with local exchange}.
\newblock {\em Stochastic Processes and their Applications\/}~{\em 127\/}(8),
  2508--2541.

\bibitem[\protect\citeauthoryear{Higuchi}{Higuchi}{1997}]{jcgs:h97}
Higuchi, T. (1997).
\newblock {Monte Carlo filter using the genetic algorithm operators}.
\newblock {\em Journal of Computational and Graphical Statistics\/}~{\em
  59\/}(1), 1--23.

\bibitem[\protect\citeauthoryear{Jin, Netrapalli, and Jordan}{Jin
  et~al.}{2017}]{arxiv:jnj17}
Jin, C., P.~Netrapalli, and M.~Jordan (2017).
\newblock {Accelerated gradient descent escapes saddle points faster than
  gradient descent}.
\newblock arXiv e-prints, arXiv:1711.10456.

\bibitem[\protect\citeauthoryear{Jordan, Lee, and Yang}{Jordan
  et~al.}{2018}]{jasa:jly18}
Jordan, M., J.~Lee, and Y.~Yang (2018).
\newblock {Communication-efficient distributed statistical inference}.
\newblock {\em Journal of the American Statistical Association\/}~{\em
  114\/}(526), 668--681.

\bibitem[\protect\citeauthoryear{Kitagawa}{Kitagawa}{1996}]{jcgs:k96}
Kitagawa, G. (1996).
\newblock {Monte Carlo Filter and Smoother for Non-Gaussian Nonlinear State
  Space Models}.
\newblock {\em Journal of Computational and Graphical Statistics\/}~{\em
  5\/}(1), 1--25.

\bibitem[\protect\citeauthoryear{Kong, Liu, and Wong}{Kong
  et~al.}{1994}]{jasa:klw94}
Kong, A., J.~Liu, and W.~Wong (1994).
\newblock {Sequential Imputations and Bayesian Missing Data Problems}.
\newblock {\em Journal of the American Statistical Association\/}~{\em
  89\/}(425), 278--288.

\bibitem[\protect\citeauthoryear{Lee and Whiteley}{Lee and
  Whiteley}{2016}]{sadm:lw16}
Lee, A. and N.~Whiteley (2016).
\newblock {Forest resampling for distributed sequential Monte Carlo}.
\newblock {\em Statistical Analysis and Data Mining: The ASA Data Science
  Journal\/}~{\em 9\/}(4), 230--248.

\bibitem[\protect\citeauthoryear{Lee, Yau, Giles, Doucet, and Holmes}{Lee
  et~al.}{2010}]{jcgs:lygdh10}
Lee, A., C.~Yau, M.~Giles, A.~Doucet, and C.~Holmes (2010).
\newblock {On the utility of graphics cards to perform massively parallel
  simulation of advanced Monte Carlo methods}.
\newblock {\em Journal of Computational and Graphical Statistics\/}~{\em
  19\/}(4), 769--789.

\bibitem[\protect\citeauthoryear{Lindsten, Johansen, Naesseth, Kirkpatrick,
  Sch{\"o}n, Aston, and Bouchard-C{\^o}t{\'e}}{Lindsten
  et~al.}{2017}]{jcgs:dc17}
Lindsten, F., A.~Johansen, C.~Naesseth, B.~Kirkpatrick, T.~Sch{\"o}n, J.~Aston,
  and A.~Bouchard-C{\^o}t{\'e} (2017).
\newblock {Divide-and-conquer with Sequential Monte Carlo}.
\newblock {\em Journal of Computational and Graphical Statistics\/}~{\em
  26\/}(2), 445--458.

\bibitem[\protect\citeauthoryear{Liu and Chen}{Liu and Chen}{1998}]{jasa:lc98}
Liu, J. and R.~Chen (1998).
\newblock {Sequential Monte Carlo Methods for Dynamic Systems}.
\newblock {\em Journal of the American Statistical Association\/}~{\em
  93\/}(443), 1032--1044.

\bibitem[\protect\citeauthoryear{Minsker, Srivastava, Lin, and Dunson}{Minsker
  et~al.}{2014}]{icml:ssld14}
Minsker, S., S.~Srivastava, L.~Lin, and D.~Dunson (2014).
\newblock {Scalable and Robust Bayesian Inference via the Median Posterior}.
\newblock In E.~Xing and T.~Jebara (Eds.), {\em Proceedings of the 31st
  International Conference on Machine Learning}, Volume~32, pp.\  1656--1664.
  PMLR.

\bibitem[\protect\citeauthoryear{Murray, Lee, and Jacob}{Murray
  et~al.}{2016}]{jcgs:mlj16}
Murray, L., A.~Lee, and P.~Jacob (2016).
\newblock {Parallel resampling in the particle filter}.
\newblock {\em Journal of Computational and Graphical Statistics\/}~{\em
  25\/}(3), 789--805.

\bibitem[\protect\citeauthoryear{Neiswanger, Wang, and Xing}{Neiswanger
  et~al.}{2013}]{arxiv:n13}
Neiswanger, W., C.~Wang, and E.~Xing (2013).
\newblock {Asymptotically Exact, Embarrassingly Parallel MCMC}.
\newblock arXiv e-prints, arXiv:1311.4780.

\bibitem[\protect\citeauthoryear{Nesterov}{Nesterov}{2013}]{bk:nesterov}
Nesterov, Y. (2013).
\newblock {\em {Introductory lectures on convex optimization: A basic course}},
  Volume~87.
\newblock Springer Science \& Business Media.

\bibitem[\protect\citeauthoryear{Papaspiliopoulos, Roberts, and
  Taylor}{Papaspiliopoulos et~al.}{2016}]{aap:prt16}
Papaspiliopoulos, O., G.~Roberts, and K.~Taylor (2016).
\newblock {Exact sampling of diffusions with a discontinuity in the drift}.
\newblock {\em Advances in Applied Probability\/}~{\em 48\/}(A), 249.

\bibitem[\protect\citeauthoryear{Pollock}{Pollock}{2013}]{phd:p13}
Pollock, M. (2013).
\newblock {\em {Some Monte Carlo Methods for Jump Diffusions}}.
\newblock Ph.\ D. thesis, Department of Statistics, University of Warwick.

\bibitem[\protect\citeauthoryear{Pollock, Fearnhead, Johansen, and
  Roberts}{Pollock et~al.}{2020}]{jrssb:pfjr20}
Pollock, M., P.~Fearnhead, A.~Johansen, and G.~Roberts (2020).
\newblock {Quasi-stationary Monte Carlo methods and the ScaLE algorithm (with
  discussion)}.
\newblock {\em Journal of the Royal Statistical Society, Series B (Statistical
  Methodology)\/}~{\em 82}, 1--59.

\bibitem[\protect\citeauthoryear{Pollock, Johansen, and Roberts}{Pollock
  et~al.}{2016}]{b:pjr16}
Pollock, M., A.~Johansen, and G.~Roberts (2016).
\newblock {On the exact and $\varepsilon$-strong simulation of (jump)
  diffusions}.
\newblock {\em Bernoulli\/}~{\em 22\/}(2), 794--856.

\bibitem[\protect\citeauthoryear{Rendell, Johansen, Lee, and Whiteley}{Rendell
  et~al.}{2018}]{jcgs:rjlw20}
Rendell, L., A.~Johansen, A.~Lee, and N.~Whiteley (2018).
\newblock {Global Consensus Monte Carlo}.
\newblock arXiv e-prints, arXiv:1807.09288.

\bibitem[\protect\citeauthoryear{Rogers and Williams}{Rogers and
  Williams}{2000}]{rogers2000diffusions}
Rogers, L. and D.~Williams (2000).
\newblock {\em {Diffusions, Markov processes and martingales: Volume 2, It{\^o}
  calculus}}, Volume~2.
\newblock Cambridge University Press.

\bibitem[\protect\citeauthoryear{Scott}{Scott}{2017}]{bjps:s17}
Scott, S. (2017).
\newblock {Comparing consensus Monte Carlo strategies for distributed Bayesian
  computation}.
\newblock {\em Brazilian Journal of Probability and Statistics\/}~{\em
  31\/}(4), 668--685.

\bibitem[\protect\citeauthoryear{Scott, Blocker, Bonassi, Chipman, George, and
  McCulloch}{Scott et~al.}{2016}]{ijmsem:setal16}
Scott, S., A.~Blocker, F.~Bonassi, H.~Chipman, E.~George, and R.~McCulloch
  (2016).
\newblock {Bayes and big data: the consensus Monte Carlo algorithm}.
\newblock {\em International Journal of Management Science and Engineering
  Management\/}~{\em 11\/}(2), 78--88.

\bibitem[\protect\citeauthoryear{Smith, Spiegelhalter, and Thomas}{Smith
  et~al.}{1995}]{sim:sst95}
Smith, T., D.~Spiegelhalter, and A.~Thomas (1995).
\newblock {Bayesian approaches to random-effects meta-analysis: a comparative
  study}.
\newblock {\em Statistics in Medicine\/}~{\em 14\/}(24), 2685--2699.

\bibitem[\protect\citeauthoryear{Srivastava, Cevher, Tan-Dinh, and
  Dunson}{Srivastava et~al.}{2016}]{aistats:sctd16}
Srivastava, S., V.~Cevher, Q.~Tan-Dinh, and D.~Dunson (2016).
\newblock Wasp: Scalable bayes via barycenters of subset posteriors.
\newblock In {\em Artificial Intelligence and Statistics}, pp.\  912--920.

\bibitem[\protect\citeauthoryear{Stamatakis and Aberer}{Stamatakis and
  Aberer}{2013}]{ieee:sa13}
Stamatakis, A. and A.~Aberer (2013).
\newblock {Novel Parallelization Schemes for Large-Scale Likelihood-based
  Phylogenetic Inference}.
\newblock In {\em 2013 IEEE 27th International Symposium on Parallel and
  Distributed Processing}, pp.\  1195--1204.

\bibitem[\protect\citeauthoryear{Verg{\'e}, Dubarry, Moral, and
  Moulines}{Verg{\'e} et~al.}{2015}]{sc:vdmm15}
Verg{\'e}, C., C.~Dubarry, P.~Moral, and E.~Moulines (2015).
\newblock {On parallel implementation of sequential Monte Carlo methods: the
  island particle model}.
\newblock {\em Statistics and Computing\/}~{\em 25}, 243--260.

\bibitem[\protect\citeauthoryear{Vono, Dobigeon, and Chainais}{Vono
  et~al.}{2019}]{ieee:vdc19}
Vono, M., N.~Dobigeon, and P.~Chainais (2019).
\newblock Split-and-augmented gibbs sampler—application to large-scale
  inference problems.
\newblock {\em IEEE Transactions on Signal Processing\/}~{\em 67\/}(6),
  1648--1661.

\bibitem[\protect\citeauthoryear{Wang, Pollock, Roberts, and Steinsaltz}{Wang
  et~al.}{2019}]{aap:wprs20}
Wang, A., M.~Pollock, G.~Roberts, and D.~Steinsaltz (2019).
\newblock {Regeneration-enriched Markov processes with application to Monte
  Carlo}.
\newblock arXiv e-prints, arXiv:1910.05037.

\bibitem[\protect\citeauthoryear{Wang and Dunson}{Wang and
  Dunson}{2013}]{arxiv:wd13}
Wang, X. and D.~Dunson (2013).
\newblock {Parallelizing MCMC via Weierstrass Sampler}.
\newblock arXiv e-prints, arXiv:1312.4605.

\bibitem[\protect\citeauthoryear{Wang, Guo, Heller, and Dunson}{Wang
  et~al.}{2015}]{neurips:wghd15}
Wang, X., F.~Guo, K.~Heller, and D.~Dunson (2015).
\newblock {Parallelizing MCMC with random partition trees}.
\newblock In {\em Advances in Neural Information Processing Systems}, pp.\
  451--459.

\bibitem[\protect\citeauthoryear{Xu, Lakshminarayanan, Teh, Zhu, and Zhang}{Xu
  et~al.}{2014}]{neurips:xltzz14}
Xu, M., B.~Lakshminarayanan, Y.~Teh, J.~Zhu, and B.~Zhang (2014).
\newblock {Distributed Bayesian Posterior Sampling via Moment Sharing}.
\newblock In {\em Advances in Neural Information Processing Systems}, pp.\
  3356--3364.

\bibitem[\protect\citeauthoryear{Xue and Liang}{Xue and Liang}{2019}]{sc:xl19}
Xue, J. and F.~Liang (2019).
\newblock {Double-parallel Monte Carlo for Bayesian analysis of big data}.
\newblock {\em Statistics and Computing\/}~{\em 29}, 23--32.

\bibitem[\protect\citeauthoryear{Y{\i}ld{\i}r{\i}m and
  Ermi{\c{s}}}{Y{\i}ld{\i}r{\i}m and Ermi{\c{s}}}{2019}]{sc:ye19}
Y{\i}ld{\i}r{\i}m, S. and B.~Ermi{\c{s}} (2019).
\newblock {Exact MCMC with differentially private moves}.
\newblock {\em Statistics and Computing\/}~{\em 29\/}(5), 947--963.

\bibitem[\protect\citeauthoryear{Zhou, Johansen, and Aston}{Zhou
  et~al.}{2016}]{jcgs:zja16}
Zhou, Y., A.~Johansen, and J.~Aston (2016).
\newblock {Toward Automatic Model Comparison: An Adaptive Sequential Monte
  Carlo Approach}.
\newblock {\em Journal of Computational and Graphical Statistics\/}~{\em
  25\/}(3), 701--726.

\end{thebibliography}


\end{document}